\documentclass[12pt]{iopart}
\usepackage{graphicx}
\newcommand{\hc}{{\rm h.c.}}
\newcommand{\cc}{{\rm c.c.}}
\newcommand{\ie}{\emph{i.e.}\,}
\newcommand{\eg}{\emph{e.g.}\,}
\newcommand{\barz} {\bar{z}}
\newcommand{\baru} {\bar{u}}
\newcommand{\barw} {\bar{w}}

\newcommand{\barv} {\bar{v}}
\newcommand{\barpart}{\bar{\partial}}

\newcommand{\figwidth}{0.6\columnwidth}
\begin{document}
\title[Quasi-1D interacting Bose gases]{Interacting Bose gases in quasi-one dimensional optical lattices}
\author{M A Cazalilla$^1$, A F Ho$^{2,3}$, and T Giamarchi$^4$}
\address{$^1$ Donostia International Physics Center (DIPC), \\
Manuel de Lardizabal 4, 20018 San Sebastian, Spain}
\ead{waxcagum@sq.ehu.es}
\address{$^2$ School of Physics and Astronomy, The University of Birmingham,\\
Edgbaston, Birmingham B15 2TT, UK, \\
$^3$ CMTH Group, Dept. Physics, Blackett Laboratory, \\
Imperial College, Prince Consort Road, London SW7 2BW, UK}
\ead{andrew.f.ho@imperial.ac.uk}
\address{$^4$ University of Geneva, 24 Quai Enerst-Ansermet, \\
CH-1211 Geneva 4, Switzerland}
\ead{Thierry.Giamarchi@physics.unige.ch}
\date{\today}
\begin{abstract}

We study a two-dimensional array of coupled 
one-dimensional (1D)  tubes of interacting bosons. Such systems can be produced by loading  ultra-cold atoms in  anisotropic optical lattices. We investigate the effects of coupling the tubes via hopping of the bosons (\ie Josephson coupling). In the absence of a periodic potential along the tubes, or when such potential is incommensurate with the
boson density, the system undergoes a transition from an array of incoherentTomonaga-Luttinger liquids at high temperature to an  anisotropic Bose-Einstein condensate (BEC), at low temperature. We determine the transition temperature and long wave-length excitations of the BEC.  In addition to the usual gapless (Goldstone) mode found in standard  superfluids, we also find a gapped mode associated with fluctuations of the amplitude of the order parameter. When a commensurate periodic potential is applied along the tubes, they can become 1D Mott insulators. Intertube hopping leads to a deconfinement quantum phase  transition between the 1D  Mott insulators and the  anisotropic BEC.  We also take into account the finite size of the gas tubes as realized in actual experiments. We map out the phase diagram of the quasi-1D lattice and compare our results with the existing experiments on such systems.

\end{abstract}
\submitto{\NJP}
\maketitle
\section{Introduction}

One dimension (1D) provides fertile grounds for studying the physics
of strongly correlated quantum many-body systems. It is a well
established theoretical result that, because of the enhanced role of
quantum fluctuations in low dimensionality, there is  no broken
continuous symmetry  and therefore long range order, even at zero
temperature, in a 1D quantum many body system. Instead,  in a large
class of systems possessing a gapless spectrum, power laws
characterize the various correlation functions. This class of
systems is known as a Tomonaga-Luttinger liquid (TLL).

One of the simplest examples of a TLL is the Lieb-Liniger
model~\cite{lieb_bosons_1D}, which describes a system of
non-relativistic bosons interacting via a contact (Dirac
delta-function) potential. This model has been solved exaclty using
the Bethe ansatz~\cite{lieb_bosons_1D}, which has provided valuable
insights into its thermodynamic properties and  excitation spectrum.
However, calculation of the asymptotics of correlation functions is
not an easy task using this method, and in this regard it is nicely
complemented by the harmonic-fluid approach introduced by
Haldane~\cite{haldane_bosons}. The combination of these two methods
has allowed us to understand that, as the the interaction is
increased, the system exhibits no quantum phase transition. Instead,
it smoothly crosses over from a weakly interacting (often called
quasi-condensate~\cite{popov_functional_book,tlho_low_d_bosons,petrov_trapped_bosons})
regime, where phase fluctuations decay very slowly, to a strongly
interacting regime, the Tonks regime~\cite{girardeau_tonks_gas}. In
the latter,  repulsion between the bosons is so strong  that it
leads to an effective exclusion principle in position space, and
makes the system resemble  a gas of non-interacting spinless
fermions in many properties. Thus it is sometimes stated that the
bosons in the Tonks regime ``fermionize''.

This very simple model solved by Lieb and Liniger in 1963 has
recently received a great deal of attention thanks to spectacular
advances in confining quantum degenerate gases of alkali atoms in
low dimensional traps~\cite{Moritz_oscillations}. While at the
beginning only weakly interacting 1D Bose gases were 
available~\cite{gorlitz_lowD_bosegas,greiner_2dlattice},
the application of a deep optical potential parallel to
the 1D axis to arrays of 1D Bose gases has recently allowed the group
at ETH (Z\"urich)~\cite{stoferle_tonks_optical,koehl_SFMI} 
to demonstrate a quantum phase
transition from a superfluid TLL to a 1D Mott insulator, which is
the low dimensional analog of the superfluid to Mott insulator
transition first demonstrated in the 3D optical lattice by Greiner
and coworkers~\cite{greiner_mott_bec}. 
Furthermore, Kinoshita \emph{et al.}~\cite{kinoshita_tonks_continuous} 
have succeeded in reaching the
Tonks regime by creating an array of  tightly confined 1D Bose
gases. A different path~\cite{cazalilla_hcbose_gases,cazalilla_tonks_gases} 
to realize a strongly interacting 1D
Bose gases has been followed by Paredes \emph{et
al.}~\cite{paredes_tonks_optical}, who applied a deep periodic
potential to an array of (otherwise weakly interacting) 1D Bose
gases. Thus they were able to dramatically  increase the ratio of  the
interaction to the kinetic energy of the atoms. In both cases~\cite{kinoshita_tonks_continuous,paredes_tonks_optical}, the
observation of   fermionization signatures was reported.

In the past, much experimental effort  focusing on low
dimensional systems has been spent on studying electronic systems
such as spin chains and quasi-1D metals (for a review see
\cite{giamarchi_book_1d}). Theoretically at least, electrons in 1D
become a TLL (provided there are no gap-opening perturbations) as soon as
interactions are taken into account in a non-perturbative way.
However, for the electronic systems comparison between theory and
experiment has proved problematic partly because of
material-specific issues such as the existence of many competing
phenomena at similar energy scales (\eg in Bechgaard salts
\cite{jerome_review_chemrev,giamarchi_review_chemrev}) or the
always-present disorder in solid state
systems~\cite{giamarchi_book_1d}. Furthermore, one major
complication stems from the fact that these materials are indeed
three dimensional (3D), as there is always a small amplitude for
electrons to  hop from chain to chain,  whereas Coulomb interactions
(often poorly screened because of bad metallicity) also enhance 3D
ordering tendencies. These inter-chain couplings cannot be neglected
at low temperatures, which leads to the interesting physics of
\emph{dimensional crossover}: at temperatures high compared to the
energy scale determined by these inter-chain couplings, the system
effectively behaves as a collection of isolated 1D chains. However,
when the temperature is lowered below this scale, 3D coherence
develops between the chains and, at zero temperature, a complete  3D
description of the system is required. Much exciting physics is
expected to occur in this crossover from the exotic 1D TLL  state to
a conventional 3D (but strongly anisotropic) Fermi liquid, or an
ordered 3D phase~\cite{giamarchi_book_1d}. Indeed, these phenomena
may well have relevance to the pseudogap (and other anomalous)
behavior of the quasi-2D high temperature superconductor materials.
For instance, one of the many interesting questions that arise in
this context  is  the following: how \cite{giamarchi_review_chemrev}
do the exotic low-energy excitations (bosonic collective modes) of
the 1D system transform into the Landau quasi-particles
characteristic of Fermi liquids and which resemble more the
constituent electrons?

The dimensional crossover phenomena described above has also a
counterpart in Bose gases loaded in quasi-1D optical lattices. From a
theoretical point of view, the fact that bosons can undergo
Bose-Einstein condensation and, therefore, can be collectively
described using a macroscopic quantum field makes more amenable
these systems to a theoretical analysis relying on a mean field
approximation. Such a mean-field analysis is undertaken in this
work: here we treat the intertube hopping via a mean field approximation,
while quantum fluctuations within a tube are treated accurately at low
energies using the harmonic-fluid approach.

 From the experimental point of view, the recent availability
of gases of ultra-cold bosonic atoms  in optical lattices has
revived the interest in understanding the properties of coupled
systems of bosons in 1D. Furthermore, the already demonstrated  high
degree of tunability of these systems can provide  a clean setup to
study the physics of  dimensional crossover. Thus, as we show
in~\sref{sec:nomott} of this work, the phenomenology of
Bose-Einstein condensation in the most anisotropic version of these
lattices is very different from the one exhibited by Bose gases
condensing in the absence of a lattice or even in shallow isotropic
optical lattices. 

  In addition to studying the properties of Bose condensed systems in
lattices, some experimental groups  have applied  a tunable optical
potential  along the longitudinal direction of an array of
effectively uncoupled 1D Bose gases. It has been thus
possible~\cite{stoferle_tonks_optical}, as mentioned above, to drive
a quantum phase transition  from the 1D superfluid (TLL) state to a
1D Mott insulator. Under these circumstances, the effect of a small
hopping  amplitude (\ie Josephson coupling) between neighboring tubes in the
array becomes most interesting: as we show in~\sref{sec:mott}, for a
finite value of the hopping amplitude, the
system undergoes a new type of quantum phase transition,  known as
``deconfinement'' transition,  from the 1D Mott insulator to an
anisotropic Bose-Einstein condensate. Thus  the tendency of 
bosons to develop phase coherence overcomes the localization effect
of the  optical potential. As explained in \sref{sec:comparison},
we believe that some features of this deconfinement  transition may
have been observed in the experiments of
Ref.~\cite{stoferle_tonks_optical}.

This paper is organized as follows: \sref{sec:model} defines our
model of a 2D array of 1D tubes of bosons coupled by weak intertube tunneling,
 and describes the harmonic-fluid approach that allows us to
derive a low-energy effective Hamiltonian. Then we discuss
separately the phenomenon of Bose-Einstein condensation and the
properties of the condensate in an anisotropic lattice
(\sref{sec:nomott}), and the deconfinement quantum phase transition
(\sref{sec:mott}). In the former case, we obtain the
zero-temperature condensate fraction, the condensation temperature,
and the excitations of the condensate using the mean field
approximation  and including gaussian fluctuations about the mean
field state. We also show that  some of the results obtained by
these methods are recovered from the self-consistent harmonic
approximation (\sref{sec:varap}). These calculations are described
in pedagogical detail in  \ref{app:rpa}, \ref{app:finitechi}, and
\ref{app:variat}. Moreover, in \sref{sec:exbec} we discuss the
excitation spectrum of the Bose-condensed phase, and in particular
the existence of a gapped mode related to oscillations of the
amplitude of the condensate fraction, in the light of a
phenomenological Ginzburg-Landau theory. Differences between
this theory and the more conventional Gross-Pitaevskii theory, where
the gapped mode is absent, are exhibited. The transition from
the (anisotropic) regime described by the Ginzburg-Landau theory to
the one where Gross-Pitaeskii theory holds is also outlined there.

The deconfinement transition is described in \sref{sec:mott}. There
we show how the renormalization group can be used  to understand the
competition between the localization in the optical potential and a
small tunneling amplitude between neighboring 1D Bose systems. To
assess the nature of the deconfinement transition, we map (see
\sref{sec:mftdecon}) the mean-field Hamiltonian,  at a specific
value of the system's parameters,   to a  spin-chain model.  Finally,
\sref{sec:finitesize} discusses the effect of the finite size of the
1D system that are realized in actual experiments, and how it
modifies the phase diagram obtained from the renormalization-group
analysis.  We  argue  that, for an array of finite 1D Bose gases,
there is a extra energy scale below which the tunneling of atoms
between tubes is effectively blocked.  Finally, in
\sref{sec:comparison}  we present a discussion of the experimental
observations and how they compare with the predictions of our
theory. A brief account of some of our results has appeared
elsewhere~\cite{ho_deconfinement_coldatoms}.

\section{Models and Bosonization} \label{sec:model}

In this section we introduce the models for an interacting Bose gas in 
a quasi-1D optical lattice  
that we study in the rest of the paper.
We also briefly describe the harmonic-fluid approach (henceforth
referred to  as ``bosonization"),  which allows us to obtain
a convenient effective low-temperature description of these models.

\subsection{Models of a Bose gas in a quasi-1D optical lattice}

\begin{figure}
\begin{center}
\includegraphics[width=\figwidth]{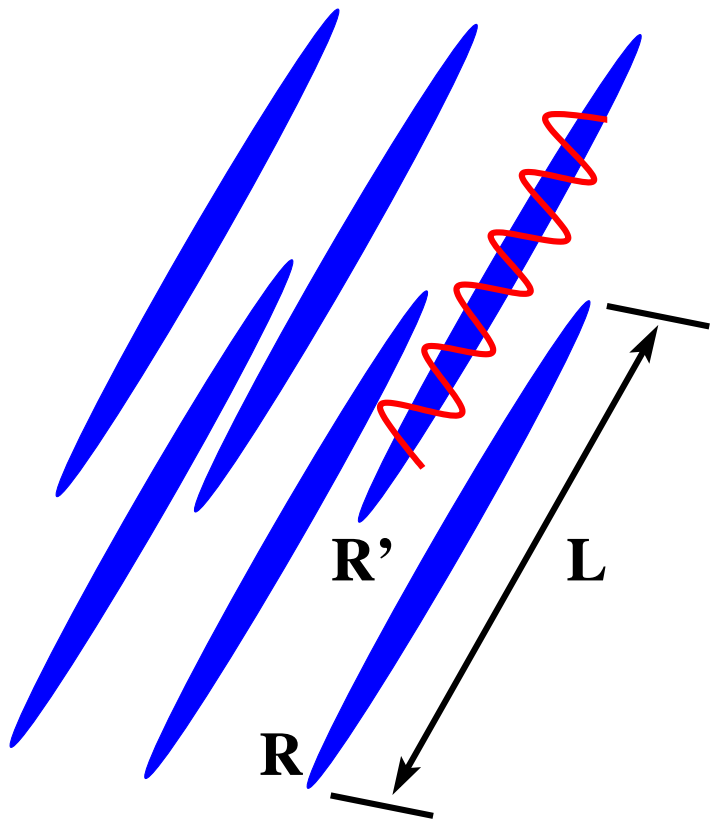}
\end{center}
 \caption{Sketch of a two-dimensional array of coupled tubes
 (also called quasi-1D optical lattice in this work). Bosons can hop from tube
 to tube with amplitude $J$, which can be controlled by the
 height of an  optical potential transverse to the tube axis. Along the tube, a periodic
 optical potential (for clarity, shown here for only one of the tubes) can be also applied. 
 This potential can drive a quantum phase transition where each tube transforms 
 into a  1D Mott insulator.}
 \label{arraydia}
\end{figure}

 In this work we study the  low-temperature equilibrium properties
of  systems  recently realized by
several experimental groups~\cite{Moritz_oscillations,kinoshita_tonks_continuous,stoferle_tonks_optical,
paredes_tonks_optical, Munich_experiment,NIST_experiment,koehl_nofk} 
which have succeeded in loading
ultracold atomic gases in highly anisotropic (quasi-1D)  optical lattices.
In the experiments carried out by these groups,
a Bose-Einstein condensate of ultracold alkali atoms (in particular,
$^{87}$Rb) is placed in a region of space where
two pairs of non-coherent and mutually orthogonal
counter-propagating lasers intersect. The intensity of the lasers
is increased adiabatically, and because of the
AC stark effect, the atoms experience
a periodic  potential~\cite{jaksch_bose_hubbard} $V_{\perp}(y,z) = V_{0 y}  \sin^2
\left(2\pi y/\lambda \right) +    V_{0 z} \sin^2 \left(2 \pi z/\lambda\right)$,
where usually $V_{0 y} = V_{0 z}= V_{0\perp}$, with  $V_{0\perp}
\propto E^2_0$,   $E_0$ being the maximum intensity of the
electric field of a laser whose wavelength is $\lambda$.
The strength of the optical potential, $V_{0 \perp}$,
 is frequently given in units of the recoil energy,
$E_R = \hbar^2 \pi^2/(2M a^2)$, where $a = \lambda/2$
is the  period of $V_{\perp}(y,z)$ and $M$ the atom mass
\footnote{For $^{87}$Rb, $E_R/\hbar = \alpha/\lambda^2$,
where $\alpha \simeq 14.424 \,  \mu {\rm m}^2\,  {\rm kHz}$ and
$\lambda$ is the laser wavelength in $\mu {\rm m}$. Thus, for
$\lambda = 0.826 \, \mu {\rm m}$ used in the experiments of
Ref.~\cite{stoferle_tonks_optical}, $E_R/\hbar =  21.14 \, {\rm kHz}$}.
For $V_{0\perp} \gg \mu_{3D}$, where
$\mu_{3D} \approx 4\pi \hbar^2 a_s \rho^{3D}_{0}/M$~\cite{pitaevskii_becbook}
is the chemical
potential of the BEC ($a_s$ being the s-wave scattering length of the atom-atom
interaction potential,
and $\rho^{3D}_0$ the mean atom density), the gas atoms are mainly confined to
the minima of $V_{\perp}(x,y)$,  thus forming long, effectively 1D,
gas tubes as shown in~\fref{arraydia} (there are additional harmonic  potentials along the $x,y,z$-axes that trap the atoms).   Under these conditions,
we can project the boson field operator $\Psi(x,y,z)$  onto the lowest (Bloch) band of the periodic potential,
\begin{equation}
\Psi(x,y,z) \simeq \sum_{{\bf R}} W_{\bf R}(y,z) \: \Psi_{\bf R}(x),
\end{equation}
where ${\bf R} = (m_y , m_z) \: a$, with integers $m_i = -M_{i}/2,
\ldots , +M_i/2$ ($i = y,z$), is the  location of the different potential
minima; $W_{\bf R}(x,y)$ is  the Wannier function  of the lowest
Bloch band centered around $\bf R$.  The system just described is
called a 2D (square) optical lattice. In typical experiments
$M_{y} \sim M_{z} \sim 10^2$, and thus the number of gas tubes
is of the order of a few thousand separated  half  a laser of wave-length,
$\lambda/2 \simeq 0.4 \, \mu{\rm m}$.

In addition, in several of the
experiments~\cite{stoferle_tonks_optical,NIST_experiment,paredes_tonks_optical}
an extra pair of  lasers was applied to define a longitudinal
periodic potential $V_{||}(x) = V_{0||} \sin^2\left(2\pi
x/\lambda\right) = \frac{1}{2} V_{0||} + u(x)$, and $u(x) = u_0\:
\cos (G x)$, where $u_0 = -V_{0||}/2$ and $G =4 \pi/\lambda =
2\pi/a$. Since,  for a commensurate potential, \ie when the period
$a$ matches the mean interparticle distance, the variation of the
strength of $u(x)$ is responsible for the
transition~\cite{haldane_bosons,giamarchi_mott_shortrev,giamarchi_book_1d}
to a Mott insulating state where the bosons are localized, we shall
henceforth refer to $u(x)$ as the ``Mott potential''.  In this work
we are interested in the extreme anisotropic case where the strength
of this potential $ |u_0| \sim V_{0||} \ll V_{0\perp}$. In 
experiments, this was achieved by increasing $V_{0\perp}$ so that
the motion of the atoms effectively becomes confined to a
one-dimensional tube, where they undergo only zero-point motion in
the transverse directions. This picture is equivalent to the
projection onto the lowest Bloch band described above. Just like in
three dimensional space, the atomic interactions in the tube are
still well described by a Dirac-delta pseudo-potential, $v_a(x) =
g_{1D} \: \delta(x)$, where the effective coupling $g_{1D}$ has been
obtained in Ref.~\cite{olshanii_potential_bec}:
\begin{equation}
 g_{\rm 1D} = \frac{2 \hbar^2 a_{s}}{(1- {\cal C} a_{s}/\sqrt{2}\ell_{\perp})M
 \ell^{2}_{\perp}}.
\end{equation}
The constant ${\cal C} \simeq 1.4603$,  and the oscillator length in
the transverse direction,  $\ell_{\perp} =
\left(\hbar/(M\omega_{\perp})\right)^{1/2}$ (where
$\hbar\omega_{\perp}/E_R \simeq 2 (V_{0\perp}/E_R)^{1/2}$ for  deep
lattices~\cite{greiner_mott_bec}, is the transverse oscillation
frequency).

 Taking into account the above considerations, we are
led to consider the following Hamiltonian:
\begin{eqnarray}\label{ham1}
\fl
 H = \sum_{\bf R}   \int^L_0 dx\: \left[ \frac{\hbar^2}{2M}
 \partial_x\Psi^{\dag}_{\bf R}(x) \partial_x\Psi_{\bf R}(x)
 + u(x)\rho_{\bf R}(x)
 +  \frac{1}{2}  \int^{L}_0 dx' \,
 v_a(x-x')  \rho_{\bf R}(x) \rho_{\bf R}(x') \right]  \nonumber \\
- \frac{J}{2} \sum_{\langle{\bf R},{\bf R}' \rangle} \int^L_0 dx
 \left[ \Psi^{\dag}_{\bf R}(x) \Psi_{{\bf R}'}(x) + \hc \right],
\end{eqnarray}
where $\rho_{\bf R}(x) = \Psi^{\dag}_{\bf R}(x)
\Psi_{\bf R}(x)$ is the density operator at point $x$ and
lattice site $\bf R$;  the field  operator $\Psi_{\bf R}(x)$
obeys $[\Psi_{\bf R}(x),\Psi_{\bf R'}^{\dag}(x')]
= \delta_{\bf R R'} \: \delta(x-x')$, and commutes
otherwise as corresponds to bosons.
The last term in equation~(\ref{ham1})
describes the hopping of bosons between two neighboring tubes. This
term is also known as Josephson coupling. The hopping
amplitude~\cite{Zwerger_review_Mott}:
\begin{equation}
J\simeq \frac{4 E_R}{\sqrt{\pi}}   \left( \frac{V_{0\perp}}{E_{R}} \right)^{\frac{3}{4}}
\: e^{-2 \left( V_{0\perp}/E_{R} \right)^{\frac{1}{2}}},
\end{equation}
for $V_{0\perp}\gg E_R$. We shall assume
a very deep  transverse potential
so that   $J \ll \mu_{1D}$, where
$\mu_{1D} = \mu_{1D}(\rho_0)$ is the chemical
potential in each tube, as can be obtained
from the exact solution of Lieb and Liniger~\cite{lieb_bosons_1D}
(or measured experimentally). Since
experimentally $\mu_{1D} < 0.1 \:
\hbar \omega_{\perp}$ ~\cite{Moritz_oscillations}, assuming $J\ll \mu_{1D}$ requires
at least that $V_{0\perp} > 10 \: E_ R$. Such limit allows us to
regard the system as an array of weakly coupled
1D interacting Bose gases.

In the experimental
systems, the Mott potential  $u(x)$, besides the periodic part described above,
also contains a slowly varying harmonic term that confines the atoms longitudinally.
Additionally, there are also harmonic traps in the transverse directions.
This makes the array, not only finite, but also inhomogeneous.
Whereas we shall discuss finite-size effects further below
(see \sref{sec:finitesize}), it
has proven so far  intractable in 1D (except for a
few limiting cases~\cite{Gangardt_correlations}) to deal with the
harmonic confinement
\emph{analytically} and only numerical
results are available~\cite{papenbrock_hcbosons_continuum,rigol_lattice_hcbosons,rigol_groundstate_hcbosons,
kollath_dmrg_bose_hubbard_trap,
wessel_MC_confined_bosons,wessel_MC_confined_bosons2}.
Thus, with the exception of~\sref{sec:conclusion}
where we discuss the experimental consequences of our work,
we shall neglect the effect of the harmonic potential,
and treat the system as an array of 
gas tubes of length $L$ and hard-wall box (open) boundary 
conditions~\cite{cazalilla_finitesize_luttinger,cazalilla_correlations_1d,Batchelor_1dbosegas_inbox}, each one containing $N_0 = \rho_0 L$ bosons
(in typical experiments, $N_0$ ranges from a few tens to a few
thousand particles). This makes sense, however, as
much of the interest of this work is focused
on computing the  phase diagram as well as some of
the thermodynamic properties of the
different phases, for which one is required to take the thermodynamic
limit $L\to \infty$ at constant density $\rho_0 = N_0/L$.

  Finally, when the axial potential $u(x)$ is made so deep that
 $|u_0| \gg \mu_{1D}$ (but still $|u_0| \ll V_{0\perp}$),
 it becomes  convenient to perform a further projection of  $\Psi_{\bf R}(x)$
 onto the lowest Bloch band of the Mott potential $u(x)$. Thus,
 \begin{equation}
 \Psi_{\bf R}(x) \simeq \sum_{m=1}^M w_m(x) \: b_{m {\bf R}},
 \end{equation}
 where $[b_{m {\bf R}}, b_{m' {\bf R}}] = \delta_{{\bf R}, {\bf R}'} \:
 \delta_{m,m'}$, commuting otherwise, and $w_m(x)$ is the
 Wannier orbital centered around site $m$.  This projection leads to the so-called
 Bose-Hubbard model~\cite{fisher_boson_loc}:
\begin{eqnarray}
 H_{\rm BH} &=&  \sum_{\bf R} H_{\bf R} + H_J, \label{bosehubbard} \\
 H_{\bf R} &=&  -\frac{J_{x}}{2} \sum_{\langle m, m' \rangle}
 b^{\dagger}_{m \bf R} b_{m' {\bf R}}  +
  \frac{U}{2}  \sum_{m} n^2_{m {\bf R}},\\
H_{J} &=&  - \frac{J}{2} \sum_{\langle{\bf R},{\bf R}'\rangle, m}
 \left[ b^{\dag}_{m \bf R} b_{m \bf R'} + \hc \right],
\end{eqnarray}
where $n_{m \bf R} = b^{\dagger}_{m \bf R}
b_{m \bf R}$ is the boson number operator at site $(m,{\bf R})$.
Note that $|u_0| \ll V_{0\perp}$ implies that $J_x \gg J$,
and therefore we are dealing with a very anisotropic (quasi-1D)
version of the Bose-Hubbard model.

\subsection{Bosonization and low-energy effective theory} \label{sec:bosometh}

In order to treat  the interactions in a non-perturbative way, it is
convenient to rewrite (\ref{ham1}) in terms of collective variables
that describe phase and density fluctuations. The method of
bosonization~\cite{haldane_bosons,giamarchi_book_1d,gogolin_1dbook,cazalilla_correlations_1d}
allows us to obtain such a description. The description is accurate
as long as one is interested in the low-temperature and
low-frequency properties of the system. In this subsection we shall
briefly outline this method, but we refer the reader to the vast
literature on the
subject~\cite{haldane_bosons,giamarchi_book_1d,gogolin_1dbook,cazalilla_correlations_1d}
for more technical details.

To study  the model of equation~(\ref{ham1}), we use the density-phase
representation of the  field operator. To leading order,
\begin{equation} \label{eq:psibos}
 \Psi_{\bf R}(x) =  ({\cal A}_B \rho_0)^{1/2} \: \exp\left[i\theta_{\bf R}(x)\right],
\end{equation}
where is ${\cal A}_B$ is a model dependent (\ie
non-universal) prefactor (see below).
The density operator,
\begin{equation} \label{eq:rhobos}
 \rho_{\bf R}(x) = \left[\rho_0-\frac1\pi\partial_x\phi_{\bf R}(x)\right]
 \sum_{m=\infty}^{+\infty} e^{2im[\pi \rho_0 x - \phi_{\bf R}(x)]}.
\end{equation}
The  fields $\theta_{\bf R}(x)$ and $\phi_{\bf R}(x)$ describe
collective fluctuations (phonons) of the phase and the density,
respectively~\footnote{We follow here the notations of
\cite{giamarchi_book_1d}. Note that compared to
\cite{cazalilla_correlations_1d,ho_deconfinement_coldatoms} $\phi$
and $\theta$ are exchanged and the density field, $\phi$,  in this
work differs by a minus sign from the density field, $\theta$, in
those references.}. Since the density and the phase are canonically
conjugate variables, it is possible to
show~\cite{cazalilla_correlations_1d,giamarchi_book_1d} that
$\theta_{\bf R}(x)$ and $\phi_{\bf R}(x)$ obey the commutation
relation: $[\phi_{\bf R}(x), \partial_{x'} \theta_{\bf R'}(x') ] = i
\delta_{\bf R R'} \delta(x-x')$. Furthermore, it is  assumed that
$\partial_x\phi_{\bf R}(x)$ is  small compared to the equilibrium
density $\rho_{0}$. Note that this approach does \emph{not} assume
the existence of a condensate and thus properly takes into account
the quantum and thermal fluctuations, which are dominant in 1D.
Indeed, it is the particle density and not a ``condensate'' density
that appears in the density-phase representation of
equations~(\ref{eq:psibos},\ref{eq:rhobos}) (unlike usual mean field
approaches). Furthermore, the terms of equation~(\ref{eq:rhobos}) with $m
\neq 0$ account for the discrete \emph{particle} nature of the
bosons, which is crucial for obtaining a correct  description for
the transition from the 1D superfluid phase  to the 1D Mott
insulator~\cite{haldane_bosons,giamarchi_mott_shortrev,giamarchi_book_1d,buchler_cic_bec,iucci_absorption}.

Using  equations (\ref{eq:psibos}) and (\ref{eq:rhobos}), the Hamiltonian
in  (\ref{ham1}) becomes
\begin{eqnarray}\label{ham2}
 H_{\rm eff}&=& \frac{\hbar v_s}{2\pi} \sum_{\bf R} \int^{L}_{0} dx\:
 \left[K  \left(\partial_x\theta_{\bf R}(x)\right)^2 +
 \frac1K \left(\partial_x\phi_{\bf R}(x)\right)^2 \right] \nonumber \\
 &&+ \frac{\hbar v_s g_u}{2\pi a^2} \sum_{\bf R}
 \int^{L}_{0} dx \: \cos \left(2\phi_{\bf R}(x) + \delta \pi x \right) \nonumber\\
 &&- \frac{\hbar v_s g_J}{2\pi a^2} \sum_{\langle {\bf R}, {\bf R}' \rangle} \int^{L}_{0} dx\:
 \cos\left(\theta_{\bf R}(x) - \theta_{\bf R'}(x)  \right),
\end{eqnarray}
where $\delta = (G - 2\pi \rho_0)/\pi = 2(a^{-1} - \rho_0)$
is the mismatch between the density $\rho_0$ and
the periodicity of the Mott potential. The two last
terms are related  to the Mott potential and the Josephson
coupling, respectively. We use the following dimensionless couplings  to
characterize their strength:
\begin{equation}
 g_J = \frac{2\pi J (\rho_0 a_0)^2}{\hbar v_s \rho_0} \: {\cal A}_{B},\quad
 g_u = \frac{2\pi u_0 (\rho_0 a_0)^2}{\hbar v_s \rho_0}
\end{equation}
where $a_0 \approx \hbar v_s/\mu_{1D}$ is a short-distance cut-off.

  The dimensionless parameter $K$ and the sound velocity $v_s$
depend both on the density, $\rho_0$, and on the strength of the boson-boson
interaction. Indeed, for the Lieb-Liniger model, they depend on a single dimensionless parameter, the gas parameter~\cite{lieb_bosons_1D}
$\gamma = M g_{1D}/\hbar^2 \rho_0$. Analytical results
for these parameters are only available in the small and large $\gamma$
limit (see ref.~\cite{cazalilla_correlations_1d} for results in the
intermediate regime):
\begin{eqnarray} \label{eq:asymplut}
 v_s K^{-1} &\simeq& v_{F} \left[1 - 8 \gamma^{-1}\right] \quad
 (\gamma  \gg 1), \\
 v_s K^{-1} &\simeq& v_{F} \gamma /\pi^2 \quad\quad\quad (\gamma \ll 1),
\end{eqnarray}
whereas Galilean invariance fixes the product~\cite{haldane_bosons,cazalilla_correlations_1d}:
\begin{equation}\label{galilean}
 v_s K = v_{F} = \frac{\hbar \pi \rho_{0}}{M}
 \end{equation}
Therefore,  $1\leq K < +\infty$ for the Lieb-Liniger model,
$K = + \infty$ corresponding to free bosons and
$K = 1$ to the Tonks limit. For the Lieb-Liniger
model, the formula~\cite{cazalilla_correlations_1d}:
\begin{equation}\label{interpolation}
{\cal A}_B(K) \simeq \left(\frac{K}{\pi} \right)^{\frac{1}{2K}}
\end{equation}
provides a reasonably good interpolation (accurate to
within $10\%$ of the exact results) for the non-universal
prefactor of the field operator.

 It  is worth stressing that in (\ref{ham2}) we have retained only
 terms which can  become  dominant at low temperatures.
In the language of the rernormalization group (RG) these correspond
to marginal or relevant  operators. Whenever, for reasons to be
given below, any of the terms retained in equation~(\ref{ham2}) becomes
irrelevant in the RG sense, we shall drop it and consider the
remaining terms only. We emphasize that the validity of
(\ref{ham2}) is restricted to energy scales smaller than $\min\{ T,
\mu_{1D}\}$, and therefore its analysis will allow us to determine
the nature of the ground state and low-energy excitations of the
system.

  Interestingly, the effective Hamiltonian of equation~(\ref{ham2}) also describes
the low energy properties of the anisotropic   Bose-Hubbard
model of~(\ref{bosehubbard}). This result can be derived by a lattice
version of the bosonization procedure~\cite{giamarchi_book_1d},
followed by a passage to
the continuum limit. However, the relationship of the
effective parameters $K$,  $v_s$, $g_u$, and $g_J$
to microscopic parameters   $J_x$, $J$, and $U$
is no longer easy to obtain analytically  (except for  $U$
small~\cite{giamarchi_book_1d} and
certain limits in  1D~\cite{cazalilla_tonks_gases}).  Thus, in this case
$K$, $v_s$,  $g_u$, and $g_J$ must be regarded as
phenomenological parameters, which must be extracted either
from the experiment or from a numerical calculation.

\section{Bose-Einstein Condensation} \label{sec:nomott}

 Let us first discuss the phenomenon of Bose-Einstein condensation
in anisotropic (quasi-1D) optical lattices. This  is the phase
transition from a normal Bose gas to a Bose-Einstein condensate
(BEC), which exhibits some degree of spatial coherence throughout
the lattice. Coherence arises thanks to the Josephson term, which
couples the 1D interacting Bose gases together. In the absence of
the Josephson coupling and the Mott potential (\ie for $J = 0$ and
$u(x) = 0$), the system behaves as an array of independent
TLL's~\cite{haldane_bosons,giamarchi_book_1d,gogolin_1dbook,cazalilla_correlations_1d}:
the excitations are 1D sound waves (phonons) and, at zero
temperature and for $L \to \infty$, all correlations decay
asymptotically as power-laws (\emph{e.g.}  phase correlations
$\langle e^{i \theta_{\bf R}(x)} e^{-i \theta_{{\bf R}'}(0)} \rangle
\sim \delta_{{\bf R}, {\bf R}'} \: |\rho_0 x|^{-\frac{1}{2K}}$ for
$x \gg \rho^{-1}_0$), which implies the absence of long-range order.
However, as soon as the Josephson coupling is turned on, bosons can
gain kinetic energy  in the transverse directions, and
therefore become delocalized  in more than one tube. This process
helps to build phase coherence throughout the lattice and leads to
the formation of a BEC. Nevertheless, the building  of  coherence
just described can be suppressed by two kinds of phenomena: one is
quantum and thermal fluctuations, which prevent  the bosons that hop
between different gas tubes from remaining phase coherent. The other
is the existence of a Mott potential that is \emph{commensurate}
with the boson density. For sufficient repulsion between the bosons,
this leads to the stabilization of a 1D Mott insulator, where bosons
become localized. Whereas we shall deal with the latter phenomenon
in~\sref{sec:mott}, we deal with the fluctuations in this section.

In this section we have in mind two different experimental
situations: one is a 2D optical lattice where the Mott potential is
(experimentally) absent. The other  is when the Mott potential is
irrelevant in the RG sense. The precise conditions for this to hold
will be given in~\sref{sec:mott} but,  by inspection of
(\ref{ham2}), we can anticipate that one particular such case is
when the Mott potential is incommensurate with the boson density
(\ie $\delta \neq 0$). Under these circumstances,   the  cosine term
proportional to $g_u$ oscillates rapidly in space and therefore its
effect on the ground state and the long wave-length excitations is
averaged out.

\subsection{Mean field theory}\label{sec:mft}

To reduce the Hamiltonian (\ref{ham1}) to a tractable problem, we
treat the Josephson coupling in a mean-field (MF) approximation
where the boson field operator $\Psi_{\bf R}(x)$ is replaced by
$\psi_c + \delta \Psi_{\bf R}(x)$, where $\psi_c  = \langle
\Psi_R(x)\rangle$ and $\delta \Psi_{\bf R}(x) = \Psi_{\bf R}(x) -
\psi_c$. We then neglect the term $\propto \delta \Psi_{\bf R}(x)
\delta \Psi_{\bf R'}(x)$, which describes the effect of fluctuations
about the mean field state characterized by the order parameter
$\psi_c$. This approximation decouples the 1D systems at  lattice
sites $\bf R \neq {\bf R}'$ so that the problem reduces to a single
``effective'' 1D system in the presence of a (Weiss) field
proportional to $|\psi_c|$. The bosonized form of the Hamiltonian
for this system is (we drop the lattice index $\bf R$):
\begin{eqnarray}\label{sg}
 H^{\rm MF}_{\rm eff} &=& \frac{\hbar v_s}{2\pi} \int^{L}_{0}\left[ K
 \left(\partial_x\theta(x) \right)^2
 + K^{-1} \left(\partial_x \phi(x)\right)^2 \right] \nonumber \\
 &&  -2 J z_C \sqrt{{\cal A}_{B} \rho_0} |\psi_c|  \int^{L}_{0}dx\: \cos
  \theta(x)  +  J z_C  L  |\psi_c|^2,
\end{eqnarray}
where $z_C$ is the coordination number of the lattice ($z_C = 4$ for the
square lattices realized thus far  in the experiments).
The phase of  the order parameter $\theta_c = \arg \psi_c$ has been eliminated
by performing a (global) gauge transformation $\theta(x) \to \theta(x) + \theta_c$,
and henceforth we shall take $\psi_c = |\psi_c|$  to be real.
We note that, up to the last constant term, $H^{\rm MF}_{\rm eff}$ defines
an exactly solvable model, the sine-Gordon model, of which
a great deal is known from  the theory of integrable
systems as well as from various approximation schemes~\cite{gogolin_1dbook,
giamarchi_book_1d,zamolodchikov_energy_sg}.  The mean-field Hamiltonian
must be supplemented by a self-consistency condition, which relates
the effective 1D system to the original lattice problem:
\begin{equation} \label{self-consist}
\psi_c =  \langle \Psi_{\bf R}(x)\rangle = 2
 \left({\cal A}_{B} \rho_0\right)^{1/2}\,  \langle \cos\theta(x)\rangle_{\rm MF}
\end{equation}
Note that the average over the cosine  must be
computed using the eigenstates of $H^{\rm MF}_{\rm eff}$.

\subsubsection{Condensate fraction at $T=0$:}\label{sec:becfrac}

To obtain the condensate fraction at zero temperature,
we first compute the ground-state energy of $H^{\rm MF}_{\rm eff}$
using the following result for the ground state energy density
of the sine-Gordon (sG) model~\cite{zamolodchikov_energy_sg}:
\begin{equation}\label{Edens}
 {\cal E}_{sG}(|\psi_c|) = - \frac{\Delta_s^2(|\psi_c|)}{4 \hbar v_s} \tan{\frac{\pi p}{2} },
\end{equation}
where $p = \beta^2/(8\pi-\beta^2) = 1/(8K-1)$, with $\beta^2 = \pi/K$.  The
excitations of the sG model are gapped solitons and anti-solitons
(and bound states of them for $\beta^2 <  4\pi$)~\cite{gogolin_1dbook,giamarchi_book_1d}.   The soliton energy
gap (also called  soliton  ``mass") is~\cite{zamolodchikov_energy_sg}:
\begin{equation} \label{smass}
\Delta_s(|\psi_c|) = \hbar v_s  \left[  \frac{m_s(|\psi_c|)}
 {\kappa(p)}\right]^{\frac{p+1}{2}} ,
\end{equation}
and
\begin{eqnarray}\label{ms-kappa}
 m_s(|\psi_c|) &=& \frac{z_C J ({\cal A}_{B} \rho_0)^{1/2}}{ \hbar v_s a_0^{-2 (p+1)/p}}\,  |\psi_c|, \\
 \kappa(p) &=& \frac{1}{\pi} \frac{\Gamma\left(\frac{p}{p+1}\right)}{\Gamma\left(\frac{1}{p+1} \right)}  \left[ \frac{\sqrt{\pi} \Gamma\left(\frac{p+1}{2}\right)}{2\Gamma(\frac{p}{2})} \right]^{\frac{2}{p+1}},
\label{eq:kappa}
\end{eqnarray}
where $\Gamma(x)$ is the gamma function. Hence, the ground state
energy  of $H^{\rm MF}_{\rm eff}$ is given by:
\begin{equation}
 E^{\rm MF}_0(|\psi_c|) / L =  {\cal E}_{sG}(|\psi_c|) +  J z_C  |\psi_c|^2.
\end{equation}
 The  self-consistent condition~(\ref{self-consist})
is equivalent to minimizing the mean-field free energy with respect to the order parameter. At $T = 0$,  the entropy is zero and
the condensate fraction is thus obtained
by minimizing $E^{\rm MF}_0 (|\psi_c|)$ with respect to
$\psi_c = |\psi_c|$, which yields:
\begin{equation} \label{psi0}
 \psi_c = \rho^{1/2}_0\:  \left[ \frac{p+1}{8 (\kappa(p))^{p+1}}
 \tan\left(\frac{\pi p}{2}\right)  \eta^{\frac{p+1}2} \left( \frac{ J
 z_C}{\hbar v_s \rho_0} \right)^{p} \right]^{\frac1{1-p}}.
\end{equation}
In the above expression, we have introduced the dimensionless  parameter $\eta =
{\cal A}_{B}(K) (\rho_0 a_0)^{1/2K}$.  We note that the condensate fraction
at $T= 0$ has a power-law dependence on the amplitude of the
Josephson coupling, \ie
\begin{equation}
\psi_c \sim \rho^{1/2}_0  \left(\frac{J}{\mu_{1D}} \right)^{\frac{p}{1-p}} = \,
\rho^{1/2}_0  \left(\frac{J}{\mu_{1D}} \right)^{\frac{1}{8K-2}}. \label{eq:becfrac}
\end{equation}
The dependence of the  exponent on the parameter $K$,
which varies continuously with the strength of the interactions
between the bosons in the tubes,  is a consequence of the fact that the appearance
of a BEC at $T =0$, as soon as $J$ is made different from zero, is affected by
the characteristic 1D quantum fluctuations of the TLL's.
This  behavior is markedly different from what can be obtained by a
Gross-Pitaevskii mean-field  treatment of the system that
overlooks the importance of 1D quantum fluctuations. Further
below we shall also encounter other instances where the present
results strongly deviate from Gross-Pitaevskii theory.

    To show the effectiveness of the fluctuations to cause depletion
of the BEC, let us estimate  the fraction  of bosons in the condensate at $T = 0$
using equation~(\ref{psi0}).  For $K = 1$ (\ie the Tonks limit, where
phase fluctuations are the strongest), we take $a_0 \rho_0 =  1$ and assume
that  $\mu_{1D} = 0.1\,  \hbar \omega_{\perp}$~\cite{Moritz_oscillations}.
Therefore, for $V_{0\perp} = 20\:
E_R$,  the hopping amplitude $J \simeq 3 \times 10^{-3} \, \mu_{1D}$, and
hence, using equation~(\ref{psi0}), $\rho^{-1}_0 \psi^2_c \approx 20\: \%$.
We also note that for weakly interacting bosons, \ie  as $K \to +\infty$,
$\psi^2_c \to \rho_0$, as expected.

  In the experiments of Ref.~\cite{stoferle_tonks_optical}, very small
coherence fractions were observed. The \emph{experimental}
coherence fraction is defined as the fraction of the total number
of atoms that contributes to the peak around ${\bf k} = 0$ in
the momentum distribution,
as measured by time of flight. The fraction was quantified
by fitting a gaussian distribution to the peak~\cite{stoferle_tonks_optical}.
We note that such a procedure yields a non-vanishing coherence fraction
even for the case of uncoupled 1D tubes, where the momentum distribution exhibits
a peak with a power-law
tail $\sim |k|^{\frac{1}{2K}-1}$ and a width $\sim \max\{L, \hbar v_s/T\}$~\cite{cazalilla_correlations_1d,cazalilla_finitesize_luttinger}.
Thus, identifying the condensate fraction with the experimental
coherence fraction may not be appropriate as there is no condensate in
such a 1D limit. Furthermore,
as we show below, in the anisotropic BEC phase discussed above,
this experimental procedure  would also pick up a
non-condensate contribution to the coherence fraction.
Nevertheless, provided heating effects and thermal depletion of the condensate are not
too strong,  one can  regard the experimental coherence fraction as an
upper bound to the condensate fraction, and since a strong decrease of the
coherence fraction was reported in the experiments of Ref.~\cite{stoferle_tonks_optical}, our results showing a strong depletion of the
condensate are consistent with the experimental
findings.

\subsubsection{Condensation temperature:}\label{sec:bectemp}

 At finite temperatures, thermal (besides quantum) fluctuations cause
 depletion of the condensate. Ultimately, at a critical temperature $T = T_c$,
the condensate is destroyed by the fluctuations. To compute  the critical temperature
we follow Efetov and Larkin~\cite{efetov_coupled_bosons} and,
taking into account that near $T_c$ the order parameter  $\psi_c(T)$ is small,
consider a perturbative expansion of the self-consistency condition (\ref{self-consist})
in powers of $\psi_c$. To lowest order,
\begin{eqnarray}\label{stoner}
\psi_c(T) &=& \langle \Psi(x,\tau) \rangle_{\rm MF} =
\frac{\langle {\cal T} \left\{ \Psi(x,\tau)  e^{-\frac{1}{\hbar}
\int^{\hbar \beta}_0 d\tau'\,   H_{\rm W}(\tau')}
\right\} \rangle_0}{ \langle {\cal T} e^{-\frac{1}{\hbar}\int^{\hbar \beta}_0
d\tau' \,   H_{\rm W}(\tau')} \rangle_{0}} \nonumber\\
 &=&  - J z_C \psi_c(T) \int^{L}_0 dx' \int^{\hbar\beta}_0 d\tau'
 g_1(x',\tau') +  O(\psi^2_c).
\end{eqnarray}
where $\Psi(x,\tau) = (\rho_0 {\cal A_B})^{1/2} \, e^{i\theta(x,\tau)}$, $\tau$ being
the imaginary time, ${\cal T}$ the ordering symbol in (imaginary) time, and
the Weiss field,
\begin{equation}
H_{\rm W} =  -J z_C \psi_c(T) \int^{L}_0 dx \: \left[ \Psi(x) + \hc \right].
\end{equation}
The correlation function:
\begin{equation}\label{chieq}
g_1(x,\tau) =  -\frac{1}{\hbar}
 \langle {\cal T} \left[ \Psi(x,\tau) \Psi^{\dag}(0,0) \right] \rangle_{0},
\end{equation}
where the average $\langle \ldots \rangle_0$ is performed using
the quadratic part of $H^{\rm MF}_{\rm eff}$. Hence, analytically continuing
to real time, equation~(\ref{stoner}) reduces to:
\begin{equation}
1 + J z_C\,  g^R_1(q = 0, \omega = 0; T_c) = 0,
\end{equation}
which determines $T_c$; $g^R_1(q,\omega)$ is Fourier transform
of the retarded version of (\ref{chieq}). This function is computed in the
\ref{app:finitechi} (cf. equation~(\ref{g0})), and leads to the result:
\begin{equation} \label{tceq}
 \left(\frac{2\pi T_c}{\hbar v_s \rho_0}\right)^{2-\frac{1}{2K}} = F(K)
 \left(\frac{Jz_C}{\hbar v_s \rho_0}\right),
\end{equation}
where
\begin{equation}
F(K) =  \pi^2 {\cal A}_B(K) \sin\left(\frac{\pi}{4K} \right)
 B^2\left(\frac{1}{8K},1-\frac{1}{4K}\right),
\end{equation}
with $B(x,y) = \Gamma(x) \Gamma(y)/\Gamma(x,y)$ the beta function.
Thus, the dependence of $T_c$ on $J$ is a power-law whose
exponent is determined by $K$.
 It is interesting to analyze the behavior of $T_c$ in the limit
of weak interactions, \ie for $K\gg 1$. Using that $F(K \gg 1) \sim K$,
along with Galilean invariance, equation~(\ref{galilean}), $v_s = v_F/K$,
we find that
\begin{equation}\label{tclargek}
\left( \frac{T_c}{\epsilon_F} \right)^2 \sim \left( \frac{J}{\epsilon_F} \right),
\end{equation}
where $\epsilon_F = \hbar \pi \rho_0 v_F/2 = (\hbar \pi
\rho_0)^2/2M$. Thus, we see that as $K \to
+\infty$, $T_c \sim J^{1/2}$. This result comes  with two caveats.
The first one is that, as we turn off the interactions  (\ie  for
$g_{1D} \to 0$) so that $K$ can diverge, the chemical potential
$\mu_{1D} \simeq g_{1D} \rho_0\to 0$. Therefore since the effective
Hamiltonian, equation~(\ref{ham2}), is valid only for energy scales below
$\mu_{1D}$, the range of validity of the result is going to lower
and lower $J$ in this limit. Indeed one cannot simply let $K\to
\infty$ and keep $J$ fixed. One would get
 $T_c/\mu_{1D} \sim (\epsilon_F/\mu_{1D})^{1/2}
(J/\mu_{1D})^{1/2} \to +\infty$. On the other hand, we know that as
$g_{1D} \to 0$ $T_c$ must remain finite because a non-interacting
Bose gas in a 2D optical lattice indeed undergoes Bose-Einstein
condensation (see  below). Therefore, if $T_c/\mu_{1D}$ diverges as
$\mu_{1D}\to 0$ for $K \gg 1$, it is because of this fact. The
second caveat is that the behavior $T_c \sim J^{1/2}$  implied by
(\ref{tclargek}) is not the correct one in the non-interacting
limit. To see this, consider the condition for the Bose-Einstein
condensation of a non-interacting Bose gas in a 2D optical lattice
(we take $L \to \infty$ but keep the number of lattice sites $M_0 =
M_y \times M_z$ finite):
\begin{equation}\label{becnonint}
L \sum_{{\bf Q}} \int^{+\infty}_{-\infty} \frac{dq}{2\pi} \,
\frac{1}{e^{\left[\epsilon(q) + \epsilon_{\perp}({\bf Q}) \right]/T_c} - 1} = N_{\rm ex} =
N_0 M_0
\end{equation}
where $\epsilon(q)$ and $\epsilon_{\perp}({\bf Q}) = - 2 J  \left(
\cos Q_y a + \cos Q_z a  - 2\right)$ are the longitudinal and
transverse dispersion of the bosons, respectively. In the latter
case, we have chosen the zero-point energy (\ie the bottom of the
band) to be zero. In the $J\to 0$ limit the 1D systems become
decoupled and $T_c$  vanishes (\ie there is no BEC in 1D).
Therefore, we expect $T_c$ to be small for small $J$. It is thus
justified to expand about ${\bf Q} = (0,0)$ the transverse
dispersion $\epsilon_{\perp}({\bf Q}) \simeq J a^2 {\bf Q}^2$.
Furthermore, $\epsilon(q) = \hbar^2q^2/2M$, for free bosons, and
thus (\ref{becnonint}) leads to the law $T_c \sim J^{2/3}$, which is
different from (\ref{tclargek}). The difference stems from the
assumption of a linear longitudinal dispersion of the excitations
implied by the quadratic part of (\ref{ham2}) and (\ref{sg}). Had we
insisted in assuming a linear longitudinal dispersion of the bosons
$\epsilon(q) = \hbar v_0 |q|$, where $v_0$ has dimensions of
velocity, we would have found that $T_c \sim J^{1/2}$ as implied
by~(\ref{tclargek}). This argument implicitly assumes that in the
weakly interacting limit, the difference between particles and
excitations vanishes. This seems quite reasonable, as the spectrum
of excitations of the non-interacting Bose gas coincides with the
single-particle spectrum.

Nevertheless, we can expect equation~(\ref{tceq}) to provide a good
estimate of the condensation temperature as long as $J/\mu_{1D} \ll
1$. This is of course particularly true in the strongly interacting
limit where $K \sim 1$. The interpolation formula for the prefactor
${\cal A}_B$, equation~(\ref{interpolation}), allows to be more
quantitative than Efetov and Larkin and to obtain an estimate of
$T_c$ for  $K = 2$, for instance. For the Lieb-Liniger model this
value of $K$ corresponds to  $\gamma \simeq
3.5$~\cite{cazalilla_correlations_1d} and to $\mu_{1D}/\epsilon_F
\simeq 0.367$. For $J \simeq
 3 \times 10^{3} \, \mu_{1D}$ (\ie $V_{0\perp} = 20 \, E_R$),  equation~(\ref{tceq}) yields $T_c/\mu_{1D} \simeq 0.6$ and therefore $T_c/\hbar \simeq 10 \, {\rm kHz}$ or $T_c \simeq
 70 \, {\rm nK}$. We note this temperature seems to be
well above the experimental estimate for the temperature of the
cloud, $T < 10^{-3} \: \hbar \omega_{\perp} \approx
10^{-2} \mu_{1D}$~\cite{Moritz_oscillations}. Thus, we expect the experimentally realized
systems to be in the anisotropic BEC phase described in this section,
and the optical lattice to exhibit 3D coherence.
\subsubsection{Excitations of the BEC phase:}\label{sec:exbec}
Let us next compute the excitation spectrum of the BEC.
Due to the existence of coherence between the 1D systems of
the array, this turns out to be very different from the ``normal"
Bose gas (TLL) phase, where the gas tubes are effectively decoupled
 (\ie $\psi_c = 0$) because fluctuations destroy the
 long range order. The most direct way of finding  the excitation
 spectrum is to consider configurations where the order parameter
 varies slowly with $x$ and $\bf R$ and  time. Thus the order parameter
becomes  a quantum field, $\Psi_c(x.{\bf R},t)$.
One then asks what is the effective  Lagragian, or Ginzburg-Landau (GL), functional,
${\cal L}_{\rm GL}(x,{\bf R},t)$,
that describes the dynamics of $\Psi_c(x,{\bf R},t)$. There are
at least two ways of finding this functional: one is to perform
a Hubbard-Stratonovich transformation and to
integrate out the original bosons to find an effective action for the fluctuations
of the order parameter about a uniform configuration. To gaussian order
(random phase approximation, RPA) this is carried out in ~\ref{app:rpa}.
This method  has the advantage of allowing us to make contact with the
microscopic theory (the Hamiltonian~(\ref{ham2}) in our case) and thus to relate  the
parameters that determine the dispersion of the excitation modes to those
of the microscopic theory. Another method, which we shall follow in this
section, is to guess ${\cal L}_{\rm GL}$ using symmetry principles
only. To this end, we first write ${\cal L}_{\rm GL} =  {\cal T} - {\cal  V}$,
where the kinetic energy term ${\cal T}$ depends
on gradients of $\Psi_c(x,{\bf R},t)$, whereas ${\cal V}$ depends only on
$\Psi_c(x,{\bf R},x,t)$. Both ${\cal T}$ and ${\cal V}$ are
subject to invariance under global Gauge transformations:
$\Psi_c(x,{\bf R},t) \to e^{i\theta_0} \:
\Psi_c(x,{\bf R},t)$.

 Using global gauge invariance, to the lowest non-trivial order in $\Psi_c$, we can write
${\cal V} = a |\Psi_c(x,{\bf R},t)|^2 + b |\Psi_c(x,{\bf R},t)|^4$. Since we
are interested in the ordered phase where $\Psi_c = \psi_c$,  we must have $a <0$.
Thus,  we can write $\cal V$ as follows:
\begin{equation}
{\cal V} = \frac{\hbar \lambda}{2} \left(|\Psi_c(x,{\bf R},t)|^2 - |\psi_c|^2 \right)^2,
\end{equation}
where $\lambda$  is a phenomenological parameter. On the other hand,
in order to guess the form of the kinetic term ${\cal T}$ global gauge invariance
alone is not sufficient. Further insight can be obtained  by close inspection of the effective
low-energy theory  described by (\ref{ham2}). If we write it down in Lagragian form:
\begin{eqnarray}
\fl
{\cal L}_{\rm eff}(x,t) = \frac{\hbar K}{2\pi} \sum_{\bf R}  \left[ \frac{1}{v_s}
\left( \partial_{t} \theta_{\bf R} \right)^2 - v_s
\left( \partial_{x} \theta_{\bf R} \right)^2 \right]
+ \frac{\hbar v_s g_J}{2\pi a_0^2} \sum_{\langle {\bf R},{\bf R'}\rangle}
\cos\left(\theta_{\bf R} - \theta_{{\bf R'}} \right),\label{eq:lagrangian}
\end{eqnarray}
we see that it is invariant under Lorentz transformations  of the coordinates $(x,t)$
where:
\begin{equation}
\left(
\begin{array}{c}
v_s t' \\
x'
\end{array} \right) =
\left(
\begin{array}{cc}
 \cosh \beta & \sinh \beta \\
 \sinh \beta & \cosh \beta
 \end{array}
\right)
\left(
\begin{array}{c}
v_s t \\
x
\end{array}
\right),
\end{equation}
being $\beta$ a real number. Thus,  ${\cal T}$ can only contain
Lorentz invariant combinations of  $|\partial_t \Psi_c(x,{\bf R},t)|^2$, $|\partial_x\Psi_c(x,{\bf R},t)|^2$, and
$|\nabla_{\bf R} \Psi_c(x,{\bf R},t)|^2$, which are the lowest order
derivative terms allowed by global gauge invariance. This leads to
\begin{equation}
\fl
{\cal T}(x,{\bf R}, t) = \frac{\hbar Z_2}{2} \left[
|\partial_t \Psi_c(x,{\bf R},t)|^2 - v^2_{||}
|\partial_x \Psi_c(x,{\bf R},t)|^2  -  v^2_{\perp} |\nabla_{\bf R} \Psi_c(x,{\bf R},t)|^2 \right],
\end{equation}
where $v_{||} = v_{s}$ and
$Z_2$ and $v_{\perp}$ are again phenomenological parameters.
Higher order terms and derivatives are also allowed, but the above terms
are the ones that dominate at long wave-lengths and low frequencies.
Therefore, the GL functional reads:
\begin{eqnarray}\label{ginzburg}
{\cal L}_{\rm GL}(x,{\bf R},t) &=&
 \frac{\hbar Z_2}{2} \Big[  |\partial_t \Psi_c(x,{\bf R},t)|^2 -  v^2_{||}
|\partial_x \Psi_c(x,{\bf R},t)|^2 \nonumber\\
&&-  v^2_{\perp}  |\nabla_{\bf R} \Psi_c(x,{\bf R},t)|^2   \Big]
- \frac{\hbar\lambda}{2}  \left(|\Psi_c(x,{\bf R},t)|^2 - |\psi_c|^2 \right)^2.
\end{eqnarray}

We find the excitation spectrum by setting
$\Psi_c(x,{\bf R},t) =
\left[\psi_c + \eta_c(x,{\bf R},t) \right]\:  e^{i\theta_c(x,{\bf R},t)}$ and keeping only terms up to quadratic order, which yields:
\begin{eqnarray}
{\cal L}_{\rm GL}(x,{\bf R},t) &\simeq& \frac{\hbar Z_2}{2\hbar} \left[ \left( \partial_t \eta_c \right)^2  -
v^2_{||} \left( \partial_x \eta_c \right)^2  - v^2_{\perp} \left( \nabla_{\bf R}\eta_c \right)^2 - Z^{-1}_2 \lambda \psi^2_c \eta^2_c \right]  \nonumber\\
&&+ \frac{\hbar Z_2}{2\hbar} \psi^2_c \left[ \left( \partial_t \theta_c \right)^2 -
v^2_{||} \left( \partial_x \theta_c \right)^2
- v^2_{\perp} \left(\nabla_{\bf R} \theta_c \right)^2 \right]
\end{eqnarray}
Hence $\omega^2_{-}(q,{\bf Q}) = (v_{||} q)^2 + \left(
v_{\perp} {\bf Q}\right)^2$ gives the
dispersion of the \emph{Goldstone} mode $\theta_c$, and
$\omega^2_{+}(q,{\bf Q}) = \Delta^2_{+} +
(v_{||} q)^2 + \left(v_{\perp} {\bf Q}\right)^2$, where $\Delta^2_{+} = Z^{-1}_2 |\psi_c|^2  \lambda$,  the dispersion of the  \emph{longitudinal} mode $\eta_c$.
Hence we conclude that the BEC phase has two excitation branches: a gapless
mode, which is expected from Goldstone's theorem, and a gapped
mode, the longitudinal mode, which describes oscillations of the
magnitude of the order parameter about  its ground state value $\psi_c$.
It is worth comparing these phenomenological results with
those of the analysis of gaussian fluctuations (RPA + SMA) about the mean-field state
described in \ref{app:rpa}. There we find, in the small $(q,{\bf Q})$ limit:
\begin{eqnarray}
\omega^2_{-}(q,{\bf Q}) &\simeq& (v_{||} q)^2 + (v^{(-)}_{\perp} {\bf Q})^2, \label{RPAgoldstone} \\
\omega^2_{+}(q,{\bf Q}) &\simeq&  \Delta^2_{(+)}  + (v_{||} q)^2  + (v^{(+)}_{\perp} {\bf Q})^2, \label{RPAgapped}
\end{eqnarray}
where $v^{(-)}_{\perp} = \Delta_1 a/\hbar$,  $\Delta^2_{+} = \Delta_2 (8K-1)^{-2}/2\hbar^2$, and $v^{(+)}_{\perp} = \Delta_2 a/\hbar$, being $\Delta_p = \Delta_s \sin(p \pi)$ the breather energy gaps and
\begin{equation} \label{soliton-mass}
 \Delta_s = \hbar v_s \rho_0 \left[  \eta \frac{p+1}{8 \kappa^2(p)}
 \tan\left(\frac{\pi p}{2}\right)  \frac{4 J}{v_s
 \rho_0}\right]^{\frac{1+p}{2(1-p)} } \sim \left( \frac{J}{\mu_{1D}}\right)^{\frac{2K}{4K-1}}.
\end{equation}
is the soliton gap. One noticeable difference of these results
from those of the GL theory is that the values of $v_{\perp}$ predicted
by the RPA+SMA for the longitudinal and Goldstone modes differ, \ie $v^{(+)}_{\perp} \neq v^{(-)}_{\perp}$.  This is due to a poor treatment of the Landau-Ginburg theory that neglected  terms of ${\cal L}_{GL}$ of order higher than quadratic in the fields $\eta_c$ and $\theta_c$.  These terms represent interactions between the
Goldstone and longitudinal field, as well as self-interactions of the longitudinal field.
On the other hand, the RPA treatment of  \ref{app:rpa}  takes into account
some of these interactions and thus yields a correction to the dispersion of the Goldstone and longitudinal modes, which leads to different values of $v_{\perp}$.

    It is interesting to consider the behavior of the system in the limit of
strong Josephson coupling where $J > \mu_{1D}$. Under these conditions the
motion of the atoms is not mainly confined to 1D and thus they can more
easily avoid collisions. Therefore,  the effect of fluctuations is expected to
be less dramatic and assuming the existence of a 3D condensate
is a good starting point. This leads to Gross-Pitaevskii (GP) theory,
which, assuming a slowly varying configuration of the order parameter,
has  the following Lagrangian:
\begin{eqnarray}\label{GP}
{\cal L}_{\rm GP}(x,{\bf R},t) &=&
\hbar Z_1 \Big[ i\Psi^{*}_c(x,{\bf R},t) \partial_t \Psi_c(x,{\bf R},t)
-  \frac{\hbar}{2M} |\partial_x \Psi_c(x,{\bf R},t)|^2 \nonumber\\
&&-  \frac{y_{\perp}}{2}   |\nabla_{\bf R} \Psi_c(x,{\bf R},t)|^2   \Big]
- \frac{\hbar\lambda}{2}  \left(|\Psi_c(x,{\bf R},t)|^2 - |\psi_c|^2 \right)^2.
\end{eqnarray}
where $Z_1 = a^{-2}$, $y_{\perp} = a^2 J/\hbar$, and  $\lambda = Z_1 \tilde{g}/\hbar$, $\tilde{g} = 4\pi \hbar^2 a_s \int dy dz \, |W_{\bf R}(y,z)|^4/M$
($W_{\bf R}(y,z)$ is the Wannier function at site $\bf R$).
The condensate fraction $\psi_c$  is the solution to the (time-independent) GP
equation:
\begin{equation}
\tilde{g} |\psi_c|^2 =   z_C J + \mu,
\end{equation}
where $\mu$ is the chemical potential. Formally, and ignoring the different
values of the couplings, ${\cal L}_{GP}$ looks almost identical to ${\cal L}_{GL}$
except for the fact that  $ i \hbar Z_1 \Psi^*_c \partial_t  \Psi_c$ replaces $Z_2
|\partial_t  \Psi_c|^2$. However, this change has a dramatic effect on the
excitation spectrum as we show in what follows. Setting
$\Psi_c(x,{\bf R},t) = \left[ \psi_c  + \eta_c(x,{\bf R},t) \right]e^{i\theta_c(x,{\bf R},t)}$
and carrying out the same steps as above, we arrive at:
\begin{eqnarray}
{\cal L}_{\rm GP}(x,{\bf R},t) &=&  \hbar Z_1\Big[ -2  \psi_c \eta_c \partial_t \theta_c
- \frac{\hbar}{2M} \left( \partial_x \eta_c \right)^2 - \frac{y_{\perp}}{2} \left( \nabla_{\bf R}
\eta_c \right)^2 - 2 \lambda \psi^2_c \eta^2_c \nonumber\\
&& - \frac{\hbar \psi^2_c}{2M} \left(\partial_x \theta_c \right)^2  - \frac{y_{\perp}}{2}
\psi^2_c \left( \nabla_{R} \theta_c\right)^2 \Big].
\end{eqnarray}
From the above expression, we see that the amplitude $\eta_c$ has no dynamics,
\ie there is not term involving $\partial_t \eta_c$ that is not a total
derivative~\footnote{total derivatives have been dropped from both ${\cal L}_{\rm GP}$
and ${\cal L}_{\rm GL}$.}. The term $-2\psi_c \eta_c \partial_t \theta_c$, however, couples the dynamics of the phase $\theta_c$ and the amplitude $\eta_c$. This is very different from the situation with the GL theory encountered above.
We can integrate out $\eta_c$, \eg by  using the equations of motion for $\eta_c$,
which yield $\eta_c(x,{\bf R},t) = - Z_1
\partial_t \theta_c(x,{\bf R},t)/(2\lambda \psi_c) +$ gradient terms.
Thus, the lagrangian becomes:
\begin{equation}
{\cal L}_{\rm GP}(x,{\bf R},t) = \frac{Z^2_1}{\hbar \lambda} \left[ \left( \partial_t
\theta_c \right)^2 - v^2_{||} \left( \partial_x \theta_c \right)^2 -
v^2_{\perp} \left( \nabla_{\bf R}\theta_c\right)^2 \right],
\end{equation}
being $v^2_{||} = \hbar |\psi_c|^2 \lambda/(2 M Z_1)$ and
$v^2_{\perp} = y_{\perp} \lambda |\psi_c|^2/(2 Z_1)$. This means
that a BEC described by the GP theory has a
single low-energy mode, whose dispersion, $\omega(q,{\bf Q}) =
\sqrt{v^2_{||}q^2 + v^2_{\perp}{\bf Q}^2}$, vanishes in the long
wave-length limit. This is precisely the  Goldstone mode, which corresponds to the
one that we have already encountered above when analyzing the excitation
spectrum of the GL
theory that describes the extremely anisotropic limit where $J \ll \mu_{1D}$.
The presence of the Goldstone
mode as the lowest energy excitation in both theories is a direct consequence of
the spontaneous symmetry breaking of global gauge invariance. Thus both
kinds of superfluids are adiabatically connected as $J$ is increased.
On the other hand, the latter theory also exhibits a mode, the longitudinal mode,
that is gapped for $(q,{\bf Q}) = (0,{\bf 0})$.
However, the fate of the longitudinal mode as $J$ increases is not clear from the above discussion. To gain some insight into this issue, let us reconsider the derivation of the
GL Lagrangian in the light of the previous discussion of GP
theory. Apparently, the Lorentz invariance discussed above forbids the
existence of a term  $ i\hbar Z_1 \Psi^*_c \partial_t \Psi_c(x,{\bf R},t)$.
Nevertheless, we have to remember the emergence of such an invariance
in equation~(\ref{eq:lagrangian}) is a consequence of the linear dispersion of the
TLL modes of the uncoupled tubes. The dispersion is linear for energies
$\hbar \omega \ll \mu_{1D}$, but for $J \sim \mu_{1D}$ non-linear
corrections must be taken into account.
These break the effective Lorentz invariance of (\ref{eq:lagrangian})
and thus a term of the form
$ i\hbar Z_1  \Psi^*_c \partial_t \Psi_c$ is allowed. Phenomenologically,
we can consider a modified GL theory of the form:
\begin{eqnarray}\label{ginzburgp}
{\cal L}^{\prime}_{\rm GL}(x,{\bf R},t) &=& i \hbar Z_1 \Psi^*_c \partial_t \Psi_c +
 \frac{\hbar Z_2}{2} \Big[  |\partial_t \Psi_c(x,{\bf R},t)|^2 -  v^2_{||}
|\partial_x \Psi_c(x,{\bf R},t)|^2 \nonumber\\
&&-  v^2_{\perp}  |\nabla_{\bf R} \Psi_c(x,{\bf R},t)|^2   \Big]
- \frac{\hbar\lambda}{2}  \left(|\Psi_c(x,{\bf R},t)|^2 - \psi^2_c \right)^2.
\end{eqnarray}
where  $Z_1$ is very small for $J/\mu_{1D} \ll 1$ but grows with $J$
and should grow large for $J/\mu_{1D} > 1$. We can perform the same
analysis as above to find the excitation spectrum, which leads to
the following quartic equation for the dispersion of the modes:
\begin{equation}
\left[ \omega^2 - \omega^2_{+}(q,{\bf Q}) \right] \left[ \omega^2 - \omega^2_{-}(q,{\bf Q}] \right] + \frac{4 Z^2_1}{Z^2_2} \omega^2 = 0.
\end{equation}
where $\omega^2_{\pm}(q,{\bf Q})$ are the dispersion of the Goldstone
and longitudinal modes for $Z_1 = 0$, which have been found above.
We see that for small $q$
and ${\bf Q}$  there is a solution such that  $\omega^2 = A q^2 + B {\bf Q}^2$.
Furthermore, there  is also a gapped mode, for which $\omega^2 \to
\Delta^2_{+} + (Z_1/Z_2)^2$ as $q$ and $\bf Q)$ tend towards zero.
Thus we see that the gap of the longitudinal mode depends on the ratio
$Z_1/Z_2$. For $Z_{1} \ll Z_2$, the dispersion of the longitudinal mode is
essentially the same as that obtained from the GL theory. However, as $Z_1/Z_2$ grows with $J$,
the gap increases and therefore the exciting the longitudinal mode becomes
more costly energetically. At this point is useful to recall that the RPA analysis
in \ref{app:rpa} indicates the existence of a continuum of soliton-antisoliton
excitations above a threshold  $\hbar\omega = 2\Delta_s$. The stability of
the longitudinal mode is ensured if $\Delta_{+} < 2\Delta_s$, which
seems to be the case provided one trusts the single-mode approximation
employed to estimate $\Delta_{+}$. However, as the gap is pushed
up towards higher energies by an increasing $Z_1/Z_2$, the longitudinal
mode is likely to acquire a large finite lifetime by coupling to this continuum. This
will eventually lead to its disappearance from the spectrum.

\subsection{Self-consistent harmonic approximation (SCHA)} \label{sec:varap}

An alternative approach to obtain the properties of the BEC phase is
the self-consistent Harmonic approximation (SCHA) (see \eg
Ref.~\cite{giamarchi_book_1d}). The idea behind this approximation
is to find an \emph{optimal} (in the sense to be specified below)
quadratic Hamiltonian that approximates the actual Hamiltonian of
the system. The method is more suitably formulated in the
path-integral language, where the role of the Hamiltonian is played
by the euclidean action.  The latter is obtained by performing, in
the Lagrangian of equation~(\ref{eq:lagrangian}), a Wick rotation to
imaginary time, $t \to  -i \tau$, and defining the action as $S = -
\int^L_0 dx \int^{\hbar\beta}_0  d\tau \, {\cal L}_{\rm
eff}(x,i\tau)/\hbar$. Hence, $S[\theta] = S_0[\theta] +
S_{J}[\theta]$, where
\begin{eqnarray}\label{action-noMott}
S_0[\theta] & =&   \frac{K}{2\pi}\sum_{\bf R} \int_0^L dx \int_0^{\beta} d\tau \left[
 \frac{1}{v_s} \left( \partial_{\tau} \theta_{\bf R} (x,\tau) \right)^2
 + v_s  \left( \partial_{x} \theta_{\bf R}(x,\tau) \right)^2  \right] , \\
 S_{J}[\theta] & =&  - \frac{1}{\hbar}  {\cal A}_B  \rho_0 J \,  \sum_{\langle {\bf R R'} \rangle} \int^{L}_0 dx \int^{\hbar\beta}_0
 d\tau \,  \cos(\theta_{\bf R}(x,\tau)-\theta_{\bf R'}(x,\tau)).\label{eq:sint}
\end{eqnarray}
The partition function  reads:
\begin{equation}
Z = \int \left[ d\theta \right] \, e^{-S[\theta]}.
\end{equation}
In the SCHA,   $S[\theta]$  is approximated by a gaussian (\ie quadratic)
action:
\begin{equation}
 S_{\rm v}[\theta] = \frac{1}{2M_0 \hbar\beta L} \sum_{q, {\bf Q},\omega_n}  G^{-1}_{\rm v}(q,{\bf Q},\omega_n) \left| \theta_{\bf Q}(q,\omega_n) \right|^2,
\end{equation}
where $\theta_{\bf Q}(q,\omega_n)$ is the Fourier transform of
$\theta_{\bf R}(x,\tau)$. This gaussian form is motivated by
considering the action, equation~(\ref{eq:sint}), in the limit of large $J$.
We then expect that the replacement
$\cos \left(\theta_{\bf R} - \theta_{\bf R'} \right) \simeq 1 -\frac{1}{2!}
\left(\theta_{\bf R}  - \theta_{\bf R'} \right)^2$ would be reasonable.
This yields a quadratic action and a phonon propagator the form
of (\ref{eq:scha2}). To find the optimal $G_{\rm v}(q,{\bf Q},\omega_n)$,
we use Feynman's variational principle~\cite{feynman_statmech}:
\begin{equation}
F = -\frac{1}{\beta} \ln Z \leq F'[G_{\rm v}] =  F_{\rm v} - \langle \left(S[\theta] - S_{\rm v}[\theta] \right) \rangle_{\rm v},
\label{eq:Fv}
\end{equation}
where $\langle \ldots \rangle_{\rm v}$ stands for the trace over configurations of
$\theta_{\bf R}(x,\tau)$ using $e^{-S_{\rm v}[\theta]}$  and
\begin{equation}
e^{-\beta F_{\rm v}}  =   \int \left[ d\theta \right] \, e^{-S_{\rm v}[\theta]}.
\end{equation}
Therefore, by extremizing $F'[G_{\rm v}]$ with respect to $G_{\rm v}$, that is,
by solving
\begin{equation}
\frac{\delta F'[G_{\rm v}]}{\delta G_{\rm v}(q,{\bf Q},\omega_n)} = 0,
\end{equation}
we find the optimal $G_{\rm v}$:
\begin{eqnarray}
\fl
\frac{1}{G_{\rm v}(q,{\bf Q},\omega_n)} =  \frac{1}{G_0(q,\omega_n)} - \frac{{z_C \cal A}_B(K) \rho_0}{2 \hbar}  J a^2   \,  F({\bf Q}) \,\,
e^{\frac{J a^2}{2 M_0 L\hbar\beta}\sum_{q',{\bf Q'},\omega_n'}  F({\bf Q'}) \: G_{\rm v}(q',{\bf Q'},\omega_n')},
\label{eq:selfcons}
\end{eqnarray}
where $F({\bf Q}) = \frac{a^{-2}}{2} \sum_{\bf t}  \left(  1 - e^{i {\bf Q}\cdot {\bf t}} \right)$,
with $\bf t$ the  vectors that connect a lattice point to its nearest neighbors, and
$G^{-1}_{0}(q,\omega_n) = \frac{K}{\pi v_{||}} \left[ \omega^2_n + v^2_{||} q^2 \right]$
the free phonon propagator ($v_{||} = v_s$).
This self-consistency equation (\ref{eq:selfcons}) can be solved by rewriting it into
two equations as follows: defining
\begin{equation}
\fl v^2_{\perp}  =   \frac{\pi z_C {\cal A}_{B}}{2 K}  (\rho_0 a)   \left(\frac{Ja}{\hbar} \right) v_{||} \:
 \exp \left[\frac{2 a^{2}}{z_C M_0 L\hbar\beta}\sum_{q,{\bf Q},\omega_n}  F({\bf Q}) \: G_{\rm v}(q,{\bf Q},\omega_n)\right], \label{eq:scha1}
\end{equation}
we arrive at:
\begin{equation}
G^{-1}_{\rm v}(q, {\bf Q}, \omega_n) =  \frac{K}{\pi v_{||}}
\left[ \omega^2_n + (v_{||} q)^2 + v^2_{\perp} \,  F({\bf Q})\right]. \label{eq:scha2}
\end{equation}
The set of equations (\ref{eq:scha1}) and (\ref{eq:scha2}) are solved in
\ref{app:variat}), for the variational parameter $v_{\perp}^2 $ which is the
transverse phonon velocity of the 3D superfluid.

 At $T=0$, the equation for $v^{0}_{\perp} = v_{\perp}(T = 0)$ can be solved analytically.
We merely state here the result (the details of the calculation can
be found in \ref{app:variat}):
\begin{equation} \label{gamma0}
 \frac{v^{0}_{\perp}}{v_{||}} \simeq  \left[ \left( \frac{{\cal B}_0 {\cal A}_{B}(K)}{K} \right)
 \left( \frac{\hbar v_{||} \rho_0}{\mu_{1D}} \right)
 \left( \frac{z_C J}{\mu_{1D}} \right) \right]^{\frac{2K}{4K-1}}
 \end{equation}
where ${\cal B}_0  \simeq 1.2971$ for a 2D square latttice.
Note that the variational propagator $G_{\rm v}(\omega_n, q, {\bf Q})$
has a pole, which  for small $q$ and ${\bf Q}$, is located
at  $\omega^2_{-}(q,\omega) \simeq v^2_{||} q^2 + (v_{\perp}{\bf Q})^2$,
for a square 2D transverse lattice.
This is the Goldstone mode with $v_{\perp}
 \sim (J/\mu_{1D})^{2K/(4K-1)}$, a power law with
an exponent  that agrees with
the RPA+SMA result of \ref{app:rpa}, where it is shown that
$v^{(-)}_{\perp} \sim \Delta_{1} \sim \Delta_s \sim (J/\mu_{1D})^{2K/(4K-1)}$.
However, the SCHA fails to capture the existence of the longitudinal
mode. This is because, as described above,
the SCHA is tantamount to expanding the cosine
in the Josephson coupling term. Such an approximation
neglects the fact that the phase $\theta_{\bf R}(x,\tau)$ is
 $2\pi$-periodic, and that as a consequence the system supports
vortex excitations. On the other hand, this is captured by RPA
because it does not disregard the periodic aspect of the phase
since the mean-field state is described
by the sine-Gordon model. The latter has soliton, anti-soliton,
and breather excitations, which are a direct consequence of the
periodicity of $\theta_{\bf R}$.

  In the weakly interacting limit (\ie for $K \gg 1$), equation~(\ref{gamma0})
becomes:
\begin{equation}
 \frac{v^0_{\perp}}{v_{||}}  \sim \sqrt{\frac{J}{\mu_{1D}}},
\end{equation}
where we have used that $\mu_{1D}/(v_{||} \rho_0) \approx
\gamma^{1/2} = \pi/K$ for $K \gg 1$. Thus, the
requirement  that $J/\mu_{1D} \ll 1$ automatically
implies that $v^{0}_{\perp}/v_{||} \ll 1$, that is,
we deal with a very anisotropic BEC.

At $T > 0$, the variational equations must solved
numerically (see \ref{app:variat}).  The absence of
a solution to equation (\ref{eq:scha1}) such that for $T > T_c$
$v_{\perp} = 0 $ allows us to also determine the
condensation temperature.  Thus for weakly interacting
Bose gas ($K \gg 1$ or $\gamma$ small),
we find:
\begin{equation}
 \frac{T_c}{\mu_{1D}} \simeq \sqrt{\frac{z_C J}{\mu_{1D}}} \left( 0.610 + 0.698 K \right),
\end{equation}
This is the same form as that obtained for large $K$ from the RPA analysis of
\sref{sec:bectemp}, equation (\ref{tclargek}). Thus, the same caveats
as those discussed there apply to it.

 At $T = T_c$ the  velocity $v_{\perp} = v^c_{\perp}$ has finite a jump, which
for $K \gg 1$ is given by
\begin{equation}
 \frac{v^c_{\perp}}{v_{\perp}^0} \simeq \sqrt{\frac{z_C J}{\mu_{1D}}} \left( 0.376 + 0.274 K^{-1} \right).
\end{equation}
This is indeed an artifact of the SCHA, which incorrectly captures the second-order character of  the transition from the  BEC to the ``normal" (TLL) phase.

\subsection{Momentum distribution function}

 Besides the excitation spectrum, the SCHA also allows us to obtain the momentum distribution of the system in the BEC phase. The latter can be measured in a time-of-flight experiment~\cite{pitaevskii_becbook}, and it is the Fourier transform of the
one-body density matrix:
 \begin{equation}
 \fl
 n(k,{\bf K}) =  \frac{1}{M_0 L}
  \int dx dx'  \, \int d{\bf r}_{\perp} d{\bf r}^{\prime}_{\perp} \,
 e^{-i {\bf K} \cdot ({\bf r}_{\perp} - {\bf r'}_{\perp})} \: e^{-ik(x-x')}\:
 \langle \Psi^{\dag}(x, {\bf r}_{\perp}) \Psi(x', {\bf r}^{\prime}_{\perp}) \rangle.
 \end{equation}
Introducing $\Psi(x, {\bf r}_{\perp}) = \sum_{\bf R} W_{\bf R}({\bf r}_{\perp}) \: \Psi_{\bf R}(x)$, where ${\bf r}_{\perp} = (y,z)$, $n(k,{\bf K})$ can be written as:
\begin{equation}
\fl
n(k,{\bf K}) =  \frac{|W({\bf K})|^2}{M_0 L}
\sum_{{\bf R}, {\bf R'}} e^{-i {\bf K}\cdot ( {\bf R} - {\bf R}')}
\int dx dx' \, e^{-ik (x-x')} \: \langle \Psi^{\dag}_{\bf R}(x)
\Psi_{\bf R'}(x') \rangle,
\end{equation}
where $W({\bf K})$ is the Fourier transform of the Wannier
orbital $W_{\bf 0}({\bf r}_{\perp})$.  Using the bosonization
formula~(\ref{eq:psibos}) and  within the SCHA
\begin{equation}
\fl
\langle \Psi^{\dag}_{\bf R}(x) \Psi_{\bf 0}(0) \rangle =
\rho_0 {\cal A}_{\bf B}(K)  \langle  e^{-i\left[\theta_{\bf R}(x) - \theta_{\bf 0}(0) \right]} \rangle
\simeq  \rho_0 {\cal A}_{\bf B}(K)
e^{-\frac{1}{2}\langle \left[ \theta_{\bf R}(x) - \theta_{\bf 0}(0) \right]^2 \rangle_{\rm SCHA}}.
\end{equation}
Since $ \frac{1}{2}\langle  \left[ \theta_{\bf R}(x) - \theta_{\bf 0}(0) \right]^2\rangle_{\rm SCHA}
= \langle \theta^2_{\bf 0}(0) \rangle_{\rm SCHA} -  \langle \theta_{\bf R}(x) \theta_{\bf 0}(0) \rangle_{\rm SCHA}$ and $ \langle \theta_{\bf R}(x) \theta_{\bf 0}(0)\rangle =
G_{\rm v}(x ,{\bf R}, t=0)$, where
\begin{eqnarray}
\fl
G_{\rm v}(x,{\bf R},t) &=& \frac{1}{M_0 L} \sum_{q, {\bf Q}} \int \frac{d\omega}{2\pi}\:
e^{i (q x + {\bf Q} \cdot {\bf R} - \omega t)} G_{\rm v}(q,{\bf Q}, i\omega_n \to \omega + i \delta) \\
\fl
 &\simeq& \frac{1}{8\pi^2 v^2_{\perp} K} \frac{a^2}{(x/v_{||})^2 + ({\bf R}/v_{\perp})^2},
\end{eqnarray}
where  $G_{\rm v}(q,{\bf Q},\omega_n)$ is given by equation~(\ref{eq:scha2}).
The result on the last line is the asymptotic behavior for large  
$x$ and ${\bf R}$. Thus we see that for large distances $G_{\rm v}(x,{\bf R},t)$ decays very rapidly
and therefore it is justified to expand
$e^{G_{\rm v}(x,{\bf R},t=0)} \simeq 1  +  G_{\rm v}(x,{\bf R},t=0) + \ldots$ Keeping
only the lowest order term, we obtain the following expression
for $(k, {\bf K})$ near zero momentum:
\begin{eqnarray}
\fl
n(k,{\bf K}) \simeq |W({\bf K})|^2 \:\left[
 (M_0  L) |\psi_c|^2  \delta_{k,0} \: \delta_{{\bf K},{\bf 0}}  + \frac{v_{||}{\cal Z}_0}{\sqrt{(v_{||} k)^2 + (v_{\perp} {\bf K})^2}}  \right]  \label{eq:nk1}
\end{eqnarray}
where ${\cal Z}_0 = |\psi_c|^2  \pi /2K$, and
the (modulus squared of the) condensate fraction,
\begin{equation}
\fl
|\psi_c|^2 = \rho_0 {\cal A}_{B} (K) \, e^{- G_{\rm v}(x = 0,{\bf R} = {\bf 0})} =
\rho_0 {\cal A}_{B}(K)  \left(\frac{\hbar v_{\perp}e^{-{\cal C}_0}
}{2\mu_{1D} a} \right)^{1/2K}
\sim \rho_0 \left( \frac{J}{\mu_{1D}} \right)^{\frac{1}{4K-1}} ,
\end{equation}
where ${\cal C}_0 \simeq 0.2365$ for a 2D square lattice. Thus the SCHA approximation
also yields a condensate fraction that behaves as a power law of $J/\mu_{1D}$.
The exponent is the same as the one obtained in \sref{sec:becfrac}
from mean-field theory, equation~(\ref{eq:becfrac}).

  The momentum distribution can be also obtained at finite temperatures from:
\begin{eqnarray}
\fl
G_{\rm v}(k, {\bf K}, t=0) &=& \frac{1}{\hbar \beta} \sum_{i\omega_n} \frac{\pi v_{||} K^{-1}}{\omega^2_n + (v_{||} k)^2 + v^2_{\perp} F({\bf K})}   \\
&=& \frac{\pi v_{||} K^{-1}}{\sqrt{(v_{||} q)^2 + v^2_{\perp} F({\bf K})}}
\coth \left[ \frac{\hbar   \sqrt{(v_{||} k)^2 + v^2_{\perp} F({\bf K})}}{T} \right].
\end{eqnarray}
Thus for $k$ and $\bf K$ near zero, and $T > 0$ we find
\begin{equation}\label{eq:nk2}
n(k,{\bf K}) \simeq |W({\bf K})|^2 \:\left[
 (M_0  L) |\psi_c (T)|^2  \delta_{k,0} \:
 \delta_{{\bf K},{\bf 0}}  + \frac{T v_{||}{\cal Z}_0(T)/\hbar}{(v_{||} k)^2 + (v_{\perp} {\bf K})^2}  \right]
\end{equation}
with ${\cal Z}_0(T)$ defined as above with $\phi_c$ replaced by $\psi_c(T)$.

 There are several points that are worth discussion about the results of equations (\ref{eq:nk1}) and (\ref{eq:nk2}). The first is that these form are anisotropic generalizations of the result that can be obtained from Bogoliubov theory for a BEC~\cite{pitaevskii_becbook}. However, for sufficiently large $q$  we expect a crossover to the same power-law behavior of a single 1D Bose-gas $\sim |k|^{\frac{1}{2K}-1}$  (for $T = 0$ and $L\to \infty$). This is nothing but a reflection of the dimensional crossover physics:  bosons
with kinetic energy $\epsilon(k) =  \hbar^2 k^2/2M \gg J$ (but such that $\epsilon(k) \ll \mu_{1D}$ for the bosonization to remain valid) cannot ``feel'' the BEC and the intertube
coherence. However, they still ``feel'' the characteristic quantum fluctuations of the TLL state. The second observation is related to our discussion of \sref{sec:becfrac} on the experimental  ``coherence fraction'': it is clear by fitting a gaussian to $n(k,{\bf K})$
about zero momentum, one does not only pick a contribution from $|\psi_c(T)|^2$ but
also from the tail of the non-condensate fraction. The presence of a trap
(and the inhomogeneity of the system caused it) should not dramatically modify
this conclusion.  However, as both contributions are positive, we expect that
the experimentally measured coherence fraction is indeed an upper bound for
$|\psi_c(T)|^2$.

\section{Deconfinement} \label{sec:mott}

In this section we shall obtain the phase diagram of the system 
for the case where the Mott potential is relevant. One necessary
condition for this is that  $\delta =0$ and $u_0 \neq 0$ so that
$g_u \neq 0$   in equation~(\ref{ham2}). As it has been mentioned above,
under these conditions, and provided repulsion between the bosons is
strong enough so that the parameter $K < 2$, the system undergoes a
transition to a Mott insulating state for $J = 0$ and arbitrarily
small
$u_0$~\cite{haldane_bosons,giamarchi_mott_shortrev,giamarchi_book_1d,buchler_cic_bec}.
In the Mott insulating state bosons are localized about the minima
of the potential. The question thus is what happens as soon as a
small Josephson coupling between neighboring tubes is present. The
latter favors delocalization of the bosons leading to the appearance
of a BEC, and therefore competes with the localization tendencies
favored by the Mott potential. The competition between these two
opposite tendencies leads to a new quantum phase transition known as
\emph{deconfinement}. For small values of $|u_0|$ and $J$, it can be
studied by means of the renormalization group (RG).  This is a much
better approach than comparing ground state energies of mean-field
theories for each phase, as it takes into account the fluctuations,
which are entirely disregarded within any mean-field approximation.
However, although the RG approach allows us to estimate the position
of the boundaries between the Mott insulating and BEC phases,  in
the present case it cannot provide any insight into  the nature of
the deconfinement transition. For this, we shall rely on the
mean-field approximation and map the mean field hamiltonian at the
special value of $K=1/2$ to a spin-chain model.

\subsection{Renormalization Group calculation} \label{sec:rg}

 Let us return to the low-energy effective field theory defined by the
 Hamiltonian in equation~(\ref{ham2}).  Since it is a description valid for low-energies
and temperatures, the couplings $K$, $g_u$, $g_J$, etc. all depend on the
cut-off energy scale, which is typically set by the temperature for
$\Lambda = T < \mu_{1D}$. Thus, as the temperature is decreased, the couplings  get
renormalized due to virtual transitions to states with energy $> \Lambda = T$, which
now are  excluded from the description. The nature of the ground state is thus determined by the dominant coupling as $T =\Lambda\to 0$.

The change (`flow') for the couplings can be described by a set of differential
equations.  For small values of $g_J$ and $g_u$ the RG flow equations can be
obtained perturbatively to second order in the
couplings. The details of such a calculation  are given in \ref{app:rgequ}.
The result reads:
\begin{eqnarray}
 \frac{dg_F}{d\ell}  &=& \frac{g^2_J}{K},  \label{eq:rg1} \\
 \frac{dg_J}{d\ell}  &=& \left(2-\frac{1}{2K}\right)g_J + \frac{g_Jg_F}{2K},  \label{eq:rg2} \\
 \frac{dg_u}{d\ell}  &=& \left( 2 - K \right) g_u,  \label{eq:rg3}\\
 \frac{dK}{d\ell} &=& 4 g^2_J - g^2_u K^2,  \label{eq:rg4}
\end{eqnarray}
where $\ell \approx \ln \mu/T$. The coupling $g_F$ is generated by the RG
and describes an interaction between bosons in neighboring tubes. However,
since initially there is not such an interaction (\ie $g_F(0) = 0$) and as $dg_F/d\ell
= O(g^2_J)$ its effect on the flow of the other couplings is not very important.

 The phase boundary between the BEC and 1D Mott insulator phases
can be obtained by studying the asymptotic behavior of the solutions of
equations (\ref{eq:rg1}) to (\ref{eq:rg4}). As the temperature is lowered,
\ie as $\ell \to +\infty$, in the regime $\frac{1}{4} \leq K < 2$~\footnote{The regime
where $K < 1$ describes a 1D system of bosons interacting via long-range
forces~\cite{giamarchi_book_1d}. Thus it is not physical for a system of cold
atoms interacting via short-range forces.
However, the discussion in this section will be done
for a general value of $K$.} the effective couplings
$g_u(\ell)$ and $g_J(\ell)$ grow as both the Mott potential and the  Josephson
coupling are \emph{relevant} perturbations in the RG  sense.
If  $g_u(\ell)$ grows  larger, the localization tendencies of the Mott
insulator phase set in, whereas if $g_J(\ell)$  grows larger,
delocalization tendencies of the BEC phase described in \sref{sec:nomott}
dominate in the ground state.  Both phases correspond to two
different strong coupling fixed points of the RG.
When both perturbations are relevant,
one has a crossing of the two strong coupling fixed points, and since
we obtained the above equations assuming that both $g_u$ and $g_J$
are small,   the RG is not perturbatively controlled when $g_u$ and $g_J$ grow
large.  Note that, given this limitation,
we cannot entirely exclude that intermediate phases might exist. We
will come back to that particular point in the next section.

\begin{figure}
\centerline{\includegraphics[width=\figwidth]{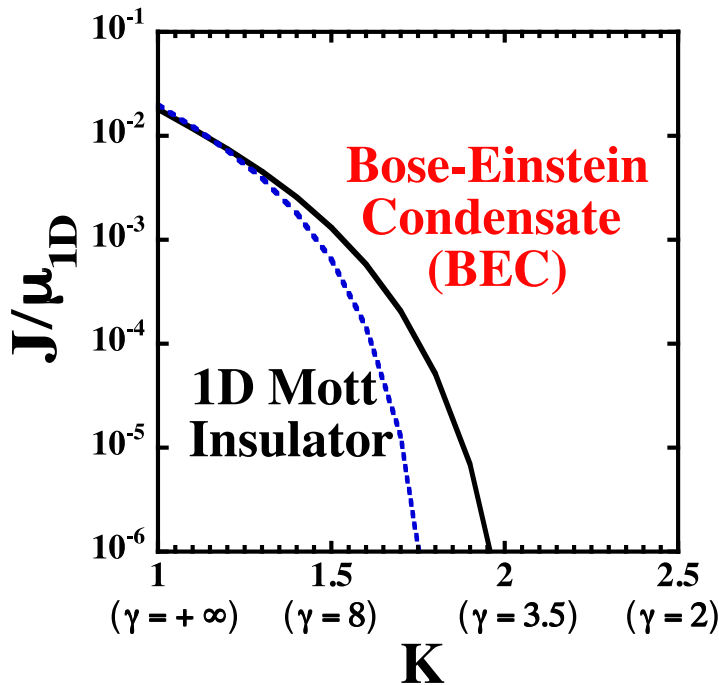}}
 \caption{Zero-temperature phase diagram of a 2D array of
 coupled, infinite,  1D  interacting boson systems. The values on the vertical axis are
 defined up to a factor of order unity. $K$ is the Luttinger-liquid parameter (see
 \sref{sec:bosometh} for explanations).
 The corresponding values of Lieb-Liniger gas parameter
 $\gamma = Mg/\hbar^2 \rho_0$ are also displayed.
 A longitudinal commensurate optical  potential of amplitude $u_0$  ($u_0\approx 0.03
 \: \mu_{1D}$ for $K=1$  and $u_0 \approx 0.02\: \mu$ for $K=2.5$) is
 assumed. The dashed line corresponds to the crude estimate for the phase
 boundary obtained from equation~(\ref{crude-jc}).}
 \label{fig1}
\end{figure}

The relative magnitude of the initial (\ie bare) values of $g_J$ and
$g_u$ determines which coupling grows faster to a value of order one,
where  the above RG equations cease to be valid. The faster growth of
one coupling inhibits the other's growth via the renormalization of the parameter
$K$: If $g_u(\ell)$ grows faster, $K$ flows towards $0$, leading to the 1D
Mott insulator. This reflects the fact that, for  large enough $g_u$, the Mott potential drives the system towards an incompressible phase, where boson number
fluctuations are  strongly suppressed by the localization of the particles.
On the other hand, if $g_J(\ell)$ grows faster, $K$ also grows
and leads to the 3D BEC phase.  A rather crude estimate of the boundary
can be obtained by ignoring the renormalization of $K$. In such a case,
dividing (\ref{eq:rg2}) by  (\ref{eq:rg3}), keeping
only the leading order terms,
and integrating this equation up to the scale
$\ell^*$, where  both $g_J(\ell^*) = g_u(\ell^*) = s$, where $s \sim 1$
we get that the critical value of $g^c_J = g_J(0)$ for a given value
of $g_u(0)$ scales as:
\begin{equation} \label{crude-jc}
 g^c_J \sim g_u(0)^{\frac{2-1/2K}{2-K}}  .
\end{equation}
In the next section we will see that this power-law is also recovered
in the mean-field treatment of the following section. However,
for a more precise estimate of the
boundary one needs  to integrate numerically the RG-flow equations
and take into account the renormalization of $K$, which can be important
especially near $K= 2$. This is shown in \fref{fig1}, where we compare the
estimate of (\ref{crude-jc}) with the phase boundary obtained from a numerical
integration of the RG equations.

  The RG equations also allow to extract the gap of the Mott
insulator as well as the soliton energy gap that characterizes
the decay of the one-body correlation functions (\ie the healing
length) in the BEC phase.  This is achieved by
integrating the equations until the fastest grown coupling is of order one.
One can then use that, at this scale $\ell^*$, the gap
$\Delta(\ell)$ is also of order one, and since gaps renormalize
like energies, \ie $\Delta(\ell) = \Delta(\ell=0) e^{-\ell}$. For example,
in the BEC phase, using equation~(\ref{eq:rg2}),
keeping only the leading term, and integrating this equation up to a scale
$\ell^* \approx  \log(\Lambda/\Delta_s)$ where $g_J(\ell^*) \sim 1$ gives
  $\Delta_s \sim (J/\mu_{1D})^{1/(2-1/2K)}$, which has the same power-law behavior
as the one  obtained from the mean-field theory, equation(\ref{soliton-mass}).
For $\frac{1}{4} < K < 2$, a better estimate (numerical) can be
obtained by accounting for the renormalization of $K$ using the full set of 
RG equations. However, for  $K > 2$, the
renormalization of $K$ is negligible and  the previous estimate of $\Delta_s$
should be accurate.

 Another instance where it can be necessary to take into account the renormalization
of $K$ is the evaluation of the condensation temperature $T_c$ in the presence of
the Mott potential. In particular, for $K < 2$, although the initial conditions are such
that the coupling $g_J$ is the dominant one, $K$ may undergo an important renormalization if $g_u$ is non-zero. The renormalized value of $K$ could be then
used in combination with the RPA formula, equation~(\ref{tceq}), for $T_c$.

\subsection{Mean field theory of the deconfinement transition}\label{sec:mftdecon}

The  RG approach explained above does not provide any insight into
the nature of the deconfinement transition, \eg it cannot answer 
questions such as whether the transition is continuous (\ie second
order) or not, or what universality class it belongs to. In order to
address, at least partially, some of these issues we will treat
again the interchain coupling in a mean field approximation. We
begin by rewriting the mean-field Hamiltonian of equation~(\ref{sg})
in the presence of the Mott potential:
\begin{eqnarray}\label{eq:doublesg}
\fl
 H^{\rm MF}_{\rm eff} &=& \frac{\hbar v_s}{2\pi} \int^{L}_{0}\left[ K
 \left(\partial_x\theta(x) \right)^2
 + K^{-1} \left(\partial_x \phi(x)\right)^2 \right] \nonumber \\
\fl
&& +  2 \rho_0 u_0  \int^{L}_0 dx \, \cos 2 \phi(x)
   -2 J z_C \sqrt{{\cal A}_{B} \rho_0} |\psi_c|  \int^{L}_{0}dx\: \cos
  \theta(x)  +  J z_C  L  |\psi_c|^2.
\end{eqnarray}
The new mean-field Hamiltonian is now a double sine-Gordon theory,
which is not exactly solvable  for general values of the parameter
$K$~\cite{lecheminant_jkkny}. However, for $K = \frac{1}{2}$ the
solution takes a particularly simple form as one can directly relate
it to the results obtained in \sref{sec:mft} for the sine-Gordon
model. As it has been pointed out above, the regime $K < 1$ is not
physical for models like the Lieb-Liniger model or the 1D
Bose-Hubbard model. However, we expect that the solution at this
point captures some of the essential features of the deconfinement
transition for $K \ge 1$. Indeed in particular in the RG flow
nothing special occurs at $K=1$ so we can expect the two limits to
be smoothly connected. The key observation behind the solution at $K
= \frac{1}{2}$ is that, as shown in~\ref{app:spinmap}, at this point
(\ref{eq:doublesg}) is the effective low-energy description of a
Heisenberg anti-ferromagnetic spin chain in the presence of a
staggered magnetic field:
\begin{equation}
H^{MF} = J_0 \sum_{m} {\bf S}_m \cdot {\bf S}_{m+1} + {\bf h} \cdot \sum_{m} (-1)^m \: {\bf S}_{m} + J z_C  L  |\psi_c|^2 ,
\label{eq:sc1}
\end{equation}
where ${\bf h} = (h_x, 0, h_z)$, $h_x = - \sqrt{8\pi} z_C  ({\cal A}_B  \rho_0)^{1/2}|\psi_c| a_0 J$ and $h_z = 2 \pi u_0 (\rho_0 a_0)$,
where $a_0$ is the short-distance cut-off. We can perform a rotation about the OY axis,
so that the spin variables transform as:
\begin{eqnarray}
S^x_m &=& \cos \varphi \: \tilde{S}^x_m + \sin \varphi \: \tilde{S}^z_m, \\
S^z_m &=& -\sin \varphi \: \tilde{S}^x_m + \cos \varphi \: \tilde{S}^z_m,
\end{eqnarray}
where $\tan \varphi = h_z/h_x$, and the mean-field Hamiltonian becomes:
\begin{equation}
H^{MF} = J_0 \sum_{m} {\bf \tilde{S}}_m \cdot {\bf \tilde{S}}_{m+1} + |{\bf h}| \sum_{m} (-1)^m \tilde{S}^x_m + J z_C  L  |\psi_c|^2.
\end{equation}
Applying the the spins ${\bf \tilde{S}}_m$, the bosonization rules  the spins given in~\ref{app:spinmap}, the above
Hamiltonian becomes:
\begin{eqnarray}
H^{\rm MF}_{\rm eff} &=& \frac{\hbar v_s}{2\pi} \int^{L}_{0}\left[ \frac{1}{2}
 \left(\partial_x\theta(x) \right)^2
 + 2 \left(\partial_x \phi(x)\right)^2 \right] \nonumber \\
 &&  + \frac{|{\bf h}|}{a_0\sqrt{2 \pi}} \int^{L}_{0}dx\: \cos
  \theta(x)  +  J z_C  L  |\psi_c|^2.
\end{eqnarray}
This is another sG model, similar to the one that  we encountered in \sref{sec:mft}.  One can
again make use of equations~(\ref{Edens}, \ref{smass}) and (\ref{ms-kappa})
to obtain the ground-state energy density,
\begin{equation}
E^{\rm MF}_0/L = {\cal E}_{sG}(|{\bf h}|) + z_C J |\psi_c|^2.
\end{equation}
The behavior of the order parameter $\psi_c$ follows from the  extremum
condition:
\begin{equation}
\frac{d}{d\psi_c} \left[ {\cal E}_{sG}(|{\bf h}|) + z_C J |\psi_c|^2  \right] = 0.
\end{equation}
The algebra is essentially very similar to that in section \ref{sec:becfrac}, and we just quote the results here.  There exists a critical tunneling
$J_c$, when the condensate fraction $\psi_c \rightarrow 0$, {\emph i.e.}
$h_x \rightarrow 0$. We find at $K=1/2$,
\begin{equation}
\left( \frac{z_C J_c}{v_{||} \rho_0} \right)^3 = \frac{\pi \rho_0 a_0
[\kappa(1/3)]^4}{2 [\frac{1}{6} \eta \tan (\pi/6)]^3}
\left(\frac{u_0}{v_{||} \rho_0} \right)^2.
\label{eq:Jc-spin}
\end{equation}
Thus the condensate fraction grows continuously from zero at the
deconfinement transition from the Mott Insulator at $J<J_c$ to the
3D anisotropic superfluid at $J>J_c$ according to:
\begin{equation}
\psi_c^2 = \rho_0 \frac{\eta^2}{[\kappa(1/3)]^4}
 \left[ \frac{1}{6} \tan \left(\frac{\pi}{6}\right) \right]^3
\left( \frac{z_C J}{v_{||} \rho_0} \right)
\left[ 1 - \left(\frac{J_c}{J} \right)^3 \right],
\label{MI-SF-psi_c}
\end{equation}
where $\kappa(1/3)$ has been defined in (\ref{eq:kappa})
and at $K=1/2$ the dimensionless ratio $\eta= \rho_0 a_0 {\cal A}_B(1/2)$.
Note that the scaling $J_c \propto u_0^{2/3}$ in (\ref{eq:Jc-spin})
agrees with that
deduced from the RG approach of equation (\ref{crude-jc}) upon setting
$K=1/2$.

It is instructive to compare this deconfinement scenario in 3D with the
case where only two 1D systems  are coupled together (via the
same Josephson coupling term), {\emph i.e.} a bosonic two-leg ladder, which has
been studied in \cite{donohue_deconfinement_bosons}. In both the quasi-1D
optical lattice and the ladder, there is the competition between the
Josephson coupling that favors superfluidity, and the
localizing tendency of the commensurate periodic potential on top of which the
interacting bosons hop. This is manifested in a very similar structure of the
bosonized Hamiltonians. For the ladder case, it is convenient use
a  symmetric and anti-symmetric combination of the fields of the two chains, which
leads to slightly different forms for the RG equations.
Nevertheless, the mentioned competition occurs for the same range of $K$:
$1/4 < K < 2$, where, in the ladder case,  $K$ characterizes to the correlations of  antisymmetric fields. Qualitatively, the critical intertube coupling
for the  deconfinement transition has the same power-law dependence
on $g_u$ as in equation~(\ref{crude-jc}). However, as the ladder  is
still effectively 1D, the nature of the deconfinement transition is
expected to be very different from the one studied here, where a fully 3D
(anisotropic) BEC results. In
Ref.~\cite{donohue_deconfinement_bosons}, the deconfinement 
transition in the ladder was found to
be in the Berezinskii-Kosterlitz-Thouless  universality class,
at least when (for fixed interaction strength $U$) 
the intertube hopping is larger than the
intratube hopping (note that this limit has not been studied in
this work as it would correspond to coupled 2D systems, with
very different physics  compared to the coupled 1D systems.)

Note that in the entire phase diagram (in the thermodynamic limit,
see figure~\ref{fig1}),
we have  found evidence  for the BEC and the 1D Mott
insulating phases only. We have not found any evidence (or hints of evidence) 
for more exotic
states such like the 
sliding Luttinger liquid phase~\cite{emeryetal_SLL,vishwanath_slide_LL,mukhopadhyay_slide_LL} or some kind of
supersolid phase. The sliding Luttinger
liquid~\cite{emeryetal_SLL,vishwanath_slide_LL,mukhopadhyay_slide_LL} is a perfect
insulator in the transverse direction (like our 2D Mott insulator
for finite size systems); along the tube direction, the system is
still a Tomonaga-Luttinger liquid, but the  parameters $K$ and  $v_s$ are
now functions that depend on the transverse momentum. This state may
be stabilized by a strong interaction of the long-wave length part of density
fluctuations between tubes~\cite{vishwanath_slide_LL}. We note that, although the RG procedure described above generates   such a density-density interaction between the tubes, in the bare Hamiltonian there are no such interactions  to start with (atoms
in different tubes have negligible interactions) and therefore, the coupling $\tilde{g}_F$ is
likely to remain small as the RG flow proceeds to lower energies, at least within 
weak coupling perturbative RG we use here. Thus it seems unlikely that the system can
enter the regime of a sliding Luttinger liquid~\cite{vishwanath_slide_LL,
mukhopadhyay_slide_LL} phase. An even more exotic possibility would be a
phase where the tubes are  1D Mott insulators (in the presence of a
commensurate periodic potential and strong enough interaction in-tube), while at the same time, there would be coherence between the tubes in the transverse 
direction. In this scenario, presumably {\it both}
the couplings $g_u$ (Mott potential) and  $J$ (Josephson coupling)
would be relevant. We cannot rule out such a possibility in the strong
coupling limit. However,  we have no evidence for it in our weak coupling 
analysis. Nevertheless, in a quasi-1D system and on physical grounds, 
it seems hard to imagine a phase where the boson density is kept commensurate  
within each tube, as required by the existence of the Mott insulator, while at the same time bosons are able to freely hop  to establish phase coherence
{\it between} tubes (but not {\it within} tubes!).
 
 As far as  the competition between the Josephson coupling and the Mott
potential is concerned, mean field theory is a useful tool
(in higher dimensions as well as when the tubes are coupled) for providing
insights into  which possible phases and what their properties are. However, it
is usually inappropriate  for predicting  the correct critical
behavior. For cold atomic systems, because the systems are finite
and of the existence of the harmonic trap that makes the samples inhomogeneous, 
it is not very realistic to study critical behavior as the latter is usually
strongly perturbed~\cite{wessel_MC_confined_bosons}. Mean
field theory seems thus perfectly adapted to the present study.
Let us however more generally comment on the universality at the
transition. Because the field $\cos(2\phi_{\bf R})$ is a vortex (instanton) creation
operator~\cite{giamarchi_book_1d} for the field $\theta_{\bf R}$, the
Hamiltonian for each tube  ${\bf R}$ can be faithfully represented by a
classical 2D XY Hamiltonian involving 
 $J_{x,\tau} \cos[\theta_{\bf R}(x',\tau')-\theta_{\bf R}(x,\tau)]$. 
It  thus becomes apparent that the coupling between the tubes 
leads to an anisotropic version of the 
$3+1$ dimensional XY model. The transition will thus be in the
universality class of the classical $D = d+z =4$ XY model in agreement with general
considerations on superfluid to insulator transitions~\cite{fisher_boson_loc}. 
In the present case, the transition
is tuned by varying the anisotropy of the model $\delta \sim J-J_c$. For such
a transition, the superfluid density of the classical model is expected to behave as
$\delta^{\nu(d+z-2)}$, where for the commensurate Mott insulator to
superfluid transition $z=1$ and, at $d=3$, which is the upper
critical dimension, the critical exponent $\nu = 1/2$ is mean-field-like. 
One would thus find that the superfluid density behaves as 
$~\delta$, in agreement with our result for $K=1/2$
of equation (\ref{MI-SF-psi_c}), upon identifying the superfluid density
of the classical XY model  with the square BEC fraction of the quantum model,
$\psi_c^2$.

Note that in the absence of the self-consistency condition~(\ref{self-consist}), the mean
field Hamiltonian is equivalent to an XY model in presence of a
symmetry breaking field~\cite{jose_planar_2d,kadanoff_modeles_2d}.
It has been recently argued that such a model can have a
sequence of two phase transitions 
\cite{fertig_XY_breaking_RG,fertig_XY_breaking_RG-long}.
Since in the present case the coefficient of the Mott potential ($\sim 
\cos(2\phi_{\bf R})$) term is small,  the system would be in the regime 
where it undergoes a single phase transition even for the model with the symmetry
breaking field.  But, more importantly, the model that we have studied has to be
supplemented by the self-consistency condition, equation~(\ref{self-consist}), 
which stems from our mean-field treatment of the original Josephson coupling
$\cos(\theta_{\bf R}-\theta_{\bf R'})$. This condition will 
modify the properties of the phase where this coupling is relevant,
and for the reasons explained above, should bring back the model in
the universality class of the $3+1$-dimensional XY model.

\section{Finite size effects} \label{sec:finitesize}

 Actual experimental systems consists of \emph{finite} 2D arrays of \emph{finite}
1D systems (tubes). The finite size of the sample has some well known
consequences in condensed matter physics such as inducing the quantization of
the excitation modes and  turning phase transitions into more
or less sharp crossovers. The latter can be understood using the
renormalization-group analysis of \sref{sec:rg}. The finite size of the tubes
introduces a length scale into the problem, namely the size of the tube $L$, which
effectively cut-offs the RG flow at a length scale of the order of $L$ (or, for that
matter, the smallest length scale characterizing the size of the sample). Thus,
flows cannot always proceed to strong coupling where the system acquires
the properties of one of the thermodynamic   phases discussed
above.  Furthermore,  as we show below, the finite size $L$ leads to
the appearance of a new phase.

\subsection{Finite size tubes and 2D Mott transition}\label{sec:2DMott}

 Typically in the experiments~\cite{greiner_2dlattice,Moritz_oscillations}
a number  $N_0$ of bosons ranging from a few a tens to a few hundreds are
confined in a tube of $\sim 10 \: \mu{\rm m}$ long. This leads to energy quantization
of the sound modes in longitudinal direction, but also to a finite energy
cost for adding (or removing) one atom to the tube.
This energy scale
is $E_C = \hbar \pi v_s/KL$~\cite{haldane_bosons,cazalilla_correlations_1d}. The tube
can be thus regarded as a small atomic ``quantum dot'' with a characteristic
``charging energy'' $E_C$. If $E_C$ is sufficiently large, it can suppress
hopping in an analogous way as the phenomenon of Coulomb blockade
suppressing tunneling through electron quantum dots and other
mesoscopic systems. In order to know when this will happen
 $E_C$ must be balanced against the hopping energy $E_J$.
 The latter is not just $J N_0$ as one would \emph{na\"ively}
 expect for non-interacting bosons. Interactions and the 
 fluctuations that they induce in the longitudinal direction alter the
 dependence of   $E_J$ on $N_0$ as we show in \ref{app:quantphase}.

\begin{figure}
\centerline{\includegraphics[width=\figwidth]{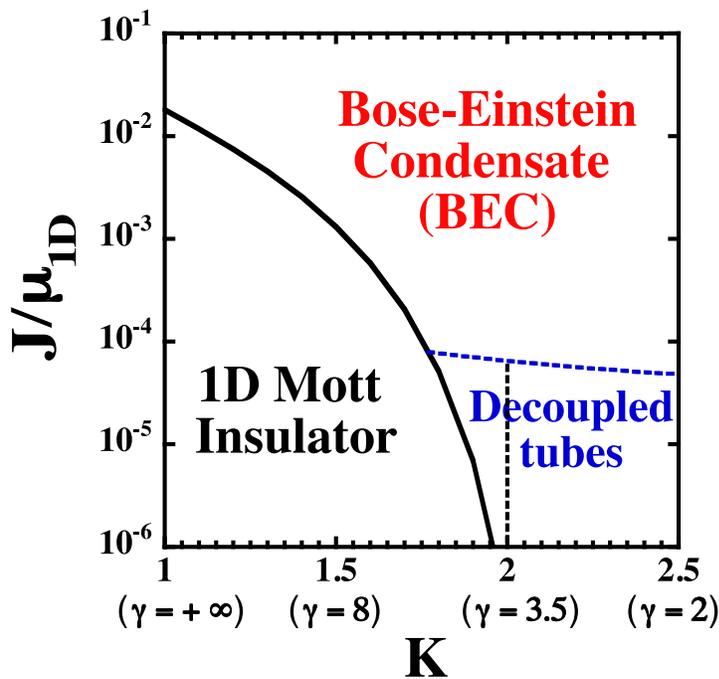}}
 \caption{Zero-temperature phase diagram of a 2D array of
 coupled, finite,  1D  interacting boson systems. The values on the vertical axis are
 defined up to a factor of order unity. $K$ is the Luttinger-liquid parameter (see
 \sref{sec:bosometh} for explanations).
 The corresponding values of Lieb-Liniger gas parameter
 $\gamma = Mg/\hbar^2 \rho_0$ are also displayed.
 A longitudinal commensurate optical  potential of amplitude $u_0$  ($u_0\approx 0.03
 \: \mu_{1D}$ for $K=1$  and $u_0 \approx 0.02\: \mu$ for $K=2.5$) is
 assumed. The vertical dashed line describes the boundary between the (finite-sized)
 Tomonaga-Luttinger phase.}
 \label{fig2}
\end{figure}
  Let us consider a system of \emph{finite} tubes in the BEC phase (\ie
the Mott potential is irrelevant or absent).
In the limit of small $J$ ($J/\mu_{1D} \ll 1$) of interest here,
the longitudinal fluctuations with wavenumber $q \neq 0$
can be  integrated out as discussed in \ref{app:quantphase}.
The resulting effective Hamiltonian  is  a quantum-phase model
identical to the one used to describe a 2D array of  Josephson junctions:
\begin{eqnarray}\label{qp}
 H_{\rm QP} &=& - E_J \sum_{\langle{\bf R}, {\bf R'}\rangle}
 \cos \left(\theta_{0{\bf R}} - \theta_{0{\bf R'}} \right) \nonumber \\
 && + \frac{E_C}{2} \sum_{\bf R} (N_{\bf R}-N_0)^2 -\mu \sum_{\bf R}
 N_{\bf R},
\end{eqnarray}
where $N_{\bf R}$ is the particle-number operator of tube at site
$\bf R$, and $\theta_{0{\bf R}}$ the canonically conjugate phase
operator: $[\theta_{0\bf R}, N_{\bf R'}] = i \delta_{{\bf R},{\bf
R'}}$. The renormalized hopping $E_J = E_J(N_0) \simeq J
N^{1-\frac{1}{2K}}_0$. The model (\ref{qp}) has been extensively
studied in the literature (see \eg Ref.~\cite{vanotterlo_2djjarrays}
and references therein). It exhibits a quantum phase transition
transition between a two-dimensional superfluid (2D SF) phase (a BEC
at $T = 0$) and a  two-dimensional Mott insulator (2D MI) at a
critical value of the ratio $E_J/E_C$. For commensurate filling
(integer $N_0 $) using Monte Carlo (MC) van Otterlo \emph{et al.}
\cite{vanotterlo_2djjarrays} found $(E_J/E_C)_c \simeq 0.15$. Using
the forms for $E_C$ and $E_J$ derived above, and assuming that the
tubes are described by the Lieb-Lininger model, the MC result
reduces to $J_c/\mu_{1D} \simeq 0.3 N^{-3/2}_0$ in the Tonks limit
($K = 1$) and to $J_c/\mu_{1D} \simeq 0.15 N^{-2}_0$  for weakly
interacting bosons ($K \gg 1$)~\cite{ho_deconfinement_coldatoms}. A
similar result has been more recently obtained in
Ref.~\cite{gangardt_finite_size_arrays} using the random phase
approximation (RPA). Our result, however, includes critical
fluctuations beyond the RPA as it relies on the numerically exact
results from MC simulations.  In figure~\ref{fig2} we show the
complete zero-temperature phase diagram including the possibility of
a phase where the tubes are effectively decoupled in a 2D MI phase
(labeled  Decoupled tubes in the diagram).
Note that in the 2D MI, only the phase coherence \emph{between}
different  tubes  of the 2D lattice is lost. However, apart from the
finite size-gap $\sim \hbar \pi v_s/L$, there is not gap for
excitations in the longitudinal direction. This makes this phase
different from the 1D MI described above, where there is a gap
(depending very weakly on the size $L$) for longitudinal
excitations. If the Mott potential is applied to the system in this
phase, it will cross over to a 1D Mott insulator for a finite, but
small value, of the dimensionless strength $g_u$. This
crossover should take place for $K <
2$~\cite{haldane_bosons,giamarchi_mott_shortrev,giamarchi_book_1d,buchler_cic_bec}
and is  thus represented by a vertical dashed line in
figure~\ref{fig2}.

\begin{figure}
\centerline{\includegraphics[width=\figwidth]{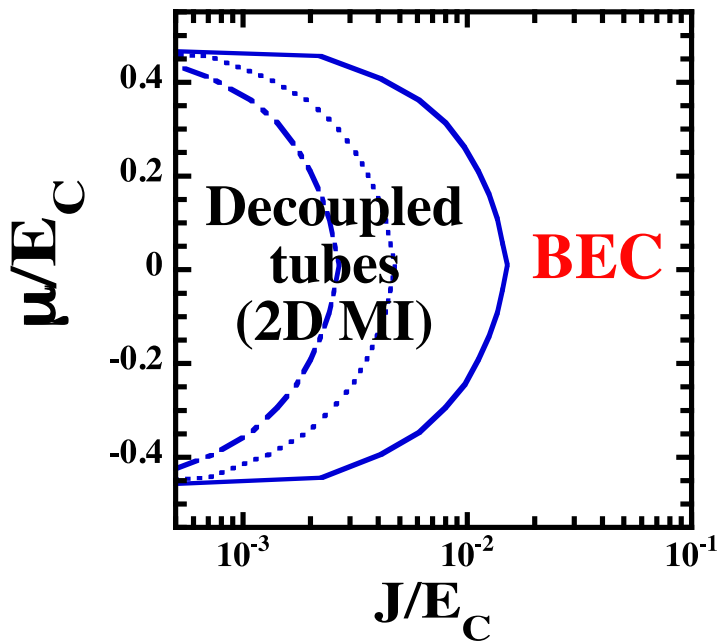}}
 \caption{Phase diagram of a 2D  array of finite 1D Bose gases as a function for
 the local chemical potential $\mu$ and the Josephson coupling $J$ for different
 values of the Luttinger parameter:  $K = 1$ ($\gamma \to +\infty$, continuous curve),
 $K =  2$ ($\gamma = 3.5$, dotted curve), $K   = 4$ ( $\gamma = 0.71$, dash-dotted curve).}
 \label{fig3}
\end{figure}

In the superfluid phase, the lattice will also exhibit a crossover
as a function of temperature: If the condensation temperature $T_c$ found
in \sref{sec:bectemp}  is larger  than the finite-size gap
$\hbar v_s /L$, the gas should behave as a 3D BEC and cross over to a
2D SF for $T \sim \hbar v_s/L$. However, if $T_c  \ll \hbar v_s/L$
the system should behave as a 2D quasi-condensate with no long range
phase coherence (and  therefore $\psi_c = 0$) at finite temperature.
Since $\hbar v_s/L \sim \hbar \omega_0$, $\omega_0$ being the
longitudinal trapping frequency, and
experimentally~\cite{Moritz_oscillations} $\omega_0 \sim 0.1$ kHz
whereas $T_c/\hbar \sim 10$ kHz, we expect to be in a regime where
the BEC is always observed. The presence of harmonic confinement
means that at least some of the tubes will have a filling deviating
from commensurate filling. The critical value of $J/\mu_{1D}$ is
reduced relative to the half-filled case, as shown in ~\fref{fig3}.

Finally, let us estimate the typical value of the critical ratio
$J_c/\mu_{1D}$. Using the above results and assuming that $\mu_{1D}
\approx 0.1 \: \hbar \omega_{\perp}$~\cite{Moritz_oscillations}
and $N_0 = 100$, we conclude that $J_c/\hbar \omega_{\perp} \simeq
10^{-4}$ for $K = 1$ whereas $J_c/\hbar \omega_c \simeq 10^{-5}$ for
$K \gg 1$. The deepest lattices used in the experiments of
Refs.~\cite{Moritz_oscillations,stoferle_tonks_optical} had
$V_{\perp} = 30 \: E_R$, which corresponds to $J/\hbar
\omega_{\perp} = 5 \times 10^{-5}$.  This is of the same order of
magnitude as  $J_c/\hbar \omega_{\perp}$ for $K \gg 1$, although a
larger number $N_0$ will further reduce this estimate, and therefore
for this lattice the system may be well in the BEC phase (see
figure~\ref{fig2}). However, in the
experiments~\cite{Moritz_oscillations,stoferle_tonks_optical} no
coherence was apparently observed between the tubes. The explanation
for this contradiction is that the value of tunneling  time
$\hbar/J$ must be compared with  the duration of the experiment
$t_{\rm exp}$ and it turns out that for a lattice depth of
$V_{0\perp} = 30 \: E_R$ this time is indeed longer than $t_{\rm
exp}$. Therefore, over the duration of the experiment almost no
tunneling process occurs and the build-up of intertube coherence
can thus hardly take place.

\section{Comparison to experiments}\label{sec:comparison}

In this section we  discuss and summarize our results in connection
with the recent experiments on anisotropic optical
lattices~\cite{Munich_experiment,Moritz_oscillations,stoferle_tonks_optical,
NIST_experiment,paredes_tonks_optical,kinoshita_tonks_continuous}.
The complete phase diagram, taking into account the finiteness of the 1D
gas tubes, is shown in figure~\ref{fig2}. At zero temperature, we
predict three phases: an anisotropic BEC, a 1D Mott insulator, and a
phase where the gas tubes are decoupled (\ie phase incoherent),
namely a 2D Mott insulator.

However, as  mentioned  in \sref{sec:model}, the experimental
systems are indeed harmonically confined. Unfortunately, so far it
has  proven very difficult to deal analytically and
simultaneously with strong correlations and the harmonic trap. On
the other hand, numerical methods can tackle this
situation~\cite{papenbrock_hcbosons_continuum,rigol_lattice_hcbosons,rigol_groundstate_hcbosons,
kollath_dmrg_bose_hubbard_trap,
wessel_MC_confined_bosons,wessel_MC_confined_bosons2}. In the
harmonic trap, the cloud becomes inhomogeneous, and thus as
demonstrated in a  number of numerical
simulations~\cite{kollath_dmrg_bose_hubbard_trap,wessel_MC_confined_bosons,wessel_MC_confined_bosons2,batrouni_domains,sengupta_domains_visibility},
for a wide range of system's parameters,  several  phases can
coexist   in the same trap. These results have confirmed the
experimental observations~\cite{greiner_mott_bec} as well as 
previous theoretical  expectations based on mean-field
calculations~\cite{jaksch_bose_hubbard}. The numerical simulations
have been mostly restricted to 1D
systems~\cite{batrouni_domains,kollath_dmrg_bose_hubbard_trap} and
isotropic lattices~\cite{wessel_MC_confined_bosons}, we expect most
of their conclusions to also apply to the anisotropic optical
lattices that interest us here. Thus, we also expect coexistence of
the phases described above when the system is confined in a harmonic
trap. Qualitatively, this  indeed can be understood using the
local density approximation, assuming that the trap potential gives
rise to a local chemical potential which varies from point to
point. Thus, from strong interactions we  expect the formation of
domains  of the 1D Mott insulating phase near the center of the
trap, whereas the superfluid phase should be located in the outer
part of the cloud. If the transverse lattice is very deep (\ie if
$V_{0\perp}$ is large) there also can be regions where the number of
boson per tube is small and  the the phase coherence is lost because
the  charging energy $E_C$ becomes locally larger than $E_J$.
Nevertheless, confirmation of this qualitative picture will require
 further  experimental and numerical investigation, which we hope this work helps to
motivate. The coexistence of different phases may be revealed by
analyzing the visibility of the interference pattern observed in an
expansion
experiment~\cite{grebier_visibility,sengupta_domains_visibility}.
Let us finally mention that the existence of the trap can also lead to
new and poorly understood phenomena such as  suppression of
quantum criticality~\cite{wessel_MC_confined_bosons}.

 Taking into account the above, we only aim at a qualitative description of
some experimental observations. Thus, we have shown in \ref{sec:becfrac}
that the strong quantum fluctuations characteristic of 1D systems strongly
deplete the condensate fraction in the BEC phase that forms for arbitrarily
small intertube  Josephson coupling.  The condensate fraction can be as low as
$\approx 10 \, \% $  for typical experimental parameters ($V_{0\perp} = 20\: E_R$).  
We have also argued that
the experimentally measured coherence fractions may also include some
of the non-condensate fraction,  and therefore, provided heating effects of the sample
in the measurement can be disregarded, the coherence fraction should be an
upper bound to the BEC fraction.
Quantum and thermal fluctuations are also responsible for the power-law behavior of the
condensation temperature with $J/\mu_{1D}$. Estimates of the latter, when
compared to the experimental estimates of the temperature of the
lattice~\cite{Moritz_oscillations} and the finite-size excitation gap
means that for $V_{0\perp} = 20 E_R$ the system
should be in the BEC phase described in \sref{sec:nomott} provided
the Mott potential is sufficiently weak.  In this phase, the system
exhibits two kinds of low energy modes in the long wave-length limit: a gapless
Goldstone mode, and a gapped longitudinal mode. Whereas the former
corresponds to fluctuations of the phase of the order parameter, the latter
is associated with fluctuations of its amplitude, namely the BEC density.

If the strength of the Mott potential is increased,
the system (or at least part of it near the center of the trap) should undergo
a deconfinement quantum phase transition  (or crossover, given the finiteness of the
sample) to a 1D Mott insulating phase.
This phase is characterized by the existence of a gap in the excitation spectrum
which is largely independent of the size of the tube. The hopping between
the tubes is suppressed for temperatures (or energies) below this gap, and
it is incoherent above it. For a deeper optical lattices, we predict that hopping
between tubes can also be suppressed by the  ``charging energy'' of the
(finite-sized) tubes.  We also notice that in some experiments the duration time
of the experiment is also a limiting factor for the establishment of phase coherence
even if hopping is not effectively suppressed by the charging energy.

St{\"o}ferle \emph{et al.}~\cite{stoferle_tonks_optical} have
studied an optical lattice system for several values of the depth of
the transverse optical lattice potential (hence $J$). By analyzing
the width of the  momentum peak near ${\bf k} = 0$ in the
interference pattern after free expansion, they were able to deduce
the energy absorption in response to a time-dependent modulation of
the Mott potential. From this analysis, they could tell when the
system behaves as a superfluid or as a Mott insulator. These 
experiments were analyzed recently~\cite{iucci_absorption,batrouni_shaking_qmc,reischl_temperature_mott,kollath_shaking_dmrg} and we will not repeat the conclusions of the analysis here. 
In the same series of experiments, they also studied the dependence of the width of the
momentum peak near ${\bf k} = 0$ in the ground state on the strength
of the Mott potential, for several values of $V_{0\perp}$. For a BEC
or a 1D superfluid, the inverse of the width is set by a geometrical
factor and it is proportional to the size of the size of the cloud
(or, in the 1D case, the thermal length $\hbar v_s/T$, whichever the
shortest). For a Mott insulator, however, the inverse of the width
is set by the correlation length $\xi_c \simeq \hbar v_s/\Delta_c$,
where $\Delta_c$ is the Mott gap. Thus, an increase of the width as
a function of the Mott potential reveal the opening of a gap in the
spectrum of the system~\cite{kollath_dmrg_bose_hubbard_trap}. The
experimental curves of the width as a function of the Mott potential
present upturns around the points where the system should enter the
Mott insulating phase in the thermodynamic limit. Our expectation,
based on the phase diagram of figure~\ref{fig2}, is that the
deconfinement phase transition should take place for values of the
Mott potential slightly  larger than in the pure 1D case
(represented by the vertical line at $K = 2$ in figure~\ref{fig2}).
It is necessary to recall that, in the Bose-Hubbard model (cf.
equation~(\ref{bosehubbard})) that describes the systems of
Ref.~\cite{stoferle_tonks_optical}, the Mott potential controls not
only the dimensionless parameter $g_u$ that enters in the RG
equations of \sref{sec:rg}, but also the ratio of the
interaction $U$ to the kinetic energy $J_x$, and therefore the
Luttinger parameter $K$ as well. The latter decreases towards the
Tonks limit ($K = 1$) with increasing $U/J_x$. The theoretical
expectation is in good  qualitative agreement with the
experimental observation.

The finite extent of trapped cloud (either longitudinally or
transversally) limits the minimum momenta of the modes. Thus  the
energy of the lowest modes in the 3D SF phase can be directly
estimated from our results by using the minimum  available momentum
in (\ref{RPAgoldstone},\ref{RPAgapped}). For instance, for a finite
2D lattice containing $M_y \times M_z$ tubes (\ie an atom cloud of
size $L \times M_y a \times M_z a$) the lowest available momentum is
$\sim \pi/(M_{i} a)$. Putting this value into
(\ref{RPAgoldstone},\ref{RPAgapped}) shows that the frequency of the
lowest transverse modes decreases with decreasing $J$. This analysis
neglects the possibility of phase coexistence  described above,
which calls for  further investigation of these issues.

\begin{figure}
\centerline{\includegraphics[width=\figwidth]{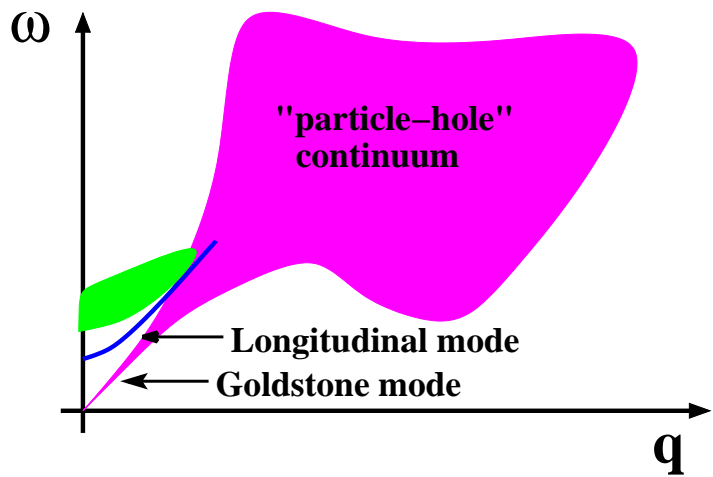}}
 \caption{Schematic excitation spectrum of the BEC phase. Shaded lines and curves represent different types of excitations. The frequency of the excitation is $\omega$ and $q$ is momentum in the longitudinal direction (\ie parallel to the tube axis). The continuum of  particle and hole excitations stems from the spectrum a 1D system, as 
found by Lieb in Ref.~\cite{lieb_excit}. At lower energies, however,
the system develops the phase coherence characteristic of a BEC and
therefore must obey Landau's criterion of superfluidity. In
particular, this means that the system must exhibit a gap at $q =
2\pi \rho_0$, where $\rho_0$ is the linear density of the 1D system.
In the long wave-length limit, we find the system exhibits two kinds
of excitations: a gapless (Goldstone) mode, and a gapped
longitudinal mode. There is also a continuum of excitations above
the longitudinal mode (see discussion at the end of \sref{sec:exbec}).}
 \label{fig4}
\end{figure}

We have been not able to calculate the finite temperature excitation spectrum
across the dimensional crossover, but it is interesting to speculate
what happens to the Tomonaga-Luttinger liquid (TLL) spectrum, especially 
when the longitudinal momentum $q$ is near $2 \pi \rho_0$. 
On general grounds, at temperatures large
compared to the dimensional crossover scale $\sim T_c$, the system
should exhibit   essentially the 1D spectrum of a TLL,  with its characteristic 
continuum of (particle-hole) excitations  extending  roughly between $q \approx 0$
and  $q \approx 2 \pi \rho_0$~\cite{lieb_excit}. As the temperature or the excitation
frequency  get  below the crossover scale, 
the low-energy excitations around $2 \pi \rho_0$ must
disappear: being the BEC a  3D superfluid, by Landau's
criterion, there can be no low energy excitation that  contributes
to dissipation of superfluid motion. The gap about $q = 2\pi \rho_0$ 
in the spectrum of the anisotropic BEC phase would be thus 
reminiscent of the roton gap in liquid Helium, which may be regarded as 
the remnant of the crystallization tendency at the  momentum of where the 
roton minimum occurs~\cite{nozieres_roton}. In our case, 
because of the underlying 1D physics,
the roton gap  should have a continuum  of excitations above it, 
which is the remnant of the 1D continuum of excitation in the
TLL (see figure~\ref{fig4}). This may account for the ``relatively broad''
energy absorption spectrum observed in the ``1D to 3D crossover''
regime in the experiment of St{\"o}ferle
\emph{et al.}~\cite{stoferle_tonks_optical,iucci_absorption}

\section{Conclusions and discussion} \label{sec:conclusion}

In summary, through a variety of methods, we have arrived at the
following  picture For the case where  the Mott
potential is either absent, incommensurate, or generally, irrelevant in the RG , 
the existence of an arbitrarily weak intertube tunneling immediately turns the
1D arrays into a 3D BEC at zero temperature, albeit with
anisotropic properties. At finite temperature, thermal fluctuations
cause a phase transition at $T = T_c$ where the 3D coherence (and the BEC)
is lost. Since $T_c \ll \mu_{1D}$ in the quasi-1D optical lattices studied here,
for $T_c < T \ll \mu_{1D}$ the system behaves as a 2D array of phase-incoherent
Tomonaga-Luttinger liquids. We have obtained explicit expressions for $T_c$, the
condensate fraction at $ T = 0$,  the excitation spectrum, and momentum distribution
function, which could be compared to measurements in the experimental systems. 
In particular, the power-law behaviors of 
$J/\mu$ found  for $T_c$ and condensate fraction are consequences of the
strong  fluctuations that dominate the properties of the uncoupled 1D systems.
In the presence of a relevant (in the RG sense)  Mott potential, a finite intertube tunneling ($J_c$) is required to overcome the localization tendencies of the interacting bosons. This process takes place in the form of a deconfinement quantum phase transition.
 Our main result in this case is the phase diagram shown in~\fref{fig2} as a
function of $J/\mu_{1D}$ and the parameter $K$
characterizing the interactions in the tube.  For the deconfinement transition,  
we also find a power-law dependence of $J_c$ on the strength of the Mott potential
$u_0$, and a power-law dependence of the condensate fraction near the 
transition (indeed, these two results are
strictly valid for $K=1/2$, which is outside the range of values for
bosons with short-range interactions, but we speculate that these results
are adiabatically connected to the physical (for bosons) regime of
$K > 1$ as we have found no evidence of any quantum phase
transitions for $1/2 \geq K \geq 1$ in our RG calculation.) Finally,
for comparisons to experiments, we have looked into trap effects by
studying tubes of finite length and described the transition where
the system goes from a system of uncoupled 1D Luttinger liquids to
the 3D superfluid as (renormalized) $J$ overcomes the ``charging
energy'' of the finite length tube. As discussed in details in the
previous section we can compare our predictions with the experiments
of St{\"o}ferle \emph{et al.} \cite{stoferle_tonks_optical}. Some of our
predictions, in particular the large reduction of the coherent
fraction is in qualitative agreement with the experiment.

Clearly, further experimental and theoretical studies are needed to 
provide a more thorough and quantitative picture  of
the dimensional crossover and the deconfinement transition. From the theoretical
point of view, taking into account the harmonic trap in a more
quantitative way would be a step towards a more quantitative
description of the experiments. From the
experimental point of view, investigation of larger systems with more
atoms per tube and more tubes, and in longer experiments that
allow to probe smaller values of the Josephson coupling $J$, would
permit a more complete exploration of the phase diagram of the quasi-1D lattices. 
Also, most of the current experiments in these  lattices
have been carried out using rather deep Mott potentials in order
to reduce the ratio of kinetic to interaction energy (this is the regime
where the Bose-Hubbard model of equation~(\ref{bosehubbard}) is applicable). However,  It would be very interesting to conduct experiments  where the interaction energy between the atoms is tuned independently of the longitudinally applied (Mott)
potential. Increasing the interaction strength in the absence of Mott potential 
was recently achieved by Kinoshita \emph{et al.}~\cite{kinoshita_tonks_continuous} 
by confining the atoms in tighter 1D traps.  Under these conditions, and 
for interaction strength $\gamma \approx 3.5$~\cite{haldane_bosons, buchler_cic_bec} 
in 1D,  the application of the Mott potential of strength much smaller than the 
chemical potential $\mu_{1D}$ should suffice to cause localization
of the bosons in the 1D Mott insulating phase. To the best of our knowledge,
this ``weak lattice'' Mott insultator has not yet really been realized.
In this phase quantum effects are stronger, and it would be then very interesting to 
drive the system across the deconfinement transition by changing $J$. 

\section{Acknowledgments}

We thank  T. Esslinger, A. Nersesyan, and M. Gunn for useful
conversations and/or correspondence. MAC and AFH thank the hospitality of
the University of Geneva where this work was started. AFH is funded
by EPSRC (UK), MAC by \emph{Gipuzkoako Foru Aldundia} and MEC
(Spain) under Grant FIS2004-06490-C03-01. This work was supported in
part by the Swiss National Science Foundation through MaNEP and
division II.
\appendix

\section{Mean-field theory (MFT) and gaussian fluctuations (RPA)} \label{app:rpa}
In this section we shall employ the functional integral to derive the mean-field Hamiltonian theory of \sref{sec:mft}, and to go further by taking into account gaussian fluctuations about the mean-field state. The method is known as the random-phase approximation (RPA). Our treatment follows  Ref.~\cite{donohue_thesis}.

  Let  us consider the model defined by equation~(\ref{ham2}). Using the coherent-state path-integral, the partition function  can be written as follows:
\begin{equation}
Z = \int \left[d\psi d\psi \right] \, e^{-S\left[\psi^*,\psi\right]},
\end{equation}
where the (euclidean) action,
\begin{equation}
S\left[\psi^*,\psi \right] = \frac{1}{\hbar} \int^{\hbar \beta}_0 d\tau \int^L_0 dx \left[  \sum_{\bf R} \psi^*_{\bf R} \left( \hbar\partial_\tau \psi_{\bf R} - \mu \psi_{\bf R} \right)+  {\cal H}(\psi^*_{\bf R},\psi_{\bf R})\right],
\end{equation}
${\cal H}(\psi^*_{\bf R},\psi_{\bf R})$ being the Hamiltonian density corresponding to
equation~(\ref{ham2}).  Since the
Mott potential plays no role in the BEC phase (\ie it is irrelevant in the RG sense), we drop it from the Hamiltonian ${\cal H}(\psi^*_{\bf R},\psi_{\bf R})$. However,
the action  $S\left[\psi^*,\psi\right]$ still contains the non-trivial
Josephson coupling between neighboring 1D systems.
\begin{eqnarray}
S_J\left[ \psi^*_{\bf R}, \psi_{\bf R} \right] &=& -\frac{J}{\hbar} \sum_{\langle {\bf R}, {\bf R}' \rangle} \int^{\hbar \beta}_0 d\tau \int^L_0 dx \,  \psi^*_{\bf R}(x, \tau) \psi_{\bf R'}(x, \tau) \\
&=& -  \sum_{{\bf R}, {\bf R}'} \int^{\hbar \beta}_0 \frac{d\tau}{\hbar}  \int^L_0 dx \, \psi^*_{\bf R}(x, \tau)
J_{{\bf R}-{\bf R}'} \psi_{\bf R'}(x,\tau).
\end{eqnarray}
In the above expression we have introduced the following notation:
\begin{equation}
J_{{\bf R}-{\bf R}'}  = J\:
\sum_{\mathbf{t}} \delta_{{\bf R}'-{\bf R}. \mathbf{t}},
\end{equation}
where $\mathbf{t}$ runs over set of vectors that connect a lattice point to its nearest neighbors. We perform a Hubbard-Stratonovich (HS) decoupling of
$S_J\left[\psi^{*},\psi\right]$:
\begin{equation}
\fl
e^{-S_J\left[\psi^{*},\psi\right]} = {\cal N}
\int \left[ d\xi^* d\xi\right] \,
e^{- \int^{\hbar\beta}_0 \frac{d\tau}{\hbar} \int^L_0 dx \, \left\{ \sum_{{\bf R}, {\bf R}'}\
\xi^*_{\bf R}(x,\tau) J^{-1}_{{\bf R},{\bf R}'} \xi_{{\bf R}'}(x,\tau)  + \sum_{\bf R}
\left[ \psi^{*}_{\bf R}(x,\tau)\xi_{\bf R}(x,\tau) + \cc  \right] \right\} }
\end{equation}
To make sense of $J^{-1}_{{\bf R}-{\bf R}'}$ the Fourier transform of
$J_{{\bf R},{\bf R}'}$ is needed,
\begin{equation}
J({\bf Q}) = J \sum_{\mathbf{t}} \: e^{i{\bf Q}\cdot  \mathbf{t}} =  2J\: \left[ \cos  Q_y a  + \cos  Q_z a \right].
\end{equation}
The last result applies to a square lattice (in the OYZ plane) with lattice parameter
 $a$. Hence,
\begin{equation}
J^{-1}_{{\bf R}-{\bf R}'} = \frac{1}{M_0}\sum_{{\bf Q} \in 1BZ} \:\frac{ e^{i {\bf Q} \cdot\left({\bf R}-{\bf R}' \right)}}{J({\bf Q})}.
\end{equation}
Thus, after performing the HS decoupling, the action becomes:
\begin{equation}
\fl
S\left[\xi^*,\xi,\psi^*,\psi \right] = S_{B} \left[\psi^*,\psi \right] + S_J\left[\xi^*,\xi \right] + \int^{\hbar\beta}_0  \frac{d\tau}{\hbar} \int^L_0 dx\: \sum_{\bf R} \left[\xi^*_{\bf R}(x,\tau) \psi_{\bf R}(x,\tau) + \cc  \right], \label{sdec}
\end{equation}
where
\begin{eqnarray}
\fl
S_{B}\left[\psi^*,\psi \right] =  \int^{\hbar \beta}_0 \frac{d\tau}{\hbar} \int^L_0 dx   \sum_{\bf R}  \left[\psi^*_{\bf R} \left(\hbar \partial_\tau-\mu\right) \psi_{\bf R} + \frac{\hbar^2}{2M} \left| \partial_x\psi_{\bf R}\right|^2 + \frac{g}{2} |\psi_{\bf R}|^4\right],\\
\fl
S_J\left[\xi^*,\xi \right] = \int^{\hbar\beta}_0 \frac{d\tau}{\hbar} \int^L_0 dx \sum_{{\bf R},{\bf R}'}\,  \xi^*_{\bf R}(x,\tau)J^{-1}_{{\bf R}-{\bf R}'} \xi_{\bf R'}(x,\tau).
\end{eqnarray}
To perform the mean field approximation we formally integrate out the boson degrees of freedom described by $\psi_{\bf R}(x,\tau)$ and $\psi^*_{\bf R}(x,\tau)$, and obtain the effective action for the HS fields:
\begin{eqnarray}
S_{GL}\left[\xi^*,\xi  \right] &=& S_{J}\left[\xi^*,\xi  \right] + S_{\psi}\left[\xi^*,\xi\right],\\\
S_{\psi}[\xi^*,\xi] &=& - \ln \langle
e^{-\int^{\hbar\beta}_0 \frac{d\tau}{\hbar} \int^{L}_{0} dx \left[\xi^{*}_{\bf R}(x,\tau)\psi_{\bf R}(x,\tau) +
\psi^{*}_{\bf R}(x,\tau)\xi_{\bf R}(x,\tau)  \right]} \rangle_{\psi^*,\psi}. \label{spsi}
\end{eqnarray}
We have introduced the notation $S_{GL}$ for the effective action to indicate that it is indeed the GL functional.

  To obtain the mean-field equations we perform a saddle-point expansion. The
saddle-point is determined by the following equations:
\begin{eqnarray}
\frac{\delta S_{GL}\left[\xi^*,\xi  \right]}{\delta\xi^*_{\bf R}(x,\tau)} =  \frac{1}{\hbar}\sum_{{\bf R}'}  J^{-1}_{{\bf R} -{\bf R}'} \xi_{{\bf R}' c}(x,\tau)  + \frac{\delta S_{\psi}[\xi^*_c,\xi_c]
}{\delta\xi_{\bf R}(x,\tau)} = 0 \label{mf},\\
\frac{\delta S_{GL}\left[\xi^*,\xi  \right]}{\delta\xi({\bf R},x,\tau)}=  \frac{1}{\hbar}\sum_{{\bf R}'} \xi^*_{\bf R'}(x,\tau) J^{-1}_{{\bf R}' -{\bf R}}   + \frac{\delta S_{\psi}[\xi^*_c,\xi_c]}
{\delta\xi^*_{\bf R}(x,\tau)} = 0 \label{mf2}.
\end{eqnarray}
We look for  uniform solutions to the above equations, \ie  $\xi_{\bf R}(x,\tau) = \xi_c$ and $\xi^{*}_{\bf R}(x,\tau) = \xi^*_c$. Furthermore, because of global gauge invariance, the particular choice for the phase of $\xi_0$ should not matter, that is, if $(\xi^*_c,\xi_c)$ is a saddle point then so is $(e^{-i\theta_c} \xi^*_c,
e^{i\theta_c}\xi_c)$.  Thus $S_{\psi}\left[\xi^*_c,\xi_c\right] = S_{\psi}(|\xi_c|^2)$.
This implies that the above equations reduce to:
\begin{equation}\label{saddle}
\frac{1}{M_0 L\hbar\beta} S'_{GL}(|\xi_c|^2) = \frac{1}{\hbar} \sum_{\bf R} J^{-1}_{\bf R} + \frac{1}{M_0L\hbar\beta} S'_{\psi}(|\xi_c|^2) = 0,
\end{equation}
where the prime stands for the derivative with respect to $\rho_{\xi}=|\xi|^2$.
Furthermore, using  equation~(\ref{spsi}), equation~(\ref{mf})  can also be written as:
\begin{equation}\label{mf3}
\sum_{{\bf R}'} J^{-1}_{{\bf R} -{\bf R}'} \xi_c  = -\langle \psi_{\bf R}(x,\tau) \rangle = -\psi_c,
\end{equation}
where it is understood that  the expectation value of $\psi_{\bf R}(x,\tau)$ is taken by tracing out the boson degrees of freedom in the presence of the constant HS
$(\xi^*_c,\xi_c)$, \ie. using $S[\xi^*_c, \xi_c, \psi^*, \psi]$ (cf. equation~(\ref{sdec})).
Therefore, (\ref{mf2}) must be solved self-consistently for $(\xi^*_c, \xi_c)$: This
is the self-consistency condition mentioned in \sref{sec:mft}. Upon
multiplying equation~(\ref{mf3}) by $J_{{\bf R}''-{\bf R}}$ from the left and summing
over ${\bf R}$, we arrive at:
\begin{equation}
\xi_c({\bf R},x,\tau) = \xi_c = - z_C J\psi_c,
\end{equation}
where $z_C$ is the number of nearest neighbors ($z_C = 4$ for a 2D square lattice).
The above system of equations defines the mean field theory, which we have solved,
in Hamiltonian form, in \sref{sec:mft}.

 We next take into account fluctuations by considering the gaussian corrections to the saddle point  $(\xi^*_c,\xi_c)$. Let
\begin{eqnarray}
\delta \xi_{\bf R}(x,\tau) &=& \xi_{\bf R}(x,\tau) - \xi_c,\\
\delta \xi^*_{\bf R}(x,\tau)  &=& \xi^*_{\bf R}(x,\tau) - \xi^*_c.
\end{eqnarray}
To proceed further we need to integrate out the boson degrees of freedom  and obtain $S_{\psi}[\xi^*,\xi]$ explicitly. Since we are interested in the properties at long wave-lengths and low frequencies and temperatures, we first apply bosonization and
obtain:
\begin{equation}
\fl
S[\xi^*,\xi,\theta] = \sum_{\bf R} S^{0}_{\bf R}[\theta] + S_J[\xi^*,\xi] + \sqrt{{\cal A}_{B} \rho_0}  \int^{\hbar\beta}_0 \frac{d\tau}{\hbar} \int^L_0 dx\: \sum_{\bf R}
\left[\xi^*_{\bf R}(x,\tau) e^{i\theta_{\bf R}(x,\tau)} + \cc \right].\label{baction}\\
\end{equation}
In the above expression,
\begin{equation}
S^{0}_{\bf R}[\theta] =  \frac{K}{2\pi}\int^{\hbar\beta}_0
d\tau \int^{L}_0 dx  \left[ \frac{1}{v_s}
\left(\partial_\tau\theta_{\bf R}(x,\tau) \right)^2 + v_s \left
(\partial_x\theta_{\bf R}(x,\tau) \right)^2 \right]. \label{GaussianA}
\end{equation}
In this (bosonized) form, integrating out the boson degrees of freedom amounts to tracing out the phase field $\theta_{\bf R}(x,\tau)$. This will carried out in what
follows using the cumulant expansion:
\begin{equation}
\fl
\ln \Bigl \langle e^{-A\left[\xi^*,\xi,\theta\right]} \Bigr\rangle_{sG}  = - \langle A\left[\xi^*,\xi,\theta\right] \rangle_{sG} + \frac{1}{2}
\bigl \langle \big( A\left[\xi^*, \xi,\theta\right] - \langle A\left[ \xi^*, \xi,\theta\right]
\rangle_{sG} \big)^2 \bigr  \rangle_{\rm sG}  +\cdots
\end{equation}
where $\langle \ldots \rangle_{sG}$ denotes the trace over (smooth) configurations of $\theta_{\bf R}(x,\tau)$ using the weight $e^{-S[\xi^*_c, \xi_c, \theta]}$ (cf. equation~(
\ref{baction}), note that this action defines a collection of independent sine-Gordon models, which explains the notation). The RPA corresponds to keeping the terms of the above expansion up to quadratic order in  $\delta \xi$ and $\delta \xi^*$. Thus,  using that
\begin{equation}
\fl
A\left[\xi^*,\xi,\theta\right] = \int^{\hbar\beta}_0 \frac{d\tau}{\hbar} \int^L_0 dx \sum_{\bf R} \left[\xi^*_{\bf R}(x,\tau) e^{i\theta_{\bf R}(x,\tau)} + e^{-i\theta_{\bf R}(x,\tau)} \xi_{\bf R}(x,\tau) \right],
\end{equation}
we obtain:
\begin{eqnarray}
\fl
S_{\psi}\left[ \xi^*,\xi \right] &=& \frac{1}{\hbar}\sqrt{\rho_0 {\cal A}_{B}}\:
\int^{\hbar\beta}_0 d\tau \int^L_0 dx \sum_{\bf R} \left[\delta\xi^*_{\bf R}(x,\tau) \, \langle e^{i\theta_{\bf R}(x,\tau)} \rangle + \langle e^{-i\theta_{\bf R}(x,\tau)} \rangle \, \delta\xi_{\bf R}(x,\tau) \right] \nonumber\\
\fl
&&- \frac{\rho_0 {\cal A}_{B}}{2\hbar^2}\int^{\hbar\beta}_0 d\tau\: d\tau' \int^L_0 dx \: dx'\: \sum_{{\bf R},{\bf R'}} \:\Big[
\langle\langle e^{-i\theta_{\bf R}(x,\tau)} e^{-i\theta_{\bf R'}(x',\tau')}\rangle\rangle
\:  \delta\xi^*_{\bf R}(x,\tau) \: \delta\xi^*_{\bf R'}(x',\tau')\nonumber\\
\fl
&&+   \langle\langle e^{-i\theta_{\bf R}(x,\tau)}
e^{-i\theta_{\bf R'}(x',\tau')} \rangle\rangle\: \delta\xi_{\bf R}(x,\tau)\:
\delta\xi_{\bf R'}(x',\tau')  \nonumber\\
\fl
&& + 2
\langle\langle e^{i\theta_{\bf R}(x,\tau)} e^{-i\theta_{\bf R'}(x',\tau')}\rangle\rangle\:
\delta\xi^*_{\bf R}(x,\tau)\:  \delta\xi_{\bf R'}(x',\tau') \Big] + \cdots \label{ssg}
\end{eqnarray}
We have introduced the following notation:
\begin{eqnarray}
\langle \langle F[\theta(x,\tau) G[\theta(x',\tau')] \rangle \rangle &=&
\langle \left( F[\theta(x,\tau)] - \langle F[\theta(0,0)] \rangle_{sG} \right) \nonumber\\
&&\times
\left( G[\theta(x',\tau')] - \langle G[\theta(0,0)] \rangle_{sG}  \right) \rangle_{\rm sG},
\end{eqnarray}
with
\begin{equation}
\langle F[\theta(x,\tau)] \rangle_{\rm sG} =
\frac{\int \left[ d\theta \right]\,
F[\theta(x,\tau)] \,  e^{-S_{\rm sG}[\theta,\xi^*_c,\xi_c]}}{\int \left[d\theta\right] e^{-S_{\rm sG}[\theta,\xi^*_c,\xi_c]}}
\end{equation}
and
\begin{eqnarray}
S_{sG} [\theta,\xi^*_c,\xi_c] &=&  \frac{K}{2\pi}\int^{\hbar\beta}_0  d\tau \int^{L}_0 dx \sum_{\bf R} \left[ \frac{1}{v_s} \left(\partial_\tau\theta_{\bf R}(x,\tau) \right)^2 + v_s \left(\partial_x\theta_{\bf R}(x,\tau) \right)^2 \right] \nonumber\\
&& + \sqrt{{\cal A}_B \rho_0} \: \int^{\hbar\beta}_0  \frac{d\tau}{\hbar} \int^{L}_0 dx \sum_{\bf R}\left[ \xi^*_{c} \,  e^{i\theta_{\bf R}(x,\tau)} +  \cc \right] \label{ssg2}.
\end{eqnarray}
Since in this action the different lattice sites are uncoupled, the correlation functions
are diagonal in $\bf R$ and $\bf R'$.  Thus we can write the quadratic action that describes the (gaussian) fluctuations around the saddle point as follows:
\begin{eqnarray}
\fl
S^{\rm RPA}_{GL}[\Xi^{\dag},\Xi] = \frac{1}{2\hbar}\int^{\hbar\beta}_{0} d\tau \:d\tau' \int dx\: dx' \sum_{{\bf R},{\bf R}'}\:
\Big[\Xi^{\dag}_{\bf R}(x,\tau)  {\bf J}^{-1}_{{\bf R}-{\bf R}'}(x-x',\tau-\tau') \Xi_{\bf R'}
(x',\tau') \nonumber\\
\fl
+ \left(\Xi^{\dag}_{\bf R}(x,\tau) +Q^{\dag}_{\bf R}(x,\tau)\right) {\bf G}_{{\bf R}-{\bf R}'}(x-x',\tau-\tau') \left(\Xi_{\bf R'}(x',\tau') + Q_{\bf R'}(x',\tau')\right) \Big]
\label{eqrpa}
\end{eqnarray}
Note that the terms in $S^{\rm RPA}_{GL}$ linear in $\Xi$ and $\Xi^{\dag}$ vanish because we expand around the saddle point. We have also dropped the constant term that is proportional to the mean field free energy. In the above expression the following spinors have been introduced:
\begin{eqnarray}
\Xi_{\bf R}(x,\tau) &=& \left[  \begin{array}{c} \delta\xi_{\bf R}(x,\tau) \\ \delta\xi^{*}_{\bf R}(x,\tau) \end{array}\right],
Q_{\bf R}(x,\tau) = \left[  \begin{array}{c} q_{\bf R}(x,\tau) \\ q^{*}_{\bf R}(x,\tau) \end{array} \right]\\
\Xi^{\dag}_{\bf R}(x,\tau)  &=& \left[  \begin{array}{cc}\delta\xi^{*}_{\bf R}(x,\tau)  & \delta\xi_{\bf R}(x,\tau) \end{array}\right],  \\
 Q^{\dag}_{\bf R}(x,\tau)  &=& \left[  \begin{array}{cc}
q^{*}_{\bf R}(x,\tau)  & q_{\bf R}(x,\tau) \end{array}\right],
\end{eqnarray}
where we have introduced the sources $q^*_{\bf R}(x,\tau)$ and  $q_{\bf R},x,\tau)$, which in the original  action are coupled to $\psi_{\bf R}(x,\tau)-\psi_c$ and
$\psi^*_{\bf R}(x,\tau)-\psi^*_c$, respectively. The matrices:
\begin{eqnarray}
{\bf J}^{-1}_{{\bf R}-{\bf R'}}(1-2) &=& J^{-1}_{{\bf R}-{\bf R}'} \delta(x-x')\delta(\tau-\tau') \: {\rm diag}\left[1,1\right],\\
{\bf G}_{{\bf R}-{\bf R'}}(1,2) &=& \left[ \begin{array}{cc}  g^{+-}_1(1,2) &
e^{+2i\theta_c} g^{++}_1(1,2) \\
e^{-2i\theta_c} g^{--}_1(1,2)  & g^{-+}_1(1,2)
\end{array}
 \right],
\end{eqnarray}
with $1 = ({\bf R},x,\tau)$ and $2 = ({\bf R}',x',\tau')$, and ($p,q = \pm$):
\begin{equation}
g^{pq}_{1}(1,2) =  - \frac{{\cal A}_{B} \rho_0}{\hbar}\langle \langle e^{ip \theta_{\bf R}(x,\tau)}
e^{iq\theta_{\bf R}(x',\tau')} \rangle \rangle_{\rm sG}\: \delta_{{\bf R}-{\bf R}', \mathbf{0}}
\,\, , \label{cf1}
\end{equation}
To perform the averages using $S_{sG}$ a gauge transformation is
applied $ \theta_{\bf R}(x,\tau) - \theta_c\to \theta_{\bf R}(x,\tau)$ so
that $\xi_c = - z_C J \psi_c$ can be treated as a real number.

     Let us now pause  to consider the consequences of global gauge invariance  on the correlation functions of~(\ref{cf1}). As we have found above, the GL functional is a function  of $|\xi_c|^2$ only. Thus,  if we expand  $S_{\psi}(|\xi |^2)$ around the saddle point solution:
\begin{eqnarray}
\fl
S_{\psi}(|\xi|^2) = S_{\psi}(|\xi_c|^2) + \frac{\partial S_{\psi}}{\partial\xi}(|\xi_c|^2) \delta\xi +   \frac{\partial S_{\psi}}{\partial\xi}(|\xi_c|^2) \delta\xi^*
+ \frac{1}{2!}\frac{\partial^2 S_{\psi}}{\partial\xi^2}(|\xi_c|^2) (\delta \xi)^2 \nonumber\\
+ \frac{1}{2!}\frac{\partial^2 S_{\psi}}{\partial\xi^{*2}}(|\xi_c|^2)  (\delta \xi^*)^2
 + \frac{\partial S_{\psi}}{\partial \xi \partial\xi^*}(|\xi_c|^2) |\delta \xi|^2 + \cdots \label{dv1}
\end{eqnarray}
where
\begin{eqnarray}
 \frac{\partial^2 S_{\psi}}{\partial\xi^2}(|\xi_c|^2) = \xi^2_c\: S^{\prime\prime}_{\psi}(|\xi_c|^2),\\
 \frac{\partial^2 S_{\psi}}{\partial\xi\partial\xi^*}(|\xi_c|^2) = S'_{\psi}(|\xi_c|) ^2\: + |\xi_c|^2 S^{\prime\prime}_{\psi}(|\xi_c|^2).\label{dv2}
\end{eqnarray}
Using equation~(\ref{saddle}), $S'_{\psi}(|\xi_0^2|) = -(M_0L \beta)\: \sum_{\bf R} J^{-1}_{\bf R} = -(M_0 L \beta)\: \left[ J({\bf Q} = 0) \right]^{-1}$. Thus,
by setting $\delta \xi^*,\delta\xi = {\rm const.}$ and taking $\beta \to \infty$
in  equation~(\ref{ssg}), and comparing with equation~(\ref{dv1}),
the following results  can be obtained:
\begin{eqnarray}
\fl
\tilde{g}_1(q = 0, \omega = 0) = g^{++/--}(q=0,\omega=0) = \frac{e^{2i\theta_c}}{M_0 L\beta}  \frac{\partial^2  S_{\psi}}{\partial\xi^2}(|\xi_c|^2) = \hbar \frac{\partial^2{\cal E}_{sG}}{\partial\xi^2}, \label{lim01}\\
\fl
g_1(q = 0, \omega = 0) = g^{+-/-+}(q=0,\omega=0) = \frac{1}{M_0L\beta}  \frac{\partial^2   S_{\psi}}{\partial\xi\partial\xi^*}(|\xi_c|^2) =  \hbar \frac{\partial^2{\cal E}_{sG}}{\partial\xi\partial\xi^*},
\label{lim02}
\end{eqnarray}
where we have used that $\lim_{\beta \to +\infty} S_{\psi}(|\xi_c|^2)/(M_0L\hbar\beta) = {\cal E}_{sG}(|\xi_c|^2)$ is the energy density of the sine-Gordon model, equation~(\ref{Edens}). We shall use the above identities to calculate of the energy dispersion of the modes in the BEC phase. This phase exists below the critical temperature ($T_c$ to be determined below), and is  characterized by having $\psi_c \propto \xi_c \neq 0$ so that global gauge symmetry is spontaneously broken. To compute the excitation spectrum in this phase,  we  diagonalize the matrix ${\bf J}^{-1} +  {\bf G}$ (henceforth we drop the index ${\bf R}-{\bf R}'$). ${\bf J}^{-1}$ is already diagonal, whereas the matrix ${\bf G}$ is rendered diagonal by:
\begin{eqnarray}\label{eq:matu}
{\bf U} = \frac{1}{\sqrt{2}} \left[ \begin{array}{cc} e^{i\theta_c} & e^{i\theta_c}\\
e^{-i\theta_c} & -e^{-i\theta_c} \end{array} \right].
\end{eqnarray}
Thus,
\begin{equation}
\fl
{\bf C}(q,\omega_n) = {\bf U}^{\dag} {\bf G}(k,\omega_n) {\bf U} = \left[ \begin{array}{cc}
g_1(q,\omega_n) + \tilde{g}_1(q,\omega_n) & 0 \\
0 & g_1(q,\omega_n) - \tilde{g}_1(q,\omega_n)
\end{array} \right],
\end{equation}
where $\tilde{g}_1(q,\omega_n)$ is the Fourier transform anomalous correlator $g^{++}_1(x,\tau) = g^{--}_1(x,\tau)$. In what follows we  denote the eigenvalues of ${\bf G}(q,\omega_n)$ as $c_{\pm}(q,\omega_n) = g_1(q,\omega_n)\pm \tilde{g}_{1}(q,\omega_n)$. Next we  express $S^{\rm RPA}_{GL}$ in terms of the eigenvectors of ${\bf G}(q,\omega)$ and to subsequently   integrate out the transformed HS fields.
In the following we use a matrix notation where the summations over ${\bf R}, {\bf R'}$ as well as over $q,\omega_n$ and multiplication by $\hbar^{-1}$ are implicitly understood. Hence,
\begin{eqnarray}
\fl
S^{\rm RPA}_{GL}\left[\Xi^{\dag},\Xi,Q^{\dag},Q\right] = \frac{1}{2}\left[\Xi^{\dag}\left({\bf J}^{-1} + {\bf G}\right)\Xi +
Q^{\dag} {\bf G} Q + Q^{\dag} {\bf G} \Xi + \Xi^{\dag}{\bf G}Q \right].
\end{eqnarray}
Let us make the replacement ${\bf G} = {\bf U} {\bf C} {\bf U}^{\dag}$, so that
\begin{eqnarray}
\fl
S^{\rm RPA}_{GL} \left[X^{\dag},X,P^{\dag},P \right] =\frac{1}{2} \left[X^{\dag} \left({\bf J}^{-1} + {\bf C}\right) X + P^{\dag}{\bf C}P + P^{\dag}{\bf C}X + X^{\dag}{\bf C} P\right],
\end{eqnarray}
where  $X = {\bf U}^{\dag} \Xi$ ($X^{\dag} = \Xi^{\dag}{\bf U}$), and $P = {\bf U}^{\dag} Q$ ($P^{\dag} = Q^{\dag} {\bf U}$). We shall obtain the generating
functional in terms of these new sources.  We can now integrate over the auxiliary fields
$X^{\dag},X$. Since they are related to $\Xi^{\dag},\Xi$ by a global unitary transformation the measure of the functional integral is not affected. Thus we are free to shift $X\to X - \left({\bf J}^{-1} + {\bf C}\right)^{-1} {\bf C} P$
and  $X^{\dag}\to X^{\dag} - P^{\dag} {\bf C} \left({\bf J}^{-1} + {\bf C}\right)^{-1}$, which yields
\begin{eqnarray}
\fl
Z\left[P^{\dag},P\right] = \int \left[ dX^{\dag} dX \right] \, e^{-\frac{1}{2}X^{\dag} {\bf C} X - \frac{1}{2}P^{\dag} \left( {\bf C} -  {\bf C}\frac{1}{{\bf J}^{-1}+ {\bf C}}  {\bf C}\right) P}
= {\cal N'} e^{- \frac{1}{2}P^{\dag} \left( {\bf C} -
{\bf C}\frac{1}{{\bf J}^{-1}+ {\bf C}}  {\bf C}\right) P},
\end{eqnarray}
where $\cal N'$ is an unimportant constant.
Although we have been careful to treat the matrices in the last expression as non-commuting, all of them are diagonal and the last expression can be simplified to:
\begin{equation}
Z\left[P^{\dag},P\right] = Z[0,0] \: \exp\left[-\frac{1}{2}P^{\dag} \frac{\bf C}{\mathbf{1}+{\bf J\: C}} P\right]
\end{equation}
Thus, the system has two modes, whose propagation is described by $\mathbf{\Delta} = {\bf C} ({\bf 1}+{\bf J\: C})^{-1}$. When expressed in Fourier components
this propagator reads:
\begin{eqnarray}
\mathbf{\Delta}({\bf Q},q,i\omega_n) = \left[  \begin{array}{cc} \Delta_{+}({\bf Q},q,\omega_n) & 0\\ 0 &\Delta_{-}({\bf Q},q,\omega_n) \end{array} \right],\\
\Delta_{\pm}({\bf Q},q,\omega) = \frac{c_{\pm}(q,\omega_n)}{{\bf 1}+J({\bf Q})c_{\pm}(q,\omega_n)}
\end{eqnarray}
The excitations are poles of the propagator:
\begin{equation}
1 + J({\bf Q}) c_{\pm}(q, i\omega_n \to \omega_{\pm}(q,{\bf Q}) )
1 + J ({\bf Q})c^R_{\pm}(q, \omega_{\pm}(q,{\bf Q}))= 0.\label{dispersion}
\end{equation}
Before we solve these equations in detail, let us first demonstrate the existence of the Goldstone mode, namely that there is a solution of~(\ref{dispersion})
at $\omega = 0$ for $|{\bf Q}|$ and $q \to 0$. To this end,  we use (\ref{lim01},\ref{lim02}), which are a consequence of gauge invariance, together with
~(\ref{dv1},\ref{dv2}), to arrive at
 $c_{-}(q=0,\omega=0) = g_1(q=0,\omega=0) - \tilde{g}_1(q=0,\omega=0) = S'_{\psi}(|\xi_c|^{2})/(M_0 L \beta)$. Moreover, by virtue of equation~(\ref{saddle}), $c_{-}(q=0,\omega=0) = S'_{\psi}(|\xi_0|^{2})/(M_0L \beta)
= -\sum_{\bf R} J^{-1}_{\bf R} =- [J({\bf Q}=0)]^{-1} = -1/(Jz_C)$. This
implies that $\omega = 0$ is a solution of equation~(\ref{dispersion})  for $|{\bf Q}|,q \to 0$,
as anticipated, and identifies the eigenvalue labeled as '$-$'
as  the dispersion of the Goldstone mode. We shall use the condition that $c_{-}(q=0,\omega=0) = -[J({\bf Q} = 0)]^{-1}$ in the calculation of the full dispersion of this mode below.
But before we proceed into this calculation we need to know more about the correlation functions
$c_{\pm}(k,\omega)$ for general $k,\omega$. Let us for a bit return to real space; we first notice that
\begin{eqnarray}
\fl
c_{+}(x,\tau) &=& g_1(x,\tau) + \tilde{g}_1(x,\tau) = -\frac{2}{\hbar} \left[ {\cal A}_{B}\rho_0\,  \langle\cos\theta(x,\tau) \cos \theta(0,0) \rangle_{\rm sG} - \psi^2_c\right],\\
\fl
c_{-}(x,\tau)  &=&  g_1(x,\tau) - \tilde{g}_1(x,\tau)  = -\frac{2}{\hbar} {\cal A}_{B}\rho_0 \, \langle  \sin\theta(x,\tau) \sin \theta(0,0) \rangle_{\rm sG}.
\end{eqnarray}
To make contact with the notation of Ref.~\cite{lukyanov_sinegordon_correlations},
we introduce $\Phi = \theta/\beta$, $\beta^2 = \pi/K$, ${\bf x}=(v_s\tau,x)$, and
$\mu =  \sqrt{{\cal A}_B \rho_0}\xi_c/\hbar v_s =  \sqrt{{\cal A}_B \rho_0}J z_C \psi_c/\hbar v_s  $. The excitations of such a sine-Gordon model are the solitons and anti-solitons. Furthermore, $\beta^2 < 4\pi$ (\ie $K > \frac{1}{4}$) the model also has breather excitations~\cite{gogolin_1dbook,giamarchi_book_1d}, which are soliton-anti-soliton bound states, and whose  zero-momentum energies (``mass" gaps) are~\cite{gogolin_1dbook,giamarchi_book_1d}:
\begin{equation}
\Delta_n  = 2 \Delta_s \sin(\pi p n/2),
\end{equation}
with $p = \beta^2/(8\pi-\beta^2)$, $n = 1,2,\ldots,[p^{-1}]-1$, and $\Delta_s \sim \mu^{\frac{1}{2-{1}/{4K}}}$ is energy gap of the soliton (see equation~(\ref{smass})).
The operators $\sin \beta \Phi({\bf x})$ and $\cos \beta \Phi({\bf x})$ have
zero conformal spin and therefore can only create breathers (if they exist)
or pairs of solitons and anti-solitons. Furthermore, since under
charge conjugation~\cite{gogolin_1dbook},
\begin{equation}
{\cal C}^{-1} \Phi {\cal C} = -\Phi.
\end{equation}
it follows that
\begin{eqnarray}
{\cal C}^{-1} (\cos \beta \Phi) {\cal C} &=& + \cos \beta\Phi,\\
{\cal C}^{-1} (\sin \beta \Phi) {\cal C} &=& -\sin \beta\Phi.
\end{eqnarray}
Thus the cosine operator can only have poles corresponding to breather states
that are even under conjugation, whereas the sine can only have poles for
breathers that are odd under conjugation. Taking all these facts into
account we can write the spectral function of the correlators of these
two operators~\cite{gogolin_1dbook}:
\begin{eqnarray}
\rho^{\sin}_{\rm sG}(q,\omega) &=& 2\pi Z \sum_{n \, {\rm odd}} w_{n} \:\delta(\omega^2 - v^2_s q^2 - \Delta^2_{n}) + \rho^{\sin}_{\rm s\bar{s}}(q,\omega),\\
\rho^{\cos}_{\rm sG}(q,\omega) &=& 2\pi Z \sum_{n \, {\rm even}} w_{n}\: \delta(\omega^2 - v^2_s q^2 - \Delta^2_{n}) +  \rho^{\cos}_{\rm s\bar{s}}(q,\omega),
\end{eqnarray}
where $\rho^{\sin,\cos}_{s\bar{s}}(q,\omega)$ describes the continuum of soliton and anti-soliton excitations. We have also used that the solitons with even quantum number $n$ are even  and those with odd $n$ are odd under $\cal C$~\cite{gogolin_1dbook}. To make further progress, we assume that the behavior of the above correlation functions is dominated by the lowest energy pole (single-mode approximation, SMA), \ie the lowest breather  $\Delta_1 = 2 \Delta_s \sin(p\pi/2)$. Hence, the \emph{retarded} correlation function:
\begin{equation}
c^R_{-}(q,\omega) \simeq \frac{z_{-}}
{(\omega+i\epsilon)^2 - v_s^2 q^2 - (\Delta_1/\hbar)^2},
\end{equation}
where $\epsilon \to 0^+$. Since $c_{-}(q=0,\omega = 0) = -\left[J({\bf Q} = 0)\right]^{-1} =
-1/(J z_C)$, this fixes the residue $z_{-}  =  \Delta^2_1/(J z_C \hbar^2)$. Hence, the  dispersion of the Goldstone mode can be obtained from equation~(\ref{dispersion}) (we
consider a square lattice below):
\begin{equation}
1 + \frac{\Delta^2_1 \sum_{j=y,z} \cos Q_j a}{\omega^2_{-}(q,{\bf Q}) - v_s^2 q^2 - (\Delta^2_1/\hbar^2)} = 0,
\end{equation}
so that $\omega^{2}_{-}(q,{\bf Q}) = (v_s q)^2 + \frac{1}{2}\Delta^2_1 \sum_{j=y,z} \left(1-\cos Q_j a  \right)/\hbar^2  \simeq (v_s q)^2  + (\Delta_1 a/2\hbar)^2 {\bf Q}^2$, the last expression being valid for $|{\bf Q}|\ll a^{-1}$. However, we emphasize that note that as $K\to +\infty$ (\ie the bosons become weakly interacting) the number of breather poles proliferate and the SMA may break down.  Going beyond the SMA requires
some knowledge of the spectral weights $w_{n}$. Instead, we show the same form
of the dispersion for the lowest-energy Goldstone mode can be also obtained from the self-consistent Harmonic approximation described in \sref{sec:varap}.

  Using similar methods we can obtain the energy dispersion of the longitudinal mode,
which is  the solution of the equation:
\begin{equation}
1 + J({\bf Q}) c^{R}_{+}(q, \omega_{+}(q,{\bf Q})) = 0.\label{modeplus}
\end{equation}
We  again resort to the SMA, but taking into account that this time the pole is given by the lowest energy breather that is even under $\cal C$, \ie $\Delta_2 = 2\Delta_s \sin(\pi p)$. We first fix the pole reside $z_{+}$ by studying the limiting behavior  of $c(q,\omega)$ as $q,\omega \to 0$. This can be obtained from~(\ref{dv1},\ref{dv2}), along with (\ref{lim01},\ref{lim02}), which imply:
\begin{eqnarray}
\fl
c_{+}(q=0,\omega=0) &=& g_1(q=0,\omega=0) + \tilde{g}_1(q=0,\omega=0) \\
\fl
&=& \frac{1}{M_0 L \beta} \left[ S^{\prime}_{\psi}(|\xi_c|^2) + 2 |\xi_c|^2 S^{\prime\prime}_{\psi}(|\xi_c|^2) \right]
= \frac{\hbar}{2} \frac{d^2{\cal E}_{\rm sG}}{d\xi_{c}^2} \\
\fl
&=& \frac{\hbar}{2(J z_C)^2} \frac{d^2 {\cal E}_{\rm sG}}{d \psi^2_c}  = -\frac{\hbar}{z_C J}\left(\frac{1}{8K-1} \right),
\end{eqnarray}
where we have used gauge invariance to set $\theta_c = 0$ so that $|\xi_c| = \xi_c$, together with
\begin{equation}
\frac{d{\cal E}_{\rm sG}(\psi_c)}{d\psi^2_c} =  - \frac{2 z_C J}{8K-1},
\end{equation}
which can be directly obtained from the expressions for ${\cal E}_{sG}$ and $\psi_c$ in \sref{sec:mft}. Hence, assuming that (SMA):
\begin{equation}
c^R_+(q,\omega) \simeq \frac{z_{+}}{(\omega+i\epsilon)^2 - v_s^2 q^2 - (\Delta_2/\hbar)^2 },
\end{equation}
we find that $z_{+}/(\Delta_2/\hbar)^2 = (z_C J)^{-1} \left[1/(8K-1)\right]$. If, within the SMA, we  solve equation~(\ref{modeplus}), we arrive at:
\begin{equation}
\omega^2_{+}(q,{\bf Q}) = \Delta^2_{+} + v^2_s q^2 + \frac{\Delta^2_{2}}{\hbar^2} \left( \frac{a^2}{8K-1} \right)  F({\bf Q}),
\end{equation}
where $\Delta^2_{+} =  \Delta^2_2\: (8K-2)/(8K-1)$ is the energy gap of the longitudinal
mode at $(q, {\bf Q}) = (0,\mathbf{0})$, and $F({\bf Q}) = a^{-2} \sum_{\bf t} \left(1 - e^{i {\bf Q} \cdot {\bf t}}\right)/2$.

At the transition temperature, $T = T_c$, the order parameter $\psi_c = \xi_c = 0$ and the  matrix ${\bf J}^{-1}_{{\bf R}-{\bf R'}} + {\bf G}_{{\bf R}-{\bf R'}}$ becomes singular. For $\xi_c = 0$, the anomalous correlators $g^{++}_1 = g^{--}_1 = 0$, and  $g^{+-}_1 = g^{-+}_1=g_1$, which leads to:
\begin{equation}
J^{-1}_{{\bf R}-{\bf R}'}\delta(x-x')\delta(\tau-\tau') + g_1(x-x',\tau-\tau') = 0.
\end{equation}
Upon multiplying by $J_{{\bf R''}-{\bf R}}$, summing over $\bf R$, and integrating over $x$  and $\tau$,  we obtain:
\begin{equation}
\sum_{\bf R} J_{{\bf R}-{\bf R}'} \int^{\hbar\beta}_0 d\tau \int^{L}_{0} dx \: g_1(x,\tau)
= z_C J \, g^{R}_1(q=0,\omega= 0;T_c) = -1.
\end{equation}
This is precisely the same condition for the BEC temperature derived in \sref{sec:bectemp}. The  expression for $g^R_1(q,\omega)$, which is needed
to obtain $T_c$ explicitly is evaluated in the  appendix that follows this one.

 Finally, to make connection with the discussion of \sref{sec:exbec}, we explain
that the identification of the mode labeled as `$-$' with the Goldstone mode and the
one labeled as `$+$' as the longitudinal mode is indeed quite natural. If we
perform a gauge transformation such that $\theta_c = 0$, the matrix $\bf U$
of  equation~(\ref{eq:matu}) produces the transformation:
\begin{eqnarray}
\delta \xi^{(-)}_{{\bf R}}(x,\tau) &=& \frac{1}{\sqrt{2}} \left[ \delta \xi_{\bf R}(x,\tau)
- \delta \xi_{\bf R}(x,\tau) \right], \\
\delta \xi^{(+)}_{{\bf R}}(x,\tau) &=& \frac{1}{\sqrt{2}}  \left[ \delta \xi_{\bf R}(x,\tau)
+ \delta \xi_{\bf R}(x,\tau) \right].
\end{eqnarray}
Let us consider (small) fluctuations of the  order parameter of the form considered
in \sref{sec:exbec}:
$\delta \xi(x,\tau) \propto \delta\Psi_c(x,{\bf R},\tau) = \Psi_c(x,{\bf R},
\tau) - \psi_c = \left[\psi_c + \eta_c(x,{\bf R},\tau)\right] e^{i\theta_c} - \psi_c \simeq
\eta_c(x,{\bf R},\tau)  + i \theta_c(x,{\bf R},\tau)$, and $\delta^*_{\bf R}(x,\tau)$ the complex conjugate. Hence $\delta \xi^{(-)}_{\bf R}(x,\tau) \propto i
\theta_c(x,{\bf R},\tau)$ and $\delta \xi^{(+)}_{\bf R}(x,\tau) \propto \eta_c(x,{\bf R},\tau)$.
Thus we see that $\delta \xi^{(-)}_{\bf R}(x,\tau)$ is related to fluctuations of the
phase of the BEC, whereas $\delta \xi^{(-)}_{\bf R}(x,\tau)$ is related to fluctuations
of its amplitude.

\section{Finite-temperature phase susceptibility} \label{app:finitechi}

In this appendix we compute the Fourier transform of the retarded
phase-susceptibility of a 1D system of interacting bosons. This correlation function is to
leading order the one-body boson Green's function:
\begin{equation} \label{g1m}
 g_1(x,\tau,T) = - \frac{1}{\hbar}\langle {\cal T}\left[ \Psi_{\bf R}(x,\tau) \Psi^{\dag}_{\bf R}(0)  \right] \rangle \simeq
 \frac{1}{\hbar} J \rho_0 {\cal A}_{B} \langle e^{i\theta_{\bf R}(x,\tau)} e^{i\theta_{\bf R}(0,0)} \rangle.
\end{equation}
In the above expression,  we have replaced the boson field by its bosonized form
to leading order:
\begin{equation}
 \Psi_{\bf R}(x,\tau) \simeq \rho^{1/2}_0 {\cal A}^{1/2}_{B} \: e^{i\theta_{\bf R}(x,\tau)} .
\end{equation}
where ${\cal A}_{B} = {\cal A}_{B}(K)$ is a non-universal prefactor.
To compute this correlation function,  the (gaussian) action
(\ref{GaussianA}) must be used.  Either by direct computation
using functional integrals~\cite{giamarchi_book_1d},
or by means of a conformal transformation
of the correlation function at $T=0$~\cite{cazalilla_correlations_1d},
one obtains the following result:
\begin{equation}
\fl
g_{1}(x,\tau,T) =  - \frac{1}{\hbar}\rho_0 {\cal A}_B(K)\:  \left(\frac{(\pi T/\hbar v_s\rho_0)^2}{\sin \left[\pi T (v_s \tau + i x)/\hbar v_s \right]
 \sin \left[  \pi T (v_s \tau - ix)/\hbar v_s \right] }  \right)^{\frac{1}{4K}}
\end{equation}
In order to obtain the retarded version of this correlation function,
we shall use the relationship (see \eg \cite{giamarchi_book_1d}):
\begin{equation}
 g^R_1(x,t,T) = - 2 \theta(t) \: {\rm Im}\: g^F_1(x,t,T).
\end{equation}
where $g^F_1(x,t,T)$ is  the time-ordered  ({\it \`a la Feynman})
propagator. The latter  can be obtained from (\ref{g1m}) by
analytical continuation: $\tau = it + \epsilon(t)$, where
$\epsilon(t) = {\rm sgn}(t) \epsilon$ ($\epsilon \to 0^+$). Hence,
\begin{equation}
\fl
 g^F_1(x,t,T) = - \frac{1}{\hbar}\rho_0 {\cal A}_B(K)\:
 \left(\frac{(\pi T/\hbar v_s\rho_0)^2}{\sinh
 \left[\pi T (x + v_s t_{\epsilon}/\hbar v_s \right]
 \sinh \left[  \pi T (x - v_s t_{\epsilon} )/\hbar v_s \right] }  \right)^{\frac{1}{4K}},
\end{equation}
where $t_{\epsilon} = t - i \epsilon(t)$.
We next exponentiate the above result and use that
\begin{eqnarray}
\fl
 R(x,t) &=& \sinh \left[\pi T (x + v_s t -i v_s \epsilon (t)) / \hbar v_s
 \right]\sinh \left[  \pi T (x - v_s t + i v_s\epsilon(t) )/\hbar v_s \right] \\
\fl
 &=& \sinh \left[\pi T (x + v_s t)/\hbar v_s \right] \sinh \left[  \pi
 T (x - v_s t )/\hbar v_s \right]     \nonumber\\
\fl
 &&+ i \epsilon \: \left( \frac{\pi
 T}{\hbar}\right) \left| \sinh\left( \frac{2\pi T t}{\hbar}\right)
 \right|
 + {\rm O}(\epsilon^2),
\end{eqnarray}
which follows upon expanding in powers of $\epsilon$, as well as the
identity $\ln(a + i\epsilon) = \ln|a| + i\pi \theta(-a)$ (the branch
cut of the logarithm is put on the negative  real axis).  Thus, the
imaginary part of $g^{F}_1(x,t,T)$ does not vanish provided that
${\rm Re}\: R(x,t) <0$, that is, if $ \sinh \left[\pi T (x + v_s t)/\hbar v_s \right]
 \sinh \left[  \pi T (x - v_s t )/\hbar v_s \right] =
 -\sinh  \left[\pi T \xi_{+}/\hbar v_s \right] \sinh  \left[\pi T \xi_{-}/\hbar v_s \right]< 0$,
\ie for $\xi_{+}\xi_{-} >  0$ ($\xi_{\pm} = v_s t \pm x$). Therefore,
\begin{eqnarray}
\fl
 {\rm Im}\: g^{F}_1(\xi_{+}, \xi_{-},T) = -\rho_0 {\cal A}_{B}(K)
 \left(\frac{\pi T}{\hbar v_s\rho_0}\right)^{\frac{1}{2K}} e^{ -\frac{1}{4K}\ln {\rm Re}\: R(\xi_{+},\xi_{-})} \:
 {\rm Im}\: \exp\left[-\frac{i\pi}{4K} \theta(\xi_{-})\theta(\xi_{+})  \right] \nonumber\\
\fl
= D(K)\: \left(\frac{\pi T}{\hbar v_s\rho_0}\right)^{\frac{1}{2K}}
 \left|\frac{1}{\sinh  \left[\pi T \xi_{+}/\hbar v_s \right] \sinh  \left[\pi T \xi_{-}/\hbar v_s \right]}
 \right|^{1/4K}\: \theta(\xi_{-})\theta(\xi_{+}),
\end{eqnarray}
where $D(K) =  \rho_0 {\cal A}_B(K)  \sin\left(\frac{\pi}{4K}\right)/\hbar$.
Since we are interested in the Fourier transform of $g^R_1(x,t,T)$,
it is convenient to express it in terms of the imaginary part of
$g^F_1(x,t,T)$:
\begin{eqnarray}
\fl
 g^R_1(k,\omega,T) &=& - 2\int^{+\infty}_{-\infty} dt
 \int^{+\infty}_{-\infty} dx e^{-i k x + i \omega t}\, g^{R}_1(x,t,T) \\
 \fl
 &=&  -2 \int^{+\infty}_{0} dt \int^{+\infty}_{-\infty}dx \, e^{-i k x
 + i \omega t} \:
 {\rm Im}\: g^F(x,t,T)\\
 \fl
 &=& -2 \int^{+\infty}_{0} dt \int^{+v_s t}_{-v_s t}dx \, e^{-i k x + i \omega t} \:  {\rm Im}\: g^F_1(x,t,T) \\
\fl
 &=& -\frac{1}{v_s} \int^{+\infty}_0 d\xi_{+} \int^{+\infty}_{0} d\xi_{-} \,\, e^{i (k_{+}\xi_{-} + k_{-}\xi_{+})/2} \,
 {\rm Im} \: g^F_1(\xi_{+},\xi_{-},T),
\end{eqnarray}
where we have used that $dt \: dx = d\xi_{+} d\xi_{-}/2v_s$ and have
introduced $k_{\pm} =\omega/v_s \pm k$. In the third step we
employed that ${\rm Im}\: g^{F}_1(x,t,T)$ is only non-zero for $-v_s
t < x < v_s t$. Finally, we have expressed the integral in terms of
``light-cone'' coordinates $\xi_{\pm}$ to be able to benefit from the
separability of ${\rm Im} \: g^{F}_1$ in terms of these coordinates.
Hence,
\begin{eqnarray}
\fl
 g^R_1(k,\omega,T) &=&  - \frac{D(K)}{v_s} \:
 \left(\frac{\pi T}{\hbar v_s\rho_0}\right)^{\frac{1}{2K}}
f\left[\frac{\hbar v_s}{T}\left(\frac{\omega}{v_s}+q\right),K\right]
 f\left[\frac{\hbar v_s}{T}\left(\frac{\omega}{v_s}-q\right),K\right],
\end{eqnarray}
where (see e.g.~\cite{giamarchi_book_1d}):
\begin{eqnarray}
\fl
 f\left(\frac{\hbar v_s q}{T}, K\right) = \int^{+\infty}_0  \:
\frac{ e^{i k \xi/2} d\xi}{\left|\sinh\left(\frac{\pi T\xi}{\hbar
 v_s}\right)\right|^{1/4K}}
 = 2^{\frac{1}{4K}} \left(\frac{\hbar
 v_s}{2T}\right) B\left(\frac{1}{8K} - i \frac{\hbar v_s q}{4\pi T},
 1-\frac{1}{4K}\right),
\end{eqnarray}
$B(x,y)$ being the beta function,
$B(x,y) = \Gamma(x)\Gamma(y)/\Gamma(x+y)$.
In the calculation of the critical temperature we need the value of
$g^R_1(k,\omega,T)$ for $q=0$ and $\omega = 0$:
\begin{eqnarray} \label{g0}
\fl
 g^R_1(q=0,\omega=0; T) = - \pi^2 \frac{D(K)}{v_s} \left(\frac{2\pi T}{\hbar v_s \rho_0}\right)^{1/2K-2}  B^{2}\left(\frac{1}{8K}, 1-\frac{1}{4K}\right)
\end{eqnarray}

\section{Self-consistent Harmonic Approximation calculation} \label{app:variat}

In this appendix, we fill in some of the details of the SCHA calculation.
First, note that in equation (\ref{eq:Fv}), we can rewrite
\begin{equation}
 \langle S - S_{\rm v}\rangle_{\rm v} = \frac{1}{2} \sum_{q,{\bf Q}, \omega_n}
 G_{\rm v}(q, {\bf Q}, \omega_n)
G^{-1}_{0}(q, \omega_n) +\langle S_{\rm int} \rangle_{\rm v} -{\rm const.} .
\end{equation} Then, the variational approximation to the free energy becomes
\begin{equation}
 \fl F'[G_v] = -\frac{1}{2} \sum_{q,{\bf Q}, \omega_n} \ln G_{\rm v}(q, {\bf Q}, \omega_n)
+ \frac{1}{2}
 \sum_{q,{\bf Q}, \omega_n}  G_{\rm v}(q, {\bf Q}, \omega_n)
G^{-1}_{0}(q, \omega_n) +\langle S_{\rm int} \rangle_{\rm v} ,
\end{equation} where
\begin{equation}
\fl \langle S_{\rm int} \rangle_{\rm v} = -  {\cal A}_B  \rho_0 J
L  \beta \sum_{\langle {\bf R R'} \rangle}
 {\rm Re} \; \exp \left[G_{\rm v}(0,{\bf R}-{\bf R'},0)-G_{\rm v}(0,{\bf 0},0) \right] ,
\end{equation}
and
\begin{equation}
\fl G_{\rm v }(x,{\bf R}-{\bf R'},\tau) = \frac{1}{M_0 \hbar \beta L}
\sum_{q,{\bf Q},\omega_n}
 e^{i {\bf Q} \cdot ({\bf R}-{\bf R'}) +i q x -i \tau \omega_n} \;
 G_{\rm v}(q,{\bf Q}, \omega_n) ,
\end{equation}
In evaluating $\langle S_{\rm int}\rangle_{\rm v}$, we can take advantage of
point group symmetry of the lattice in the perpendicular directions and replace
$G_{\rm v}(0,{\bf R}-{\bf R'},0)-G_{\rm v}(0,{\bf 0},0) $ inside the exponential
by the more symmetric expression $ \frac{1}{z_c}
\sum_{{\bf R'}={\bf R} + {\bf t}} G_{\rm v}(0,{\bf R}-{\bf R'},0)-G_{\rm v}(0,{\bf 0},0)$, where
${\bf t}$ are the unit lattice vectors in the perpendicular
directions. Then, the outer $\sum_{\langle {\bf R R'} \rangle}$ just gives
a factor of $z_c$. Taking the variational derivative of $F$, we get:
\begin{eqnarray}
\fl
\frac{1}{G_{\rm v}(q,{\bf Q},\omega_n)} =  \frac{1}{G_0(q,\omega_n)} -
\frac{z_C {\cal A}_B(K) \rho_0}{2 \hbar}  J   \,  F({\bf Q}) \,\,
e^{\frac{2 J}{z_C M_0 L\hbar\beta}\sum_{q',{\bf Q'},\omega_n'}  F({\bf Q'}) \: G_{\rm v}(q',{\bf Q'},\omega_n')},
\end{eqnarray} with  the free phonon propagator:
$G^{-1}_{0}(q,\omega_n) = \frac{K}{\pi v_{||}} \left[ \omega^2_n + v^2_{||} q^2 \right]\;\;$ ($v_{||} = v_s$), and $F({\bf Q}) = \frac{a^{-2}}{2} \sum_{\bf t}  \left(  1 - e^{{\bf Q}\cdot {\bf t}} \right)$. This single self-consistent equation for
$G_{\rm v}(q,{\bf Q},\omega_n) $ can be solved by
rewriting it into two equations:
\begin{equation}
v^2_{\perp}  =   \frac{\pi z_C {\cal A}_{B}}{2 K}  (\rho_0 a)
\left(\frac{Ja}{\hbar} \right) v_{||} \:
 e^{\frac{2 a^{2}}{z_C M_0 L\hbar\beta}\sum_{q,{\bf Q},\omega_n}  F({\bf Q})
 \: G_{\rm v}(q,{\bf Q},\omega_n)},
\label{eq:var1}
\end{equation}
\begin{equation}
G^{-1}_{\rm v}(q, {\bf Q}, \omega_n) =  \frac{K}{\pi v_{||}}
\left[ \omega^2_n + (v_{||} q)^2 + v^2_{\perp} \,  F({\bf Q})\right].
\label{eq:var2}
\end{equation}
Substituting equation (\ref{eq:var2}) into (\ref{eq:var1}),
we get the equation for the variational parameter, the
perpendicular phonon velocity $v_{\perp}$:
\begin{equation}
\fl \ln \frac{v^2_{\perp}}{\frac{\pi z_C {\cal A}_{B}}{2 K}  (\rho_0 a)
\left(\frac{Ja}{\hbar} \right) v_{||} } = \frac{2 a^{2}}{z_C M_0 L\hbar\beta}
\sum_{q,{\bf Q},\omega_n}
F({\bf Q}) \frac{ \pi v_{||}/K}{\omega_n^2 +
 v_{||}^2 q^2 + v_{\perp}^2 F({\bf Q})}.
\end{equation}
The Matsubara sum can be done to give:
\begin{equation}\label{gamma}
\ln \frac{v^2_{\perp}}{\frac{\pi z_C {\cal A}_{B}}{2 K}  (\rho_0 a)
\left(\frac{Ja}{\hbar} \right) v_{||} }  =
\frac{ \pi v_{||}}{K} I(T,v_{\perp})
\end{equation}
with
\begin{equation}
\fl  I(T,v_{\perp}) =  \frac{2 a^{2}}{z_C M_0 L}\sum_{q,{\bf Q}}
 \frac{F({\bf Q})}{\left[
 v_{||}^2 q^2 + v_{\perp}^2 F({\bf Q}) \right]^{1/2}}
\coth \frac{\left[ v_{||}^2 q^2 +v_{\perp}^2 F({\bf Q}) \right]^{1/2}}{2T} .
\end{equation}

We first study the variational equation (\ref{gamma}) at zero
temperature. At $T=0$, $\coth(x/2T) \rightarrow \rm{sign} (x)$.
To perform the summations over momentum, we take the continuum approximation
for both the parallel ($q$) and perpendicular (${\bf Q}$) momenta, with
a cut-off for the parallel momentum to be $\Lambda=\mu_{1D}/v_{||}$
($\mu_{1D}$ is
the 1D chemical potential of an isolated tube). Thus, with
$\frac{1}{M_0}\sum_{\bf Q}
\longrightarrow a^2 \int_{-\pi/a}^{\pi/a} \frac{dQ_y dQ_z}{(2\pi)^2}$,
and $\frac{1}{L}
\sum_q \longrightarrow \int_{-\Lambda}^{\Lambda} \frac{dq}{2\pi}$,
 we get:
\begin{equation}
 I(0,v^0_{\perp}) \approx \frac{1}{\pi v_{||}} \left[ \ln\left(\frac{2
 \mu_{1D}}{v^0_{\perp}/a}\right) -\frac{B}{2} \right]
\label{I0}
\end{equation}
where $v_{\perp}^0 = v_{\perp}(T=0)$ and
\begin{equation}
 B = \frac{a^2}{2} \int_{-\pi}^{\pi} \frac{dQ_x dQ_y}{(2\pi)^2}
\; F({\bf Q}) \; \ln a^2 F({\bf Q})
 \approx 0.836477 .
\end{equation}
Putting equations (\ref{I0}) and (\ref{gamma}) together then gives equation
(\ref{gamma0}) of Section \ref{sec:varap}.
We have assumed here $v^0_{\perp}/a \ll \mu$. Otherwise, our
starting point using the Haldane harmonic fluid construction is not
valid in principle.

When $K\rightarrow \infty$ (towards the non-interacting Bose gas limit),
${\cal A}_B(K) \approx 1$, and using the  Lieb and 
Liniger solution~\cite{lieb_bosons_1D}
we can  show that $\mu_{1D} \approx v_{||} \rho_0 \pi/K$. Thus, asymptotically
for large $K$, we get:
\begin{equation}
 \frac{v^0_{\perp}}{a \mu_{1D}} \simeq \frac{v^0_{\perp}}{v_{||}}
\frac{K}{\rho_0 a \pi} \longrightarrow \sqrt{\frac{z_C J}{\mu_{1D}}} .
\end{equation}
Thus, $v_{\perp}^0 /a < \mu_{1D}$ if and only if $z_c J < \mu_{1D}$.
But since for the nearly free
Bose gas, $\mu \propto 1/K^2$ is small, this demands a very 
hopping amplitude $J$.

In equation (\ref{gamma0}), there is a singularity as $K\rightarrow 1/4$,
which is a signature that for  $K<1/4$  the Josephson coupling  becomes 
irrelevant in the RG sense. As this
regime is beyond the reach for bosons with a Dirac-delta interaction, we
will not consider it any further.

We now look at the variational equation (\ref{gamma}) at finite
temperature. First, defining $v_{\perp}(T_c) = v_{\perp}^c$, we rewrite
(\ref{gamma}) as:
\begin{equation} \label{Tcmod}
 \left(\frac{v_{\perp}^0}{v_{\perp}^c}\right)^{1-1/4K} =
\exp \frac{\pi v_{||}}{K} \left[ I(T_c, v_{\perp}^c) - I(0,v_{\perp}^c) \right] .
\end{equation}
Following Donohue\cite{donohue_thesis}, we approximate crudely
$\coth(x) \approx 1/x$ for $|x|\leq 1$, and $\coth(x) \approx 1$
otherwise. Next, we develop a series expansion in $v_{\perp}^c/a
T_c <1$ for the integrand in $I(T_c, v_{\perp}^c) - I(0,v_{\perp}^c)$. It
will turn out that at the transition $T=T_c$, $v_{\perp}^c$ is actually
finite (i.e. first order transition) but less than $v_{\perp}^0$, and
the expansion can be justified \emph{a posteriori}, at least for $K$ not
too close to 1. With the help of Mathematica to  perform the 
expansion and to evaluate  the resulting integrals, we find:
\begin{eqnarray}
\fl v_{||} \left( I(T_c, v_{\perp}^c) - I(0,v_{\perp}^c)  \right) \approx
 c_{-1} \frac{a T_c}{ v_{\perp}^c} + c_0  -\frac{1}{\pi} \ln \frac{z_C a
 T_c}{ v_{\perp}^c} +
 \sum_{n=1} c_{2n} \left( \frac{ v_{\perp}^c}{T_c a} \right)^{2 n} ,
\end{eqnarray}
where $c_{-1} \approx 0.677473$, $c_0 \approx 0.133129 -1/\pi$, $c_2
= 5/96\pi$,  $c_4 = 21/2560 \pi$,  $c_6 = 845/344604\pi$,  $c_8 =
9415/9437184 \pi$, $c_{10} = 112203/230686720 \pi$, $c_{12} =
233695/872415232\pi$. We can show that the series converges quite
rapidly for $v_{\perp}^c/a T_c <1$, but because this expression 
appears in an exponential, we need quite a few terms to get
accurate trends, especially when $K$ gets close to $1$. Introducing
this expansion into (\ref{Tcmod}), 
\begin{equation} \label{Tc}
  \frac{v_{\perp}^c}{v_{\perp}^0} = \left(\frac{z_C a
 T_c}{v_{\perp}^0}\right)^{1/4K} e^{-\frac{\pi}{4 K} \left[ c_{-1}
 \frac{a T_c}{v_{\perp}^c} + c_0 + \sum_{n=1} c_{2n} \left(
 \frac{v_{\perp}^c}{a T_c} \right)^{2 n} \right]} .
\end{equation}
This equation is solved graphically for $(T_c, v_{\perp}^c$) (the
right hand side of (\ref{Tc}) is an s-shaped curve that intersects
the straight line of the lef hand side of (\ref{Tc}) at
0, 1, or 2 points, as a function of $T_c$, and $v_{\perp}^c$; $T_c$ is,
by definition, when the intersection is at one point only. We use
Mathematica's routine FindRoot to bracket this point to the desired
accuracy.)

  We thus find the following asymptotic results for $K\rightarrow \infty$:
\begin{eqnarray}
 \frac{a T_c}{v_{\perp}^0} &\approx& 0.610 + 0.698 K  ,\\
 \frac{v_{\perp}^c}{v_{\perp}^0} &\approx& 0.376 + 0.274/K  .
\end{eqnarray}
Both asymptotes appear to hold good when $K \geq 4$. Dividing these
equations, we get that $a T_c / v_{\perp}^c \approx 1.86 K + 0.270 +
\cdots$, thus justifying our expansion in $v_{\perp}^c / a
T_c$, at least for $K \geq 4$. Combining with the asymptotic results
for $\gamma_0$ as $K\rightarrow \infty$, we get the results quoted
in \sref{sec:varap}.

\section{Derivation of RG equations} \label{app:rgequ}

 In this appendix  we show how to compute the RG flow to second order in the Josephson and Mott potential couplings. To this purpose,
it is convenient to introduce the   chiral fields $\phi_{{\bf R}+}(x)$
and $\phi_{{\bf R}-}(x)$, which
are implicitly defined as follows:
\begin{eqnarray}
 \phi_{\bf R}(x) &=& \frac{\sqrt{K}}{2}\left[\phi_{{\bf R}+}(x) + \phi_{{\bf R} -}(x)  \right],\\
 \theta_{\bf R}(x) &=& \frac{1}{2\sqrt{K}} \left[ \phi_{{\bf R}+}(x) - \phi_{{\bf R} -}(x)  \right].
\end{eqnarray}
The effective Hamiltonian, equation~(\ref{ham2}), thus takes the form:
\begin{eqnarray}
\fl
H_{\rm eff} &=& \frac{\hbar v_s}{4\pi} \sum_{\bf R} \int dx \left[
 \left(\partial_x \phi_{{\bf R}+}(x)\right)^2 +
 \left( \partial_x \phi_{{\bf R }-}(x)\right)^2\right] \nonumber\\
\fl
&& + \tilde{g}_u \sum_{\bf R} \int dx\:
 \cos \sqrt{K}\left( \phi_{{\bf R}+}(x) +  \phi_{{\bf R}-}(x) \right)
 + \tilde{g}_F \: \sum_{\langle{\bf R},{\bf R}'\rangle} \int dx\,
 \partial_x \phi_{{\bf R}+}(x) \partial_x\phi_{{\bf R}'-}(x) \nonumber\\
\fl
&& +  \tilde{g}_J \sum_{\langle {\bf R}, {\bf R'} \rangle} \int dx \:
 \cos \left( \frac{\phi_{{\bf R}+}(x)-\phi_{{\bf R'}+}(x)}{2\sqrt{K}}
 - \frac{\phi_{{\bf R}-}(x) -  \phi_{{\bf R'}-}(x)}{2\sqrt{K}} \right).
\end{eqnarray}
We have added here a term proportional to $\tilde{g}_F$, which is
not present initially, but which is generated at second order in $\tilde{g}_J$
after integrating out short-distance degrees of freedom (see below). It
represents a density interaction between nearest neighbor 1D systems.

We next introduce the chiral \emph{vertex} operators $V^{+}_{\beta}({\bf
R},\barz) = \,\,  :e^{i\beta \phi_{{\bf R}+}(\barz)}:$ and
$V^{-}_{\beta}({\bf R},z) = \,  :e^{i\beta \phi_{{\bf R}-}(z)}:$,
and define  $z = v_s \tau + ix$ and $\barz = v_s
\tau - i x$, and $d^2z = v_s d\tau dx = dz d\barz/2i$. To explicitly
display the scaling dimensions of the various  operators in $H_{\rm eff}$, we
introduce the dimensionless couplings:
\begin{equation}
g_{u} = \pi a^{2}_0 \tilde{g}_{u}/v_s, \quad g_{J} = \pi a^{2}_0\tilde{g}_{J}/v_s
\quad g_{F} = \pi \tilde{g}_{F}/v_s, \quad g_{K} = \pi \tilde{g}_{K}/v_s.
\end{equation}
Thus, the interactions are described by
\begin{eqnarray}\label{hamRG}
\fl
S_{\rm int} &=& g_K \sum_{\bf R} \int \frac{d^2z}{2\pi} \, \barpart\phi_{{\bf R}+}(\barz) \partial\phi_{{\bf R}-}(z)
 + g_F \sum_{\langle {\bf R}, {\bf R'}\rangle}
 \int \frac{d^2z}{2\pi}\, \barpart\phi_{{\bf R}+}(\barz)\partial\phi_{{\bf R}'-}(z) \nonumber \\
\fl
&& +  \frac{g_u}{a^{2-K}_0} \sum_{\bf R}  \int \frac{d^2z}{2\pi} \,\left[
V_{\sqrt{K}}^{+}({\bf R},\barz) V_{\sqrt{K}}^{-}({\bf R},z) + \left( \sqrt{K} \to -\sqrt{K} \right)\right] \nonumber \\
\fl
&& + \frac{g_J}{a^{2-1/2K}_0} \sum_{\langle {\bf R}, {\bf R'} \rangle}  \int \frac{d^2z}{2\pi} \:
 \Big[ V^{+}_{\frac{1}{2\sqrt{K}}}({\bf R},\barz) V^{+}_{-\frac{1}{2\sqrt{K}}}({\bf R'},\barz)
 V^{-}_{-\frac{1}{2\sqrt{K}}}({\bf R},z) V^{-}_{\frac{1}{2\sqrt{K}}}({\bf R'},z) \nonumber\\
\fl
& &+  \left( \sqrt{K} \to -\sqrt{K} \right) \Big]
\end{eqnarray}
The RG is performed in real space on the perturbative expansion of the
partition function $Z(a_0) = Z_0  \langle {\cal T} \exp\left[ - S_{\rm
int} \right] \rangle$. We first consider an infinitesimal change of the short
distance cut-off $a_0 \rightarrow a'_0 = (1+\delta\ell) a_0$, where $0
< \delta\ell \ll 1$. Subsequently, we integrate out short-distance
degrees of freedom in the range $a_0 < |z| < a_0 (1+\delta \ell) a_0$.
Following Ref.~\cite{cardy_book_renormalization},
we use operator-product expansions (OPE) to evaluate expectation values
of products of operators at two nearby points in space and time. For $z \to w$
(respectively, $\barz \to \barw$):
\begin{eqnarray}
\fl
\partial\phi_{{\bf R}-}(z)\partial\phi_{{\bf R}-}(w) = -\frac{1}{(z-w)^2} + :\left[
\partial\phi_{{\bf R}-}\left(\frac{z+w}{2}\right)  \right]^2: \, +
\cdots \\
\fl
\barpart\phi_{{\bf R}+}(\barz)\barpart\phi_{{\bf R}+}(\barw) =
-\frac{1}{(\barz-\barw)^2} +
:\left[\barpart\phi_{{\bf R}+}\left(\frac{\barz+\barw}{2}\right)  \right]^2: \, +
\cdots\\
\fl
V^{+}_{\beta}({\bf R},\barz) V^{+}_{-\beta}({\bf R},\barw) = \frac{1}{(\barz-\barw)^{\beta^2}}\left[ 1 + i\beta (\barz-\barz)
\barpart\phi_{{\bf R}+}\left(\frac{\barz+\barw}{2}\right) + \cdots \right],\label{opev1}\\
\fl
V^{-}_{\beta}({\bf R},z) V^{-}_{-\beta}({\bf R},w) = \frac{1}{(z-w)^{\beta^2}}\left[ 1 + i\beta (z-w)
\partial\phi_{{\bf R}-}\left(\frac{z+w}{2}\right) + \cdots \right].\label{opev2}\\
\fl
\partial\phi_{{\bf R}-}(z)V^-_{\beta}({\bf R},w) = \frac{-i\beta}{z-w} V^-_{\beta}({\bf R},\frac{z+w}{2}) +\cdots
\label{opep1}\\
\fl
\barpart\phi_{{\bf R}+}(\barz)V^+_{\beta}({\bf R},\barw) = \frac{-i\beta}{\barz-\barw} V^+_{\beta}({\bf R},\frac{\barz+\barw}{2}) +\cdots\label{opep2}
\end{eqnarray}

   Therefore, we begin with by considering the perturbative expansion of
the partition function at the new scale $a'_0 = (1+\delta\ell) a_0$,
\begin{eqnarray}
\fl
 Z((1+\delta \ell)a_0,\left\{g_{i}(\ell+\delta\ell)\right\}) &=& 1 + Z^{(1)}((1+\delta\ell)a_0,
 \left\{g_{i}(\ell+\delta\ell)\right\}) \nonumber\\
\fl
 && + Z^{(2)}((1+\delta\ell)a_0,\left\{g_{i}(\ell+\delta\ell)\right\}) + \cdots,
\end{eqnarray}
where
\begin{eqnarray}
 Z^{(1)}((1+\delta\ell)a_0,\left\{g_{i}(\ell+\delta\ell)\right\}) &=& -\langle S_{\rm int} \rangle,\\
 Z^{(2)}((1+\delta\ell)a_0,\left\{g_{i}(\ell+\delta\ell)\right\}) &=&  \frac{1}{2!} \langle S^{2}_{\rm int}\rangle
\end{eqnarray}
The partition function must be left invariant by the RG transformation,
\ie $Z((1+\delta \ell)a_0,\left\{g_{i}(\ell+\delta\ell)\right\}) = Z(a_0,\left\{g_{i}(\ell)\right\})$,
provided that the couplings $\left\{g_{i}\right\}$ are properly transformed after integrating out the short-distance degrees of freedom in  range defined by $a_0 < |z-w| < (1+\delta \ell) a_0$.

At the lowest order (tree level), the
couplings change only because of the explicit factors of $a_0$
(\ie the scaling dimensions of the perturbing operators) in
(\ref{hamRG}). Thus, the only couplings that change are:
\begin{equation}
\fl
 g_u(\ell) = g_u(\ell+\delta\ell)(1+\delta\ell)^{-(2-K)} \\
\fl
           \simeq g_u(\ell+\delta\ell) \left(1-(2-K(\ell)\delta \ell)\right),
\end{equation}
and
\begin{equation}
\fl
 g_J(\ell) = g_J(\ell+\delta\ell)(1+\delta\ell)^{-(2-1/2K)}
    \simeq g_J(\ell+\delta\ell)\left[ 1-\left(2-\frac{1}{2K(\ell)}\right)\delta\ell\right] .
\end{equation}
At second order, there are ten terms in the expansion of
$Z^{(2)}[(1+\delta \ell)a_0,\{g_i(\ell + \delta \ell) \}]$. As the
manipulations are standard but rather long, we will just look at two of
them in detail to illustrate the procedure. To see what kind of
terms are generated from the product of two operators, we shall use
``center-of-mass'' coordinates: $u =z-w$ (respectively, $\baru = \barz-\barw$)
and $v=(z+w)/2$ (respectively, $\barv = (\barz+\barw)/2$), and we split the
integrals of the operator products appearing at second order as follows:
\begin{eqnarray}
\fl
\int \frac{d^2v}{2\pi} \int\limits_{(1+\delta\ell)a_0 < |u|} \frac{d^2u}{2\pi} \, O_1(v+\frac{u}{2},\barv+\frac{\baru}{2})
 O_2(v-\frac{u}{2},\barv-\frac{\baru}{2})  \nonumber \\
\fl
=  \int \frac{d^2v}{2\pi} \Bigg[ \int\limits_{a_0 < |u|} \frac{d^2u}{2\pi}
 -\int\limits_{a_0 < |u|<(1+\delta\ell)a_0} \frac{d^2u}{2\pi} \Bigg]\,  O_1(v+\frac{u}{2},\barv+\frac{\baru}{2})
 O_2(v-\frac{u}{2},\barv-\frac{\baru}{2})
\end{eqnarray}
The terms generated at second order by the RG transformation  stem form the second integral, where the relative coordinate is restricted to the  infinitesimal range $a_0 < |u| < a_0(1+\delta\ell)$.  To $O(g_F g_J)$, there are two possible operator pairings because the hopping term couples operators from neighboring chains:
\begin{eqnarray}
\fl
 O(g_Fg_u) &=& -\frac{1}{2!} \frac{g_F(\ell)g_J(\ell)}{a_0^{2-1/2K}}
 \sum_{\langle {\bf R},{\bf R}'\rangle, \langle{\bf T},{\bf T}'\rangle}
 \int \frac{d^2v}{2\pi} \int\limits_{a_0<|u|<(1+\delta\ell)a_0} \frac{d^2u}{2\pi}
 \Bigg\{ \Bigl \langle
\barpart\phi_{{\bf R}+}(\barv+\frac{\baru}{2})\partial\phi_{{\bf R'}-}(v+\frac{u}{2})
\nonumber \\
\fl
&& \times \Big[
 V^{+}_{\frac{1}{2\sqrt{K}}} ({\bf T},\barv-\frac{\baru}{2})   V^{+}_{-\frac{1}{2\sqrt{K}}}
 ({\bf T}',\barv-\frac{\baru}{2})
 V^{-}_{-\frac{1}{2\sqrt{K}}}({\bf T},v-\frac{u}{2}) V^{-}_{\frac{1}{2\sqrt{K}}}({\bf T}',v-\frac{u}{2}) \nonumber\\
\fl
&& +\left( \sqrt{K}\to -\sqrt{K}\right)
 \Big] \Bigr\rangle \Bigg\}
\end{eqnarray}
Focusing on the terms where ${\bf R} = {\bf T}$ and ${\bf R}'
= {\bf T}'$, or ${\bf R} = {\bf T}'$ and ${\bf R}' = {\bf T}$ and
using the OPE's (\ref{opep1},\ref{opep2}) with  $\beta = \pm
1/2\sqrt{K}$, we arrive at the following expression:
\begin{eqnarray}
\fl
 O(g_Fg_u) &=& +2\times 2\times\frac{1}{2!} \frac{g_F(\ell)g_J(\ell)}{a_0^{2-1/2K}}\left(\frac{1}{4K(\ell)}\right) \sum_{\langle {\bf R},{\bf R}'\rangle}
 \int \frac{d^2v}{2\pi}
\Bigg[ \Bigl \langle V^{+}_{\frac{1}{2\sqrt{K}}}({\bf R},\barv)   V^{+}_{-\frac{1}{2\sqrt{K}}}({\bf R}',\barv) \nonumber \\
\fl
&& \times V^{-}_{-\frac{1}{2\sqrt{K}}}({\bf R},v) V^{-}_{\frac{1}{2\sqrt{K}}}({\bf R}',v)
 +\left( \sqrt{K}\to -\sqrt{K}\right)    \Bigr\rangle \Bigg] \int^{(1+\delta\ell)a_0}_{a_0} \frac{du}{u}
 \\
\fl
 &=&  \frac{g_F(\ell)g_J(\ell) \delta\ell}{4K(\ell)a_0^{2-1/K}}\sum_{\langle {\bf R},{\bf R}'\rangle}
 \int \frac{d^2v}{2\pi}
 \Bigg[ \Bigl \langle V^{+}_{\frac{1}{2\sqrt{K}}}({\bf R},\barv)   V^{+}_{-\frac{1}{2\sqrt{K}}}({\bf R}',\barv)
 V^{-}_{-\frac{1}{2\sqrt{K}}}({\bf R},v) V^{-}_{\frac{1}{2\sqrt{K}}}({\bf R}',v) \nonumber \\
\fl
 && +\left( \sqrt{K}\to -\sqrt{K}\right)    \Bigr\rangle \Bigg]
\end{eqnarray}
where the first factor of two stems from the two terms of
$O(g_Fg_J)$ and the second from the two possible pairings of ${\bf
R},{\bf R}'$ and ${\bf T},{\bf T}'$. Thus we see that to second
order in the couplings the coupling $g_J$ renormalizes as follows:
\begin{equation}
 g_J(\ell) = g_J(\ell+\delta\ell)  \left\{1 - \left[(2 - \frac{1}{2K(\ell)})g_J(\ell) + \frac{g_F(\ell) g_J(\ell)}{4K(\ell)} \right]\delta\ell.
 \right\}
\end{equation}
In the limit $\delta \ell \to 0$, this leads to the following differential equation:
\begin{equation}
 \frac{dg_J}{d\ell} = \left(2-\frac{1}{2K}\right) g_J + \frac{g_Jg_F}{4K}.
\end{equation}
Other terms are dealt with in a similar fashion. It is also worth noticing that terms of $O(g^2_K,g_Fg_K,g_Kg_J,g_Kg_J)$ need not be taken into account because at each RG step we set $g_K = 0$ by properly renormalizing
$K$. Thus $g_K(\ell+\delta\ell) = 0$, but new terms of the form of the operator
proportional to $g_K$ are generated at $O(g^2_u, g^2_J)$. In order to get
rid of them, we perform an infinitesimal canonical
transformation. To understand this, let us consider a simplified
version of the above model, where the chain index $\bf R$ is
dropped, and the Hamiltonian reads:
\begin{eqnarray}
 H = \frac{\hbar v_s}{4\pi}\int dx \:\left[ \left( \partial_x\phi_+(x) \right)^2 +
 \left( \partial_x\phi_-(x) \right)^2\right] + \frac{\hbar v_s \delta g}{2\pi} \int \partial_x\phi_+(x)  \partial_x\phi_{-}(x) \nonumber \\
 + H_+\left[\frac{\sqrt{K}}{2} (\phi_{+} + \phi_{-})\right] + H_{-}\left[
 \frac{\left(\phi_{+} - \phi_{-} \right)}{2\sqrt{K}}\right].
\end{eqnarray}
We assume the dimensionless coupling $\delta g \ll 1$. The precise
functional forms of $H_\pm$ are of no importance in what follows. Next consider
the following infinitesimal transformation:
\begin{eqnarray}
 \phi_{+} &=& \tilde{\phi}_{+} - \frac{\delta g}{2} \tilde{\phi}_{-},\\
 \phi_{-} &=& \tilde{\phi}_{-}   -\frac{\delta g}{2} \tilde{\phi}_{+}.
\end{eqnarray}
To $O(\delta g)$ one can show that this is a canonical
transformation which, to $O(\delta g^2)$, brings the first term of
the Hamiltonian to the diagonal form:
\begin{equation}
 H_0 = \frac{\hbar v_s}{4\pi}\int dx \:\left[ \left( \partial_x\tilde{\phi}_+(x) \right)^2 +
 \left( \partial_x\tilde{\phi}_-(x) \right)^2\right] + O(\delta g^2).
\end{equation}
However, the combinations:
\begin{eqnarray}
 \phi_{+} +\phi_{-} &= (1-\frac{\delta g}{2})\left( \tilde{\phi}_{+} + \tilde{\phi}_{-}\right),\\
 \phi_{-} - \phi_{-} &=  (1+\frac{\delta g}{2})\left( \tilde{\phi}_{+} - \tilde{\phi}_{-}\right).
\end{eqnarray}
Since these combinations enter $H_{+}$ and $H_{-}$, it means that
$K$ must be renormalized as follows:
\begin{equation}
 \tilde{K} = \left(1-\frac{\delta g}{2}\right)^2 K \simeq  \left( 1- \delta g\right) K
\end{equation}
Returning to the problem of interest,  for the array of coupled 1D Bose gases
we have that $\delta g = g_K(\ell) = \left( g^2_J/K(\ell) -  g^2_v(\ell)
K(\ell)\right) \delta \ell$. Hence,
\begin{eqnarray}\label{deltag}
\fl
 K(\ell) = (1 - \delta g) K(\ell +\delta\ell) =  \left[ 1- \left( n
 g^2_J/K(\ell) -  g^2_v(\ell) K(\ell)\right) \delta \ell\right]
 K(\ell + \delta \ell),
\end{eqnarray}
which can be turned into the following differential equation:
\begin{equation}
 \frac{dK}{d\ell} = n g^2_J - g^2_v K^2
\end{equation}
With these considerations, the equations given in \sref{sec:rg} can be obtained.

\section{Spin chain mapping} \label{app:spinmap}

In this Appendix, we outline the mapping from the continuum
double sine-Gordon model to a spin chain model.
We take the continuum limit of the spin chain, and use the standard  
spin-operator bosonization formulas~\cite{giamarchi_book_1d,gogolin_1dbook}:
\begin{eqnarray}
 \frac{1}{a_0} S^z_{m} \rightarrow  S^z(x= m a_0) &=&
 \frac{1}{\pi} \partial_x \phi
 + \frac{1}{\pi a_0} (-1)^{x/a_0} \cos 2 \phi(x),  \\
 \frac{1}{a_0} S^+_m \rightarrow  S^+(x=m a_0) &=&
 \frac{(-1)^{x/a_0}}{\sqrt{2 \pi a_0}} e^{-i \theta(x)}  ,
\label{eq:spinmap}
\end{eqnarray}
where $a_0$ is the lattice spacing of the spin chain.  Using this mapping,
the continuum double sine-Gordon mean field model (\ref{eq:doublesg})
becomes the spin chain model of (\ref{eq:sc1}) in Section \ref{sec:mftdecon}.
 The mean field
self-consistency condition is still as before (\ref{self-consist}).
For simplicity, we take the isotropic Heisenberg chain.
  Then $v_s=v_{||} = \pi J_0 a_0/2$, and $K=1/2$. Now $K=1/2$ is not a
physically valid regime for bosons interacting with a contact
potential. However, from the RG approach above, there are no
indications of a phase transition from $K=1/2$ to the physical $K>1$
regime, and we expect that at least qualitatively, the results
derived here should be valid for $K>1$. Moreover, strictly speaking,
there is a \emph{umklapp} operator $\sim \cos 4 \phi(x)$
 that is marginally irrelevant at $K=1/2$, but
since it is less relevant than the Mott and Josephson terms  we drop
it.

\section{Derivation of quantum phase model} \label{app:quantphase}

Consider a 2D array of finite tubes described by the Euclidean action shown
in equations (\ref{action-noMott}) and (\ref{eq:sint}). We describe the tubes as
1D systems with open boundary conditions~
\cite{cazalilla_finitesize_luttinger,cazalilla_correlations_1d}.
In such a finite systems, the phase field operator consists
two terms:
\begin{equation} \label{repl}
 \theta_{\bf R}(x,\tau)  = \theta_{0{\bf R}}(\tau) + \Theta_{\bf R}(x,\tau),
\end{equation}
where $\Theta_{\bf R}(x,\tau)$ describes phase fluctuations (phonons)
of wave number $q\neq 0$. The operator $\theta_{0{\bf R}}$ is canonically
conjugate to the number operator $N_{\bf R}$, \ie:
$[\theta_{0{\bf R}},N_{\bf R'}] = i\delta_{{\bf R},{\bf R'}}$.
Introducing (\ref{repl}) into
(\ref{action-noMott}),  the quadratic part of $S$ reduces to
\begin{eqnarray}
 \fl S_{0}\left[\theta_0,\Theta\right] = \frac{KL}{2\pi v_s} \sum_{\bf R} \int^{\hbar\beta}_0 d\tau\:
 \left( \partial_\tau \theta_{0{\bf R}}(\tau)\right)^2 \nonumber \\
 + \frac{K}{2\pi}\sum_{\bf R} \int^{\hbar\beta}_0 d\tau\: \int^L_0 dx \:
 \left[ \frac{1}{v_s} \left(\partial_{\tau}\Theta_{\bf R}(x,\tau) \right)^2 + v_s \left(\partial_x \Theta_{\bf R}(x,\tau)  \right)^2 \right].
\end{eqnarray}
 The hopping term (\ref{eq:sint}), however, couples $\theta_{0{\bf R}}(\tau)$ and
$\Theta_{\bf R}(x,\tau)$ in a non-linear fashion:
\begin{equation}
\fl S_{J}\left[\theta_{0{\bf R}}, \Theta_{\bf R}\right] = \frac{J}{2\hbar}
{\cal A}_B \rho_0
 \sum_{\langle{\bf R},{\bf R'}\rangle}\int^{\hbar\beta}_{0} d\tau \int^{L}_0 dx \: \left[
 e^{i\left(\theta_{0{\bf R}}(\tau) - \theta_{0{\bf R'}}(\tau)\right)} \:
 e^{i\left(\Theta_{\bf R}(x,\tau) - \Theta_{\bf R'}(x,\tau)\right)} + {\rm c.c.} \right]
\end{equation}
 In order to obtain an effective action in terms of
$\theta_{0{\bf R}}(\tau)$ only we need to integrate out $\Theta_{\bf
R}(x,\tau)$. We shall do this perturbatively assuming that $J/\mu_{1D}$ is small.
Thus, let us define:
\begin{equation}
 e^{-S^{\rm eff}_J[\theta_0]} = \langle e^{-S_J} \rangle_{\Theta} =
 \exp\left[ - \langle S_J \rangle_{\Theta} + \frac{1}{2!}  \langle \left(S_J - \langle S_J \rangle_{\Theta} \right)^2 \rangle_{\Theta} + \cdots \right]
\end{equation}
where we have used the cumulant expansion and employed the following notation:
\begin{equation}
 \langle A \rangle_{\Theta}  = \frac{\int \left[d\Theta\right] \,
 A[\Theta]\: e^{-S_0\left[\Theta\right]}}{\int \left[d\Theta\right]
 \, e^{-S_0\left[\Theta \right]}}.
\end{equation}
for a given functional of $\Theta$, $A[\Theta]$.
The lowest order contribution to $S^{\rm eff}_J[\theta_0]$ comes
from the term $\langle S_J \rangle_{\Theta}$, which yields:
\begin{eqnarray}
 S^{\rm eff}_J \simeq -\frac{J}{\hbar} {\cal A}_B \rho_0 L \sum_{\langle{\bf
 R},{\bf R'}\rangle} e^{-\langle \Theta^2_{\bf R}(0)
 \rangle_{\Theta}} \int^{\hbar\beta}_{0} d\tau
 \cos \left[
 \theta_{0{\bf R}}(\tau) - \theta_{0{\bf R'}}(\tau)\right]
\end{eqnarray}
Using that $\langle \Theta^2_{\bf R}(0) \rangle_{\Theta}=
\frac{1}{2} K^{-1}\ln(L/a_0)$, where $a_0$ is the short-distance cut-off,
the above expression leads to:
\begin{eqnarray}
 S^{\rm eff}_J  =  -\frac{J}{\hbar} {\cal A}_B (\rho_0 L)
 \left(\frac{L}{a_0}\right)^{-\frac{1}{2K}} \sum_{\langle{\bf R},{\bf
 R'}\rangle} \int^{\hbar\beta}_{0} d\tau
 \cos \left[ \theta_{0{\bf
 R}}(\tau) - \theta_{0{\bf R'}}(\tau)\right]
\end{eqnarray}
Assuming that each 1D tube is described by the Lieb-Liniger model,
${\cal A}_B = (a_0 \rho_0)^{1/2K} \simeq
(K/\pi)^{1/2K}$~\cite{cazalilla_correlations_1d},
where  $a_0 \simeq \hbar v_s/\mu$ (\ie approximately the
healing length, $\xi$). Hence, the effective hopping energy of the
quantum-phase model is  $E_J = E_J(N_0) \simeq J (N_0)^{1-1/2K}$,
where we have used that $\rho_0 L = N_0$. The complete
effective action reads:
\begin{eqnarray}
 S^{\rm eff}[\theta_0] & =& \frac{\hbar}{2E_C} \int^{\hbar\beta}_0 d\tau \:
 \left( \partial_\tau \theta_{0{\bf R}}(\tau)\right)^2  \nonumber \\
 & & - \frac{E_J}{\hbar} \sum_{\langle{\bf R},{\bf
 R'}\rangle} \int^{\hbar\beta}_0 d\tau \: \cos \left[ \theta_{0{\bf
 R}}(\tau) - \theta_{0{\bf R'}}(\tau)\right] + O(J^{2}),
\end{eqnarray}
where the charging energy $E_C = \hbar \pi v_s/KL$. This euclidean action
corresponds to the following Hamiltonian (we restore the chemical
potential term):
\begin{eqnarray}
 H_{\rm eff} &= & \frac{E_C}{2} \sum_{\bf R} \left(N_{\bf R} -  N_0
 \right)^2  - \mu \sum_{\bf R} N_{\bf R}  \nonumber \\
& &-  E_J
 \sum_{\langle{\bf R},{\bf R'}\rangle} \cos \left[ \theta_{0{\bf
 R}}(\tau) - \theta_{0{\bf R'}}(\tau)\right] + O(J^2).
\end{eqnarray}
This is the quantum-phase model used in \sref{sec:2DMott}.

\vspace{1cm}

\bibliography{totphys,longbose}

\end{document}